\documentclass[12pt]{article}
\pdfoutput=1

\usepackage{amsmath}
\usepackage{amsthm}
\usepackage{amssymb}
\usepackage{graphicx}
\usepackage{bm}
\usepackage{pdfpages}
\usepackage{dcolumn}
\usepackage{listings}
\usepackage{setspace}
\usepackage{pdflscape}
\usepackage[bottom]{footmisc}
\usepackage{geometry}
\usepackage{ifpdf}
\usepackage[utf8]{inputenc}
\usepackage[T1]{fontenc}
\usepackage[english]{babel}
\usepackage{csquotes}
\usepackage{multicol}
\usepackage{tabularx}
\usepackage{makecell}
\usepackage{multirow}
\usepackage{natbib}
\usepackage{jf}
\usepackage{changepage}
\usepackage[titletoc]{appendix}
\usepackage{caption,subfig}
\usepackage{pgfplots}
\pgfplotsset{compat=1.18}
\usepackage{rotfloat}
\usepackage{float}
\usepackage{booktabs}
\usepackage{longtable}
\usepackage{tabularx}
\usepackage{array}
\usepackage{enumitem}
\usepackage{comment}
\usepackage{silence}
\usepackage[super]{nth}
\usepackage{rotating}
\usepackage{afterpage}
\usepackage{chronology}
\usepackage[nolists,nomarkers]{endfloat}
\DeclareDelayedFloatFlavor{sidewaystable}{table}
\DeclareDelayedFloatFlavor{sidewaysfigure}{figure}

\usepackage{color}
\definecolor{orange}{rgb}{1,0.5,0}
\usepackage{indentfirst}
\usepackage{etoolbox}
\usepackage{titlesec}
\usepackage[unicode=true,
  pdfusetitle,
  bookmarks=true,
  bookmarksnumbered=false,
  bookmarksopen=false,
  breaklinks=false,
  pdfborder={0 0 0},
  pdfborderstyle={},
  backref=false,
  colorlinks=true,citecolor=blue,
  breaklinks]{hyperref}

\makeatletter

\AtBeginDocument{

}
\makeatother

\date{}

\titleformat{\section}{\singlespace\normalfont\Large\bfseries}{\thesection .}{1em}{}
\titleformat{\subsection}{\singlespace\normalfont\large\itshape}{\thesubsection .}{1em}{}
\titleformat{\subsubsection}{\singlespace\normalfont\itshape}{\thesubsubsection .}{0.5em}{}
\renewcommand{\thesection}{\arabic{section}}
\renewcommand{\thesubsection}{\arabic{section}.\Alph{subsection}}
\renewcommand{\thesubsubsection}{\arabic{section}.\Alph{subsection}.\arabic{subsubsection}}
\renewcommand{\thetable}{\arabic{table}}

\hypersetup{
  pdftitle={Talking to Digital Twins: Selective Disclosure and Belief Measurement in Financial Social Media},
  pdfauthor={Bowles, Duch, and Sorescu},
  pdfsubject={Finfluencers},
  pdfkeywords={Finfluencers, Social Media, Artificial Intelligence, Digital Twins}
}

\newcolumntype{P}[1]{>{\raggedright\arraybackslash}p{#1}}
\title{\vspace{-2.0cm}\textbf{\large Talking to Digital Twins: Selective Disclosure and \\ \vspace{-0.6cm} Belief Measurement in Financial Social Media}}
\author{Boone Bowles, Raymond Duch, and Sorin Sorescu\thanks{Mays Business School, Texas A\&M University (\texttt{boone.bowles@tamu.edu}), Nuffield College, University of Oxford (\texttt{raymond.duch@nuffield.ox.ac.uk}), and Mays Business School, Texas A\&M University (\texttt{ssorescu@mays.tamu.edu}). The authors thank Avanidhar Subrahmanyam for many patient discussions and significant suggestions for improvement. This project was supported in part by the Adam C. Sinn '00 Center for Investment Management, with additional support from several Mays Business School research grants, and from the Texas A\&M University Office of the Provost. We are grateful to each of these units for supporting the data collection, computing resources, and research infrastructure required for this project. We are also grateful for the generous research funding support provided by the University of Oxford ``Talking to Machines Project,'' the Swiss National Science Foundation (Grant No. 100018M-215519), and Nuffield College, Oxford. All errors are our own. \textcopyright\ 2026 Boone Bowles, Raymond Duch, and Sorin Sorescu.}}

\begin{document}
 \newpage


\date{July 30, 2026}
\clearpage\maketitle

\begin{center}
\textbf{ABSTRACT}
\end{center}
\singlespacing
\noindent
Social media affect financial markets, but public posts by financial media personas are voluntary disclosures. What is not disclosed is therefore usually unobserved. We address this measurement problem by conducting repeated, real-time interviews of ``digital twins'' built from monitored finfluencers' X accounts under a fixed protocol. The interviews recover stock-level public-persona belief proxies even when no public recommendation is made. Because the interviews are generated and archived before the relevant return windows, the design avoids the look-ahead bias that arises when LLMs are queried ex post. The evidence shows that information obtained from these digital-twin interviews predicts the cross section of large-cap stock returns in the expected direction. Repeated real-time interviews therefore show how selective disclosure can be turned into measurable panels of market views.
\\


\pagebreak

\clearpage
\newpage

\titlepage
 
 \vspace*{.5in}

 \begin{large}
 \begin{center}
 Abstract

 \vskip .3in

{\bf Talking to Digital Twins: Selective Disclosure and \\ Belief Measurement in Financial Social Media}
\vspace{.2in}
\end{center}
\end{large}
\vspace{.2in}

\noindent   \onehalfspacing
Social media affect financial markets, but public posts by financial media personas are voluntary disclosures. What is not disclosed is therefore usually unobserved. We address this measurement problem by conducting repeated, real-time interviews of ``digital twins'' built from monitored finfluencers' X accounts under a fixed protocol. The interviews recover stock-level public-persona belief proxies even when no public recommendation is made. Because the interviews are generated and archived before the relevant return windows, the design avoids the look-ahead bias that arises when LLMs are queried ex post. The evidence shows that information obtained from these digital-twin interviews predicts the cross section of large-cap stock returns in the expected direction. Repeated real-time interviews therefore show how selective disclosure can be turned into measurable panels of market views.
\\



\newpage

\onehalfspacing
\setcounter{page}{1}
\pagenumbering{arabic}
\section{Introduction}\label{sec: Introduction}

\noindent A central premise of financial economics is that markets aggregate dispersed information through prices. In Hayek's (\citeyear{Hayek1945}) formulation, prices act as a mechanism for communicating and coordinating knowledge that is distributed across agents. \cite{GrossmanStiglitz1980} point out that when information is costly, fully revealing prices would eliminate the incentive to acquire it, implying that some degree of inefficiency must persist in equilibrium.  More recent research emphasizes that beyond cost, limited attention also causes imperfect adjustment of prices to new information, leading to predictable return dynamics \citep{hong1999gradual, Sims2003, Reis2006}. 

The rise of digital communication has further transformed the link between information and prices by enabling decentralized diffusion of price-relevant content. Investors encounter market narratives, stock recommendations, and trading rules of thumb through creators (``finfluencers'') on popular platforms such as TikTok and X.\footnote{Recent surveys and regulatory analyses document both the scale of finfluencer activity and associated concerns about advice quality, disclosures, and investor protection \citep{EspeutePreece2024,  FINRA2025, OSC2025}.}
In modern digital environments, these activities interact with amplification via algorithms and influencer popularity, further enhancing transmission of opinion.  An expanding body of work  demonstrates that such  content can also move prices, predict near-term returns, and affect market efficiency \citep{AntweilerFrank2004, AdamsEtAl2023FEDS, NunezMora2023, kakhbod2023finfluencers, Gordillo2024}.

Although the literature indicates that choices of social media creators are economically meaningful, these posts create a measurement problem.  Specifically, silence is ambiguous. A creator may be silent because she has no view, because the stock is not salient, because the view is uncertain, because the post would not attract attention, or because disclosure would be strategically costly. Thus,  a post-based measure mixes two objects: the underlying view and the decision to reveal it. The absence of a recommendation may be as important as the recommendation itself, yet silence is normally unobserved as a belief state.

In this paper, we address the challenge of selective disclosure by introducing a novel measurement approach based on structured elicitation of beliefs. Rather than passively observing what finfluencers choose to reveal in public posts, we recover their views using repeated, standardized interviews applied to AI-based ``digital twins'' constructed from their observable social-media personas. Advances in large language models enable us to create virtual representations of real personas that can be queried systematically over time.\footnote{LLMs can now condition responses on rich text supplied at the time of the query. Several developments make this possible. First, transformer models use attention mechanisms to relate different parts of a text prompt to one another, allowing the model to connect an account biography, recent posts, a ticker, and a question in the same response \citep{vaswani2017attention}. Second, large-scale pretraining and in-context learning allow a pretrained model to perform new tasks from instructions and examples in the prompt, without training a separate model for each account \citep{brown2020language}. Third, instruction tuning and human-feedback training make the model better at following detailed prompts and returning structured answers, which matters because our protocol asks for scores, confidence, speculation, horizons, and catalysts \citep{ouyang2022training}. Fourth, retrieval and context augmentation show how model outputs can account for external information supplied at query time, rather than relying only on knowledge stored in model parameters \citep{lewis2020retrieval}. Finally, recent social-science work shows that LLM agents can produce meaningful persona-conditioned responses when given rich background information, including interviews, memories, or assigned personas \citep{horton2023large, park2024generative}.} By implementing a fixed daily interview protocol, we generate a panel dataset of forward-looking beliefs that is time-stamped, comparable across influencers, and not restricted to voluntary disclosure.

Specifically, we implement a method to ask finfluencers questions instead of just waiting for them to post. Each day, we run repeated, \emph{real-time} interviews of ``digital twins'' of finfluencers on X.\footnote{The real-time aspect of this study is similar to epidemiological research in medicine such as the Framingham Heart Study \citep{mahmood2014framingham,andersson201970}, where a cohort of subjects is followed forward to ascertain outcomes, with no look-ahead bias.} A \emph{digital twin} in this setting is an artificial interviewee built from account information and monitored recent content and instructed to respond from the perspective of the associated public persona. The interview is daily and standardized; public posts are voluntary and non-uniform. The design, therefore, separates the recovery of views from the creator's decision to disclose them publicly. 

We use the term digital twin operationally. The twin does not reveal what the human finfluencer privately believes or actually owns. It is a standardized representation of the finfluencer's public persona, built from account-specific public information and interviewed under the same protocol as the other twins.

We ask each digital twin to share views about broad stock market conditions and to make specific buy, hold, or sell recommendations about each large-cap stock in our sample. The same monitored personas are queried repeatedly. The questions are fixed, and so is the cross-section of stocks. The answers are generated contemporaneously, timestamped, and aligned to market event time before the relevant stock return outcomes are observed.  The timing of these interviews is central to our research design because pre-trained LLMs create a special challenge for financial predictability tests. If a researcher asks a model today to reconstruct what a market participant would have believed last year, the model may have been trained on, or have access to, information that was not available at the time. That is look-ahead bias in a new form \citep{sul2017trading, ludwig2024large,  SarkarVafa2024}. Our interviews are not run ex post. They are generated live, timestamped, archived, and then compared with subsequent returns. The tests therefore evaluate interview records that existed before the return windows being studied.

The interview protocol has two immediate advantages. The first is a coverage gain. Because we query monitored personas directly and repeatedly, rather than relying only on voluntary public posts, the interviews recover stock-level views for many names that never receive an actual public recommendation. They also create a daily path of views around names that eventually become public recommendations, including days before and after the disclosure date. The second advantage is a measurement gain. The stock-pick interviews not only ask whether a digital twin is buy-leaning or sell-leaning on a given stock. They also ask how confident the response is, whether the recommendation is speculative, and what catalyst supports the view. This allows us to separate three distinct states of the belief panel: direction, disagreement, and uncertainty. Direction tells us which stocks the panel favors. Disagreement tells us which stocks divide the panel. Uncertainty tells us where the panel lacks grounded conviction. These distinctions are difficult to recover from public posts alone, especially when no post is made.

We first validate whether the digital twin produces responses consistent with its human finfluencer counterpart. We examine overlaps (same day, same ticker, same account) between recommendations made by human finfluencers and those made by their digital twin counterparts. Digital twins' recommendations match those of their human counterparts $91.5\%$ of the time. Moreover, before later public recommendations appear, digital twins' interviews already lean toward the direction that eventually becomes public. And, after removing common date-question effects from the macro interviews, the digital twin's own identity still explains a substantial portion of the variation in the outcome of these macro interviews. These validations show that the information produced by digital-twin interviews is not merely generic market commentary.

We then use the results from these digital-twin interviews to predict cross-sectional stock-level returns over the next ten trading days. For each stock-event, we summarize the interview output with \emph{Net Buy Share}, a simple buy-minus-sell measure: the share of digital-twin recommendations that are buys minus the share that are sells. We find that stocks with more buy-leaning interview recommendations outperform their benchmark, while stocks with more sell-leaning recommendations underperform, suggesting that recommendations elicited from finfluencers' digital twins contain relevant stock-selection information. Importantly, this predictability is strongest in the silent region, i.e., for stocks that are not explicitly mentioned by the human finfluencers. This silent region accounts for $84.8\%$ of the interview stock-event observations.

Within the silent region, in controlled specifications, a ten-percentage-point increase in this buy-minus-sell interview measure predicts $24$ basis points higher future excess return at the five-trading-day horizon and $50$ basis points at the ten-trading-day horizon. The interviews, therefore, do not merely restate public recommendations. They recover stock-level views when the public channel is silent. This evidence is useful for the measurement argument because the covered universe consists of a set of large, liquid, and heavily followed firms for which short-horizon cross-sectional predictability is usually difficult to detect \citep{fama2008dissecting}. We therefore do not interpret the result primarily as a trading-strategy claim. Rather, the fact that the interview signal has predictive content in this relatively efficient part of the market is evidence that the digital-twin instrument is recovering economically meaningful belief proxies.

This finding shapes the core contribution of our study. The paper is not primarily a horse race between scraped posts and AI interviews, and it is not mainly a claim that finfluencers can pick stocks. The core contribution is that standardized digital-twin interviews make a previously unmeasured part of selective disclosure visible. Return predictability is important because it shows that the recovered silent-region beliefs have economic content. In this sense, the return tests are corroborating evidence for the measurement strategy.

We use social media---the X platform in particular---as a laboratory for our research design because investors increasingly encounter market narratives, stock recommendations, and trading rules of thumb through social media creators.\footnote{Recent surveys and regulatory analyses document both the scale of finfluencer activity and associated concerns about advice quality, disclosures, and investor protection \citep{EspeutePreece2024, SEC2024IAC, FINRA2025, OSC2025}.} A growing empirical literature shows that social-media sentiment, online discussion, and influencer content can move prices, predict near-term returns, and affect market efficiency.\footnote{See \cite{AntweilerFrank2004, bollen2011twitter, sprenger2014tweets, ranco2015effects, AdamsEtAl2023FEDS, NunezMora2023, kakhbod2023finfluencers, Gordillo2024}.} Social networks also shape how information spreads, with influential users acting as key nodes in amplification and diffusion \citep{Hong2004social, engelberg2018social, cookson2019social}.

The existing social-media finance literature has made substantial progress by extracting sentiment, attention, disagreement, and recommendations from observed text. However, the observed text is only the part of the creators' equity research analysis that becomes public. The unpublished part of their analysis---the silent region---is not well understood. The distinction between public posts and the silent region is important because the aggregation of information into prices is neither frictionless nor complete.

Our work contributes in several dimensions. First, we contribute to the literature on investor beliefs, sentiment, and disagreement. Models of noise trading and sentiment-driven demand show that optimism and pessimism can move prices away from fundamentals, especially when limits to arbitrage prevent rational traders from correcting mispricing \citep{delong1990noise, barberis1998model, shleifer2000inefficient}. Empirical evidence shows that investor sentiment predicts returns, volatility, and trading volume, particularly for hard-to-value securities \citep{baker2006investor}. Disagreement is also central to asset pricing, with dispersion in beliefs linked to trading volume, volatility, and expected returns \citep{miller1977risk, hong2007disagreement, daniel2023dynamics}. Our setting contributes by measuring public-facing belief proxies for a class of retail-facing intermediaries whose disclosures are episodic, strategic, and incomplete.

Second, our paper contributes to the literature on attention and retail-facing information. Investors are more likely to buy stocks that attract their attention, generating predictable demand imbalances \citep{barber2008all, da2011search}. In digital environments, search intensity and online activity provide direct proxies for attention and have been shown to predict trading volume and returns \citep{da2011search}. We take a different approach. Rather than harvesting attention proxies from observed behavior, we repeatedly query the same monitored personas under a fixed protocol. That step widens coverage when attention is absent from the public record and gives the researcher a clean separation between views and disclosure decisions.

Third, the paper contributes to the literature on belief and expectation measurement. Directly elicited expectations help explain investor behavior and asset prices \citep{GreenwoodShleifer2014, BordaloGennaioliLaPortaShleifer2019, GiglioMaggioriStroebelUtkus2021}. Our setting differs from standard surveys in two ways. The respondents---the digital twins---are public financial intermediaries rather than households or institutions, and the instrument can be run daily at stock-level scale. The paper therefore shows how a survey-like design can be used to measure public-facing belief proxies for market participants who are difficult to survey directly and whose naturally observed communication is selective.

Fourth, the paper contributes to work on retail investors and non-traditional information sources. Retail order flow contains information about future returns, and retail trading has become increasingly important in modern markets \citep{barrot2016information, boehmer2021retail}. Episodes such as the GameStop trading frenzy show how coordinated retail attention, often facilitated by social media, can produce large and persistent deviations from fundamental values \citep{cookson2022gamestop}. Sentiment extracted from online forums and social-media platforms also predicts cross-sectional and aggregate market behavior \citep{sprenger2014tweets}. Our contribution to this literature is on the interpretation of retail-facing market signals \citep{barber2008all, da2011search, CooksonNiessner2020, BarberHuangOdeanSchwarz2022}. The same interview panel recovers aggregate market stance, disagreement, uncertainty, and stock-selection direction. It also shows that aggregation level matters: broad day-level optimism looks like market mood and carries a contrarian sign, while within-day ranking carries a positive stock-selection relation.

Fifth, the paper contributes to the literature on financial text analysis, large language models, and digital twins. Financial text has been used to measure firm fundamentals, product-market relations, political risk, expected returns, and news content \citep{LoughranMcDonald2011, HobergPhillips2016, HassanHollanderVanLentTahoun2019, chen2022expected, didisheim2026inefficient}. Recent work uses large language models for return prediction, financial statement analysis, financial forecasting, and digital-twin applications.\footnote{See \cite{LopezLiraTang2024, KimMuhnNikolaev2024, JadhavMirza2025, XiaoEtAl2025Retrieval, YaoYangWangZhang2023, Iliuta2024DTreview, he2025chronologically, koijen2026assessing, chenetal2025purdue, lehner2026chatgpt}.} Our method is different from a retrospective classification of observed text. The unit of observation is an LLM-elicited response to a non-selective query, not a voluntary post or document. Moreover, the responses are generated and archived in real time, so the outcomes are traced after the interviews rather than reconstructed from archival prompts.

Finally, our paper relates to the broader literature on media and financial markets. Prior research on traditional media shows that news coverage and tone are associated with trading and market outcomes \citep{Tetlock2007, tetlock2008more, EngelbergParsons2011}, while subsequent research studies analogous information in online forums and social-media content \citep{AntweilerFrank2004, sprenger2014tweets, CooksonNiessner2020, kakhbod2023finfluencers}. Our setting emphasizes a feature that is especially salient in social media: communication is voluntary and therefore selectively observed. Rather than measuring only the views that finfluencers choose to disclose, we repeatedly elicit standardized public-persona belief proxies under a fixed protocol, including for stocks about which the corresponding finfluencers remain silent. To our knowledge, ours is the first study to measure and evaluate stock-level public-persona belief proxies in this silent region.

The remainder of the paper proceeds as follows. Section \ref{sec: Digital-Twin Interviews Under Selective Disclosure} develops the selective-disclosure problem and explains why digital-twin interviews are useful for measuring the silent region. Section \ref{sec: The Interview Protocol as a Research Instrument} describes the monitored account pool, interview protocol, real-time implementation, and belief-proxy variables. Section \ref{sec: Validation: Does the Instrument Work?} validates the instrument and documents the coverage gain from the silent region. Section \ref{sec:xp} studies the stock-level information content of the interview responses, including direction, disagreement, uncertainty, and silent-region returns. Section \ref{sec: Macro Beliefs and Future Market Returns} studies the broader, market-level information content of the interviews. Section \ref{sec: Discussion and Applications Beyond Financial Social Media} discusses broader applications and limitations of our research design, and Section \ref{sec: Conclusion} concludes.



\hypertarget{Digital-Twin Interviews Under Selective Disclosure}{%
\section{Digital-Twin Interviews Under Selective Disclosure}\label{sec: Digital-Twin Interviews Under Selective Disclosure}}

A social media post is a form of disclosure. When a finfluencer posts about a stock, the researcher observes both the content of the post and the decision to make that content public. When the same finfluencer does not post, the researcher observes silence. Silence is ambiguous. It may mean that the finfluencer has no view, that the view is uncertain or weakly held, that the post would not attract attention, or that disclosure would be costly. Thus, a post-based measure mixes the underlying view with the decision to reveal it.

The disclosure decision is unlikely to be random. Finfluencers compete for attention and audience, making them more likely to post about stocks that readers are already likely to notice: firms in the news, firms with high trading activity, or firms with sharply moving prices.\footnote{This attention motive is consistent with work on online influence, influencer value, retail attention, media attention, and social diffusion \citep{fainmesser2021market, de2017marketing, barber2008all, da2011search, fang2009media, EngelbergParsons2011, berger2012makes, muchnik2013social, vosoughi2018spread}.} Recent finfluencer evidence points in the same direction. Financial influencers are more likely to recommend stocks with strong past performance, high trading volume, and high intraday activity \citep{kakhbod2023finfluencers, lalwani2025finfluencer}. Disclosure and reputation models also imply that public communicators may withhold or soften views when disclosure is costly, weakly grounded, inconsistent with audience priors, or reputationally risky.\footnote{See \cite{verrecchia1983discretionary}, \cite{dye1985disclosure}, \cite{scharfstein1990herd}, \cite{trueman1994analyst}, \cite{graham1999herding}, \cite{hong2000security}, \cite{mullainathan2005market}, and \cite{gentzkow2006media}.}

A public-post sample is therefore likely to overrepresent salient firms and high-conviction views. It is likely to underrepresent less familiar stocks, weaker views, and cases where the finfluencer has a view but no reason to disclose it. A scrape of public posts can measure what was said. It cannot measure the silent region created by selective disclosure.

Our empirical design changes the data-generating process. Instead of waiting for finfluencers to post, we ask their digital twins each day a fixed set of questions about a set of pre-determined large-cap stocks. For each monitored finfluencer, the digital twin is constructed from observable material associated with the public persona on X: profile information, recent monitored content, and finance-relevant context. The output from daily digital-twin interviews is a standardized public-persona belief proxy for a given finfluencer.

The term \emph{digital twin} comes from engineering, where virtual models of physical objects or systems are used to simulate, predict, and optimize the behavior of real-world objects \citep{tao2018digital, jones2020characterising, lattanzi2021digital, abdelrahman2025digital}. 

Digital twins of human subjects have also begun to appear in medical settings, using data such as health records and patient histories \citep{eadie2026arrival, wang2024twin, angelopoulos2023prediction, de2026efficient, broska2025mixed}. In these studies, the conditioning material is static: twins are built at the onset of the research design from baseline medical data, health histories, demographics, and patient interviews. Our setting requires a more dynamic version of this idea because the information environment changes as markets move and as finfluencers' accounts produce new content. In our setting, the same digital twin must, therefore, be interviewed on different dates using updated context.

Our research design also differs from an emerging literature that uses LLMs to simulate stock market outcomes or predict stock returns \citep{mahdavi2025integrating}. That literature adopts a generic approach to return predictability: it does not attempt to construct digital twins conditioned on the public personas of actual human agents. Instead, it aggregates information from public sources to generate buy, sell, or hold recommendations, extract sentiment, construct alpha factors, or simulate markets. \emph{Alpha-GPT 2.0}, for example, is a project in which LLM agents help identify, test, combine, and analyze trading signals \citep{yuan2024alpha}. \emph{TwinMarket} creates generic synthetic traders and lets them interact in a simulated market where they recreate well-known anomalies such as fat tails, bubbles and crashes \citep{yuzhe2026twinmarket}. Although these synthetic traders are trained on real transaction data, they are not digital twins of human traders. They are not conditioned on anyone’s specific public persona and there is no one-to-one mapping between them and corresponding human traders.\footnote{More recently, \cite{LiKimCucuringuMa2026} provide a related benchmark for LLM-based investing strategies, showing that apparent LLM trading advantages often deteriorate when evaluated over longer horizons, broader symbol universes, and with explicit controls for survivorship, look-ahead, and data-snooping biases. Their paper evaluates LLMs as trading systems; our design uses LLMs as account-conditioned interview instruments to measure public-persona belief proxies when finfluencers do not disclose a recommendation.}

Our empirical design and research purpose are different. We do not use the LLM as a trading engine or to simulate market interactions. We use it as an interview tool to recover public-persona belief proxies when the public channel is silent. Moreover, we conduct our research in \emph{real time}. Ex post LLM prompting can create look-ahead bias because pretrained models may have been exposed to future returns, future news, or later commentary \citep{glasserman2023assessing, SarkarVafa2024, he2025chronologically, LopezLiraTang2024}. Thus, our interviews are generated live, timestamped, and archived before the relevant return window begins.\footnote{\cite{koijen2026assessing} similarly emphasize the need for real-time experiments when using AI systems in asset pricing, but their study uses LLMs to explain contemporaneous stock-price reactions to observed earnings announcements and transcripts; our study uses LLMs as interview instruments to measure belief proxies generated by finfluencers' digital twins when public disclosure is absent.}

Our digital twin methodology offers three unique advantages for studying selective disclosure. First, it expands coverage because we can elicit responses even when no public recommendations exist. Second, it separates disclosure from measurement: public posts reflect both a view and a decision to reveal it, while interview responses are generated \emph{on demand} by a fixed protocol, whether or not the finfluencer decided to post. Third, it separates four different components of the public-persona belief proxies: stock-pick direction, disagreement, uncertainty, and market-level sentiment. Each of these four components is useful because it answers a different empirical question---whether a stock is favored, whether views about it are contested, whether the panel lacks grounded conviction, and whether broad market mood is high or low---and the fact that these components predict different future outcomes further corroborates that the interviews uncover economically relevant information in the silent region.

The next section describes the mechanics of our design: how we find and monitor finfluencer accounts, run the daily digital-twin interviews, and turn the responses into stock-level and macro belief-proxy variables.



\hypertarget{The Interview Protocol as a Research Instrument}{%
\section{The Interview Protocol as a Research Instrument}\label{sec: The Interview Protocol as a Research Instrument}}

Our research design begins with a monitored sample of finfluencer accounts on X. The sample was assembled in three stages. First, we compiled candidate accounts from public X-handle lists in third-party directories published by industry outlets such as The CFO Club, Investopedia, Investing.io, and Yahoo Finance. Second, we manually screened each candidate account against a topical inclusion rule, retaining only accounts whose primary content consisted of directional recommendations or analysis of individual stocks, indices, options, or other derivatives, and excluding real-estate, personal-finance, pure macroeconomic commentary, news-aggregator, and anonymous pump-and-dump accounts. Finally, remaining accounts were retained only if they had at least $5{,}000$ followers, met a minimum post-history threshold, and passed a content-consistency check. Our final sample contains $81$ finfluencer accounts. Importantly, these accounts were selected in June 2025, before any interviews or outcome data were collected, preventing contamination of the research experiment.\footnote{The aim of the finfluencer selection process was not to create a census of every finance account on X, but to build a panel of accounts whose public personas regularly speak about markets and stocks. For a full description of our data-generating process---finfluencer selection, daily interviews, timing rules, filters, and variable definitions---see Section \ref{appx: Data Appendix} of the Internet Appendix.}

We then create a digital twin for each of the $81$ finfluencers. A \emph{digital twin} in our experimental design is an account-conditioned interview respondent. For each finfluencer, the LLM is given information about the account’s public persona: profile metadata, recent monitored posts, and finance-relevant context available at the time of the interview. This information gives the model a working memory of the account: what kinds of stocks the account discusses, what language it uses, what themes it emphasizes, and how it tends to frame market views. We ask each twin the same questions under a fixed protocol. Modern LLMs make this design feasible because they can use information supplied in the prompt at the time of the query, follow detailed instructions, and return account-specific answers without training a separate model for each account \citep{brown2020language, ouyang2022training, lewis2020retrieval}. The same LLM is therefore used for all finfluencers, but the information supplied to the model differs by account and by date. Importantly, the conditioning material is updated over time as the finfluencer account produces new content and as market conditions change. The resulting responses are time-varying public-persona belief proxies of human finfluencers: they summarize what the public persona says when asked a standardized question at a known time. After creating the digital twins, we repeatedly interview them daily in real time, instructing each of them  to answer as the finfluencer would answer if asked the same question at the same time.

The daily interview has two main branches: stock-pick and macro. The \textit{stock-pick branch} asks every digital twin for a view on stocks in the S\&P 500 index, recording both a recommendation score (ranging from $0$ to $100$) and a speculation score ($0$ to $100$).\footnote{A recommendation score of $50$ is neutral; higher scores are more buy-leaning and lower scores are more sell-leaning. The speculation score---where a higher score indicates greater speculation---measures whether the twin's recommendation score is based on specific information in the twin's profile or is an unsupported guess.} The \textit{macro branch} asks every digital twin about broad market conditions, including recession risk, business conditions, current investor sentiment, stock-market direction, and expected bond-rate direction. The responses from the two interview branches produce two panels: an account $\times$ date $\times$ stock panel of stock-level views, and an account $\times$ date panel of broad-market views. The digital-twin interviews were conducted repeatedly from December 14, 2025 to March 15, 2026.\footnote{Section \ref{appx: Data Appendix} of the Internet Appendix provides additional details about the structure, content, and timing of these interviews. The interviews were intended to be conducted daily; however, due to operational backlogs and some missed days around holidays, the sample is almost daily.}

The timing of the interviews is also critical to our research design, for two reasons. First, to prevent contamination, all interviews are conducted \textit{after} finfluencer selection, so the interview responses cannot affect which accounts enter the sample. Second, studies that predict stock returns using LLMs can suffer from the look-ahead bias if the model is asked today to consider an earlier market date. Our design avoids this problem by generating and archiving every digital-twin interview in real time and assigning each response to the first trading day on which an investor could have observed and acted on it. This timing convention keeps the experiment forward-looking: the sample is fixed before the interviews, and the return tests use only information that existed before the return window begins.\footnote{Section \ref{appx: Data Appendix} of the Internet Appendix provides additional details and descriptive statistics about the timing of these interviews and their relation to the return-measurement period.}

Table \ref{tab:draft_maintext_sample_summary} provides descriptive statistics of the macro and stock-pick branches. The macro panel monitors $81$ finfluencer accounts across $53$ trading days from December 15, 2025 through March 16, 2026. The stock-pick sample uses the same $81$ accounts and contains $21{,}410$ stock-by-trading-day observations, covering $429$ unique tickers across $50$ trading days. All market data, including stock prices, returns, market capitalizations, index returns, and volatility inputs, are obtained from S\&P Capital IQ, the financial-data platform of S\&P Global Market Intelligence.\footnote{The macro and stock-pick branches are timestamped and aligned separately. Macro interviews were typically run first, while stock-pick interviews were completed later, so the $53$ macro trading days and $50$ stock-pick trading days reflect branch-specific completion times rather than missing return data.}

We also build a third data panel from the finfluencers' public stock recommendation posts. This is not an interview branch since the posts are voluntary disclosures written by the human finfluencers themselves. Instead, the public-post panel starts from observed posts that mention specific stocks and classifies the direction of the disclosed recommendation. We apply the same timing convention to these public posts, so a public recommendation and a digital-twin response are compared only when they are aligned to the same effective trading date. In the validation tests, this lets us compare public disclosures with digital-twin interview responses for the same account, ticker, and trading date.\footnote{The panel of actual recommendations was created over the same period as the interview branches: December 14, 2025 to March 15, 2026. See Section \ref{appx: Data Appendix} of the Internet Appendix for additional details.}

\begin{table}[!htbp]
\caption{Interview Sample and Analysis Coverage}
\label{tab:draft_maintext_sample_summary}
{\footnotesize{}This table summarizes the two interview branches used in the main analyses after timing alignment and final sample filters. Both branches use the 81 finfluencer accounts observed in the shared macro/stock-pick sample; the median account has 36,118 followers in the latest macro snapshot. The macro branch collapses cleaned interviews to trading-day observations. The stock-pick branch collapses account-stock interview responses to stock-by-trading-day observations in the recurring ticker universe.}
{\footnotesize\par}
\vspace{10pt}
\centering{}
{\footnotesize
\begin{tabular}{lcc}
\hline\hline
\noalign{\vskip 4pt}
 & \makebox[3.2cm][c]{Macro branch} & \makebox[3.2cm][c]{Stock-pick branch} \\
\hline
\noalign{\vskip 4pt}
Analysis observations & 53 & 21,410 \\
Main observation unit & Trading day & Stock $\times$ trading day \\
Unique stock tickers & -- & 429 \\
Analysis days & 53 trading days & 50 trading days \\
Analysis period & 2025-12-15 to 2026-03-16 & 2025-12-15 to 2026-03-16 \\
\hline\hline
\end{tabular}
}
\end{table}


From the stock-pick panel we construct six daily output variables by aggregating interview responses across the 81 digital twins. These six variables summarize the direction of the twins' recommendations as a group (buy, hold or sell), as well as the aggregate level of conviction, disagreement, and uncertainty prevailing among these twins. The six variables are as follows:\footnote{Precise construction details of these variables are provided in Section \ref{appx: Data Appendix} of the Internet Appendix.}

\begin{enumerate}

    \item \emph{Tilt} is the average recommendation score centered around 50, so positive values are buy-leaning and negative values are sell-leaning.
    
    \item \emph{Net Buy Share} is the share of meaningful buy recommendations minus the share of meaningful sell recommendations, where meaningful recommendations are at least 15 points away from neutral (50).
    
    \item \emph{Meaningful Tilt} is the \emph{Tilt} obtained by averaging only meaningful recommendations (those that are at least 15 points different from the neutral value).
    
    \item \emph{Conviction-Weighted Meaningful Tilt} multiplies \emph{Meaningful Tilt} by the share of recommendations that are meaningful. This measure preserves the direction of the meaningful recommendations, but gives more weight to stock-dates where that direction is supported by a larger fraction of the interview panel.\footnote{We view this as the cleaner intensity measure because \emph{Meaningful Tilt} alone conditions on the subset of meaningful responses and therefore can be large even when only a small number of twins express a meaningful view. We report both measures for completeness.}
    
    \item \emph{Informative Polarization} measures disagreement among finfluencers' twins for recommendations that have low-speculation scores.
    
    \item \emph{Uncertainty Score} measures the absence of firm conviction among finfluencers' twins for a particular stock. This score rises when a stock-event has fewer meaningful low-speculation responses.

\end{enumerate}

From the macro panel we construct four daily measures of broad market sentiment and disagreement among twins:\footnote{Precise construction details of these variables and others are provided in Section \ref{appx: Data Appendix} of the Internet Appendix.}

\begin{enumerate}

    \item \emph{Net Sentiment (3 of 7)} is the share of interviewed digital twins classified as bullish minus the share classified as bearish. On any given day, a digital twin is classified bullish or bearish only when the responses to at least three of seven core questions are strong and low-speculation. The seven core questions ask about recession risk, business conditions, investor sentiment, stock-market direction, and the expected direction of interest rates. 
    
    \item \emph{Net Sentiment (3 of 6)} is identical to \emph{Net Sentiment (3 of 7)}, except that we exclude the question about the expected direction of future interest rates. 
    
    \item \emph{Disagreement IQR (3 of 7)} is the cross-sectional inter-quartile range of daily Bull scores and measures macro-level disagreement among digital twins. The Bull score is a 0--100 coding of each response to a question, with higher values meaning a more bullish market view and 50 meaning neutral. This variable is high when the digital twins give meaningfully different broad-market views on the same day.
    
    \item \emph{Continuous Sentiment} is the daily average of account-level normalized Bull scores across the seven core questions. Each answer is transformed to $(\text{Bull score}-50)/50$, so the scale runs from bearish to bullish. Account-question scores are averaged by day. Then these account-day scores are averaged across accounts to get the daily average.
    
\end{enumerate}

Table \ref{tab:draft_maintext_variable_summary} reports descriptive statistics for our main variables (output of digital twin interviews, future stock returns and future volatility). Statistics related to the \emph{stock-pick} branch are reported in Panels A and B; those for the \emph{macro} branch are reported in Panels C and D. 

Panel A of Table \ref{tab:draft_maintext_variable_summary} summarizes interview outcomes for the stock-pick branch. In that panel, $15\%$ of twin-stock responses have meaningful recommendations, on average, and \emph{Net Buy Share} averages almost $11\%$, suggesting that finfluencers' twins are generally optimistic. The high average \emph{Uncertainty Score}, $0.93$, reflects the fact that many twin-stock responses are neutral or speculative. 

Panel B summarizes stock-level market outcomes for the stock-pick branch, for the same (stock $\times$ trading-day) observations as in Panel A. Future excess returns are measured relative to the S\&P 500 and are modestly positive on average, increasing from $0.067$ percentage points over one trading day to $0.619$ percentage points over ten trading days. These averages are small relative to the cross-sectional dispersion in returns, with standard deviations ranging from $2.30$ to $7.23$ percentage points. Future excess realized volatility is also positive, consistent with individual stocks being more volatile than the aggregate market index.

Panel C of Table \ref{tab:draft_maintext_variable_summary} summarizes interview outcomes for the macro branch. \emph{Net Sentiment} averages $0.374$, meaning twins are bullish on the market, on average. However, bullishness is not unanimous, since the median \emph{Disagreement IQR} is $22$ bull-score points, indicating that meaningful dispersion exists across twins. 

Panel D summarizes market outcomes for the macro branch. S\&P 500 returns are negative on average over the sample. Thus, the positive average macro sentiment in Panel C does not simply reflect a rising market; the digital twins were bullish, on average, during a period of weak broad-market performance. VIX returns are positive on average, consistent with elevated market uncertainty during the sample window, including the geopolitical uncertainty surrounding the U.S.--Iran conflict that began on February 28, 2026.

\begin{table}[!htbp]
\caption{Summary Statistics for Main Variables}
\label{tab:draft_maintext_variable_summary}
{\footnotesize{}This table reports summary statistics for the main belief proxies, stock-level outcomes, and market variables used in the paper. Panels A and B summarize the stock-pick branch at the stock $\times$ trading-day level. Panels C and D summarize the macro branch at the trading-day level. Return rows labeled Future are measured after the aligned trading day; rows labeled Lagged are measured before it. Share, return, and realized-volatility rows labeled (\%) are reported in percentage points. Min and Max are raw extrema after the displayed sample filters. RV abbreviates realized volatility. Net sentiment, polarization, and uncertainty variables that are naturally bounded between zero and one are left on their native scale; Disagreement IQR is reported in bull-score points. N can differ across rows because some variables require market-data availability, nonmissing forward windows, or variable-specific construction screens.}
{\footnotesize\par}
\vspace{10pt}
\centering{}
{\footnotesize
\setlength{\tabcolsep}{2.2pt}
\begin{tabular}{lrrrrrrrr}
\hline\hline
\noalign{\vskip 4pt}
Variable & Mean & SD & Min & P25 & Median & P75 & Max & N \\
\hline
\noalign{\vskip 8pt}
\multicolumn{9}{l}{\textit{Panel A. Stock-pick branch belief proxy variables}} \\[-1pt]
\noalign{\hrule height 0.15pt}
\noalign{\vskip 3pt}
Recommendation Count ($n_{rec}$) & 74.06 & 9.42 & 1.00 & 73.00 & 76.00 & 77.00 & 81.00 & 21,410 \\
Meaningful Rec. Share (\%) & 15.06 & 14.31 & 0.000 & 5.33 & 10.39 & 19.74 & 100.0 & 21,410 \\
Tilt & 2.95 & 3.38 & -25.00 & 0.846 & 2.40 & 4.29 & 25.00 & 21,410 \\
Net Buy Share (\%) & 10.78 & 14.23 & -100.0 & 2.53 & 7.25 & 15.58 & 100.0 & 21,410 \\
Meaningful Tilt & 12.45 & 11.21 & -35.00 & 9.00 & 16.58 & 19.00 & 29.78 & 20,823 \\
Conv.-Weighted Meaningful Tilt & 2.13 & 3.07 & -25.00 & 0.456 & 1.42 & 2.96 & 25.00 & 20,823 \\
Informative Polarization & 0.210 & 0.329 & 0.000 & 0.000 & 0.000 & 0.438 & 1.00 & 10,536 \\
Uncertainty Score & 0.932 & 0.076 & 0.150 & 0.914 & 0.956 & 0.978 & 1.00 & 21,410 \\
\hline
\noalign{\vskip 8pt}
\multicolumn{9}{l}{\textit{Panel B. Stock-level size, returns, and volatility windows}} \\[-1pt]
\noalign{\hrule height 0.15pt}
\noalign{\vskip 3pt}
Log market cap & 10.82 & 1.23 & 2.16 & 10.02 & 10.64 & 11.46 & 15.37 & 21,410 \\
Future Excess Return, 1d (\%) & 0.067 & 2.30 & -26.95 & -1.04 & 0.022 & 1.18 & 22.46 & 21,410 \\
Future Excess Return, 5d (\%) & 0.268 & 5.09 & -32.43 & -2.39 & 0.125 & 2.82 & 42.64 & 21,410 \\
Future Excess Return, 10d (\%) & 0.619 & 7.23 & -36.70 & -3.39 & 0.403 & 4.44 & 44.93 & 21,410 \\
Future Excess RV, 5d (\%) & 4.25 & 2.80 & 0.238 & 2.50 & 3.57 & 5.16 & 39.33 & 21,410 \\
Future Excess RV, 10d (\%) & 6.24 & 3.43 & 0.925 & 4.02 & 5.43 & 7.45 & 40.81 & 21,410 \\
\hline
\noalign{\vskip 8pt}
\multicolumn{9}{l}{\textit{Panel C. Macro branch belief proxy variables}} \\[-1pt]
\noalign{\hrule height 0.15pt}
\noalign{\vskip 3pt}
Contributing Influencers per Day & 76.06 & 3.25 & 63.00 & 75.00 & 77.00 & 78.00 & 80.00 & 53 \\
Net Sentiment (3 of 7) & 0.374 & 0.087 & 0.182 & 0.317 & 0.377 & 0.447 & 0.557 & 53 \\
Disagreement IQR (3 of 7) & 22.46 & 4.54 & 13.57 & 20.00 & 21.43 & 24.29 & 37.14 & 53 \\
Net Sentiment (3 of 6) & 0.366 & 0.086 & 0.169 & 0.312 & 0.361 & 0.434 & 0.544 & 53 \\
Continuous Sentiment & 0.270 & 0.060 & 0.123 & 0.231 & 0.273 & 0.322 & 0.358 & 53 \\
High-Speculation Share, Q1--Q7 & 0.238 & 0.094 & 0.173 & 0.193 & 0.202 & 0.219 & 0.501 & 53 \\
\hline
\noalign{\vskip 8pt}
\multicolumn{9}{l}{\textit{Panel D. Market return and VIX return windows}} \\[-1pt]
\noalign{\hrule height 0.15pt}
\noalign{\vskip 3pt}
Future S\&P 500 Return, 1d (\%) & -0.040 & 0.779 & -2.06 & -0.434 & -0.008 & 0.539 & 1.97 & 53 \\
Future S\&P 500 Return, 5d (\%) & -0.240 & 1.29 & -2.45 & -1.15 & -0.402 & 0.642 & 3.13 & 53 \\
Future S\&P 500 Return, 10d (\%) & -0.883 & 1.77 & -5.31 & -2.45 & -0.524 & 0.525 & 2.04 & 53 \\
Lagged VIX Return, 5d (\%) & 0.936 & 6.29 & -9.88 & -2.41 & -0.052 & 3.89 & 23.72 & 53 \\
Lagged VIX Return, 10d (\%) & 1.43 & 9.26 & -13.65 & -6.53 & 1.29 & 5.87 & 25.94 & 53 \\
\hline\hline
\end{tabular}
}
\end{table}




\hypertarget{Validation: Does the Instrument Work?}{%
\section{Validation and Coverage Gain}\label{sec: Validation: Does the Instrument Work?}}

\subsection{Validation Against Observable Benchmarks: Does the Instrument Work?}

Before using the interview output to study market outcomes, we first ask whether the digital twins produce responses that are consistent with the public personas of the finfluencers they represent. Of course, we cannot observe a finfluencer's private beliefs, and we do not claim to do so. Instead, we validate the interview output where validation is possible. Table \ref{tab:validation_main} reports four tests that address three validation questions. First, when a finfluencer makes a public stock recommendation, does the digital-twin interview point in the same direction, ranking public buys above public sells? Second, before a finfluencer's public recommendation appears, does the digital-twin already lean toward that later public disclosure? Third, after removing common daily market conditions, do digital-twin responses still uncover specific views of the corresponding finfluencers  rather than generic knowledge of the LLM?\footnote{Section \ref{app:validation_public_disclosure} of the Internet Appendix provides additional evidence in support of the validation tests presented in Table \ref{tab:validation_main}.}

\begin{table}[!htbp]
\caption{Validation of Digital-Twin Interview Belief Proxies}
\label{tab:validation_main}
{\footnotesize{}This table reports four validation tests that address three concerns about the digital-twin interview belief proxies. Panel A uses the exact-overlap sample of 4,063 same-account, same-ticker, same-date public recommendation overlaps. Sign alignment is the share of overlaps in which the interview response and public recommendation point in the same direction, either buy-buy or sell-sell. The all non-neutral row uses every overlap with a non-neutral interview response, while the meaningful low-speculation row keeps only interviews far enough from neutral and not highly speculative. The AUC row reports the area under the curve, or the probability that an interview score ranks a public buy above a public sell. Panel B asks whether interviews lean toward later public recommendations during the five prior trading days; positive signed-tilt values indicate alignment with the eventual public recommendation. Panel C uses macro questions Q1--Q7 in the 81-account interview sample and residualizes responses by removing date-question means. It reports the incremental $R^2$ from including account fixed effects, an adjacent-day same-account similarity score, and top-1 and top-5 account re-identification rates. Difference is Actual minus Benchmark. Percentage rows are in percentage points; signed tilt is in recommendation-score points; and AUC, $R^2$, and similarity are unitless. The Benchmark column reports the comparison value used for each test. The $p$-value column reports the validation-test p-value for the Actual-Benchmark difference. N is row-specific: the sign-alignment rows use the directional overlap subset shown in the row, the AUC row uses all exact overlaps, Panel B uses pre-disclosure validation cases, and Panel C uses residual macro-response rows or held-out account-days. For the incremental account $R^2$ row, N is the number of account-date-question macro bull-score responses used in the fixed-effect calculation, not the number of account-days.}
{\footnotesize\par}
\vspace{7pt}
\centering{}
{\footnotesize
\begin{tabular}{lccccc}
\hline\hline
 & (1) & (2) & (3) & (4) & (5) \\
\noalign{\vskip 6pt}
\multicolumn{6}{l}{\textit{Panel A. Public-Recommendation Overlap}} \\[-1pt]
\noalign{\hrule height 0.15pt}
\noalign{\vskip 2pt}
Measure & Actual & Benchmark & Difference & $p$-value & N \\
\cmidrule(lr){1-1}\cmidrule(lr){2-2}\cmidrule(lr){3-3}\cmidrule(lr){4-4}\cmidrule(lr){5-5}\cmidrule(lr){6-6}
All non-neutral sign alignment & 89.5\% & 50.0\% & 39.5 pp & $<0.001$ & 3,271 \\
Meaningful low-speculation sign alignment & 91.5\% & 50.0\% & 41.5 pp & $<0.001$ & 2,439 \\
Buy-sell ranking AUC & 0.776 & 0.522 & 0.255 & $<0.001$ & 4,063 \\
\hline
\noalign{\vskip 6pt}
\multicolumn{6}{l}{\textit{Panel B. Pre-Disclosure Belief Recovery}} \\[-1pt]
\noalign{\hrule height 0.15pt}
\noalign{\vskip 2pt}
Measure & Actual & Benchmark & Difference & $p$-value & N \\
\cmidrule(lr){1-1}\cmidrule(lr){2-2}\cmidrule(lr){3-3}\cmidrule(lr){4-4}\cmidrule(lr){5-5}\cmidrule(lr){6-6}
Pre-disclosure signed tilt & 5.32 & 2.60 & 2.72 & $<0.001$ & 1,500 \\
Matched pre-disclosure signed tilt & 4.99 & 2.28 & 2.71 & $<0.001$ & 1,500 \\
\hline
\noalign{\vskip 6pt}
\multicolumn{6}{l}{\textit{Panel C. Account-Specific Belief Structure}} \\[-1pt]
\noalign{\hrule height 0.15pt}
\noalign{\vskip 2pt}
Measure & Actual & Benchmark & Difference & $p$-value & N \\
\cmidrule(lr){1-1}\cmidrule(lr){2-2}\cmidrule(lr){3-3}\cmidrule(lr){4-4}\cmidrule(lr){5-5}\cmidrule(lr){6-6}
Incremental account $R^2$ & 0.338 & 0.000 & 0.338 & $<0.001$ & 27,997 \\
Adjacent-day same-account similarity & 0.600 & 0.023 & 0.577 & $<0.001$ & 3,900 \\
Top-1 account re-identification & 10.8\% & 1.2\% & 9.5 pp & $<0.001$ & 1,951 \\
Top-5 account re-identification & 35.2\% & 6.2\% & 29.0 pp & $<0.001$ & 1,951 \\
\hline\hline
\end{tabular}
}
\end{table}


Panel A of Table \ref{tab:validation_main} reports the results of the first two validation tests. In this Panel, we ask the cleanest validation question: when a finfluencer account publicly recommends a stock on a given trading date, does the digital-twin interview about that stock point in the same buy or sell direction? The exact-overlap sample contains $4,063$ account-ticker-date observations. Among meaningful low-speculation interview responses---responses far enough from neutral and not highly speculative---the interview direction aligns with the public recommendation $91.5\%$ of the time, compared with a $50$-$50$ benchmark. Using all non-neutral interview scores, the alignment rate is $89.5\%$.\footnote{Interview direction is determined from the digital-twin recommendation score after centering it at $50$: scores above $50$ are buy-leaning and scores below $50$ are sell-leaning. Public recommendation direction is classified from the observed public-post recommendation panel, as described in Section \ref{appx: Data Appendix} of the Internet Appendix. The meaningful low-speculation subset requires interview recommendation scores at least $15$ points from neutral and speculation scores of $40$ or below. The all non-neutral measure uses any non-neutral interview score.} Thus, when the public-post panel gives us a same-day benchmark, the digital-twin interview usually points in the same direction.

Panel A of Table \ref{tab:validation_main} also asks whether, in the exact-overlap sample, digital-twin interview scores rank publicly disclosed buy recommendations above publicly disclosed sell recommendations. This ranking test uses the area-under-the-curve (AUC) statistic. Let $s^{DT}_{\text{public buy}}$ denote the digital-twin interview score for a randomly chosen public-buy recommendation, and define $s^{DT}_{\text{public sell}}$ analogously for a randomly chosen public-sell recommendation. Then
\[
AUC=\Pr(s^{DT}_{\text{public buy}}>s^{DT}_{\text{public sell}}).
\]
The AUC statistic is $0.776$. In other words, if we randomly draw one public buy and one public sell recommendation from the overlap sample, the digital-twin interview score ranks the public buy above the public sell $77.6\%$ of the time. As a placebo test, we shuffle the public buy and sell labels within each event date and then recompute the same AUC statistic. This benchmark AUC is $0.522$.\footnote{Shuffling within event date preserves the number of public buys and sells on each date, but breaks the link between the public recommendation direction and the digital-twin interview score. Also, the difference between the actual AUC statistic and the placebo, $0.255$, is statistically significant (p$< 0.001$).} This second check gives the same message as the first. Public buys receive higher interview scores than public sells.\footnote{Table \ref{tab:validation_public_overlap_appendix} in the Internet Appendix provides additional validation results.}

Panel B of Table \ref{tab:validation_main} presents the results of our third validation test. We seek to determine whether the digital twin interviews merely restate public recommendations after those recommendations have already appeared. We address this question by asking whether the digital-twin interviews already lean in the direction of later public recommendations. For each account-ticker pair, we identify the first directional public recommendation, then examine the digital-twin interviews for the same account and ticker in the five prior trading days.\footnote{Public recommendations are first collapsed to the account-ticker-date level using the effective trading date of the public post. Here, directional means that the public recommendation score is at least $65$ for a buy or at most $35$ for a sell. Neutral public recommendations are excluded.} Let $t$ be the effective trading date of that first directional public recommendation. For an interview $k$ trading days earlier, define signed tilt as
\[
SignedTilt_{i,j,t-k}=D_{i,j}(RecScore^{DT}_{i,j,t-k}-50), \qquad k=1,\ldots,5,
\]
where $RecScore^{DT}$ is the digital-twin stock-pick recommendation score, $D_{i,j}=1$ if that first directional public recommendation is a buy, and $D_{i,j}=-1$ if it is a sell. Positive values therefore mean that the earlier interview leaned toward the eventual public direction. In the five trading days before the public post, signed tilt averages $5.32$ points, compared with $2.60$ points for same-account, same-date placebo tickers.\footnote{The difference between the two is statistically significant (p$<0.001$).} We also use a matched benchmark. For each future-public ticker, this comparison uses the closest available pre-disclosure interview date and compares it with other tickers from the same account and interview date that do not receive a public recommendation over the next five trading days. The matched comparison gives a similar result: future-public tickers average $4.99$ signed-tilt points, compared with $2.28$ points for matched control tickers.\footnote{Again, the difference between these two values is statistically significant (p$<0.001$).} This result shows that even before public recommendations, digital-twin interview responses lean toward those future recommendations.  

Panel C of Table \ref{tab:validation_main} presents the results of our fourth and final validation test. Here we seek to determine whether the interviews uncover specific views of the digital twins rather than generic knowledge of the LLM. For this test we use the macro-interview branch, where every digital twin answers the same recurring broad-market questions on the same dates. For each question on each date, we subtract the average response across twins. For example, if a twin answered macro question \#1 with a score of $60$, and the average answer to question \#1 across all twins that day was $50$, the residualized answer is $10$. This removes the common daily component: the part of the answer that could come from market news, broad sentiment, generic market commentary, or the wording of the question itself. We then ask whether the residualized responses still reveal which finfluencer account they came from. If the responses were primarily generic LLM output, the identity of the twin account would explain little of the remaining variation after this common component is removed. We test this first using account fixed effects.\footnote{The fixed-effects comparison is between $BullScore_{i,q,t}=\gamma_{q,t}+\epsilon_{i,q,t}$ and $BullScore_{i,q,t}=\gamma_{q,t}+\alpha_i+\epsilon_{i,q,t}$, where $\gamma_{q,t}$ absorbs common date-question effects and $\alpha_i$ captures account-specific structure.} If the residual answers are generic, then knowing the account should add little explanatory power. Instead, adding account fixed effects raises explained variation by $0.338$. This means that account identity still explains a meaningful share of the interview responses after the common date-question component has been removed.

Panel C also reports two more direct diagnostics. The first asks whether the same account has a stable residualized response pattern across nearby interview days. For each account on each date, we take the residualized responses to the recurring macro questions and treat them as that account's response pattern for the day. We then use cosine similarity to compare that pattern with the same account's pattern on the next interview day.\footnote{Each account-date response vector contains the residualized bull-score responses to macro questions Q1--Q7, where residualization subtracts the event-date-by-question mean. We keep account-dates with all seven responses observed and compute \(\cos(x,y)=x'y/(\|x\|\|y\|)\) between an account's response vector and the same account's next observed response vector. The reported statistic is the mean across these adjacent same-account pairs.} If the digital twin is not generic and is taking on the persona of a finfluencer account, we would expect the twin's views to be auto-correlated. Indeed, we find that the similarity score is $0.600$ when we compare a twin's response pattern with that same twin's pattern on the next interview day, compared with $0.023$ when we compare it with a placebo account.

The second diagnostic is an out-of-sample identification test. The idea is simple. If all digital twins are just repeating the same generic market commentary, then once we remove the common daily market component, it should be difficult to tell which account produced which answer. If the interviews are truly account-conditioned, then each account should leave some stable pattern in its answers.

To test this, we split the data into two parts. We use the first part to build a reference profile for each account. For each account, this reference profile summarizes how that account's digital twin tends to answer the recurring macro questions after removing the common response on that date. We then take the remaining held-out observations, which were not used to build the reference profiles, and ask whether each held-out response pattern can be matched back to the correct account.

For each held-out account-date observation, we compare its residualized response pattern to the reference profiles of all 81 accounts. The comparison uses cosine similarity again, measuring whether two response patterns point in the same direction across the macro questions. We then rank all 81 accounts from most similar to least similar. If the digital-twin answers were generic, the correct account should appear near the top of the ranking only by chance.\footnote{With $81$ accounts, the chance benchmarks are $1/81=1.2\%$ for the closest match and $5/81=6.2\%$ for the top-five match.} Instead, the correct account is the closest match $10.8\%$ of the time and appears in the top five matches $35.2\%$ of the time. The result shows that, even after removing the common daily market component, the interview outputs retain twin-specific structure.

Taken together, these validation tests support the main premise of our research design. We show that outcomes from digital twin interviews line up with public recommendations when we can observe them, lean toward later recommendations before they appear, and can be traced back to a specific finfluencer even after common market conditions are removed.

\subsection{The Coverage Gain from the Silent Region}\label{subsec: The Coverage Gain from the Silent Region}

Having established that the digital-twin responses are consistent with the public personas of the corresponding human finfluencers, we now want to understand the magnitude and composition of the silent region. Recall that the silent region is the set of stock-events for which the digital twin interview produces a recommendation without any matching recommendation publicly posted by the corresponding human finfluencer. The evidence, presented below in Table \ref{tab:stockpick_public_recommendation_attention}, makes two simple points. First, the silent region is large. Most stock-events in the interview panel would be absent from a sample that is based exclusively on human posts, so the digital twin interviews substantially expand the set of measured stock-events. Second, public recommendations are concentrated among a smaller set of firms, especially larger, technology-heavy firms. The coverage gain, therefore, does more than just increase the number of observations. It also uncovers parts of the stock universe that public posts underrepresent.

Over our sample period, the digital-twin interviews generate $21{,}410$ stock-date observations. Only $3{,}251$, or $15.2\%$, coincide with a same-stock, same-date public post. The remaining $18{,}159$ stock-dates, or $84.8\%$ of the sample, have no same-date public post. Thus, a post-based sample would miss most of the stock-events for which we have interview responses. The silent region is large, and the interview protocol substantially expands coverage.

We also ask whether finfluencers' public recommendations are spread evenly across the recurring stock-pick universe. Table \ref{tab:stockpick_public_recommendation_attention} shows that they are not. Panel A of Table \ref{tab:stockpick_public_recommendation_attention} groups firms by the number of distinct public-post days during the stock-pick interview window. Sixty-one firms, or $14.2\%$ of the stocks in our sample, receive no public-post days at all. More broadly, $222$ firms have no more than four public-post days, so roughly half of the covered universe receives little or no public recommendation attention. At the other end, $54$ high-attention firms receive $20$ or more public-post days. These firms average $36.7$ public-post days and $147.3$ public posts.

Panel B of Table \ref{tab:stockpick_public_recommendation_attention} shows that human finfluencers pay more attention to larger firms and those that are in the business equipment and technology industry. Thus, public recommendations are tilted toward large, technology-heavy firms rather than spread evenly across firms.

\begin{table}[!htbp]
\caption{Public Recommendation Attention and Firm Characteristics}
\label{tab:stockpick_public_recommendation_attention}
{\footnotesize{}This table describes how public stock-recommendation attention is distributed across the 429-stock universe during the interview window. The sample contains 4,063 account-stock-date events that have a same-account, same-stock, same-date interview and public post match. Panel A groups firms by the number of distinct public-post dates from December 15, 2025 through March 16, 2026; a public-post date is a date on which the firm has at least one public stock-recommendation post. Mean public days and mean public posts are per-firm averages within each bin. Mean posts per public day is the average firm-level ratio of public posts to public-post dates, computed among firms with at least one public-post date. Panel B compares low-, middle-, and high-attention firms, defined as 0--4, 5--19, and 20 or more public-post days. These groups contain 222, 153, and 54 firms. The last two columns report the high-minus-low difference and Welch two-sample $t$-statistic. Continuous characteristics are winsorized at the 1st and 99th percentiles. Market capitalization is reported in billions of dollars. Indicators ***, **, * denote statistical significance at the 1\%, 5\%, and 10\% level, respectively.}
{\footnotesize\par}
\vspace{7pt}
\centering{}
{\footnotesize
\setlength{\tabcolsep}{2pt}
\begin{tabular}{lccccc}
\hline\hline
 & \makebox[1.55cm][c]{(1)} & \makebox[1.80cm][c]{(2)} & \makebox[2.00cm][c]{(3)} & \makebox[2.05cm][c]{(4)} & \makebox[2.05cm][c]{(5)} \\
\noalign{\vskip 6pt}
\multicolumn{6}{l}{\textit{Panel A. Public Recommendation Attention Distribution}} \\[-1pt]
\noalign{\hrule height 0.15pt}
\noalign{\vskip 2pt}
 & \makebox[1.55cm][c]{} & \makebox[1.80cm][c]{Share of} & \makebox[2.00cm][c]{Mean} & \makebox[2.05cm][c]{Mean} & \makebox[2.05cm][c]{Mean posts} \\
Public-post-day bin & \makebox[1.55cm][c]{Firms} & \makebox[1.80cm][c]{firms} & \makebox[2.00cm][c]{public days} & \makebox[2.05cm][c]{public posts} & \makebox[2.05cm][c]{per public day} \\
\cmidrule(lr){1-1}\cmidrule(lr){2-2}\cmidrule(lr){3-3}\cmidrule(lr){4-4}\cmidrule(lr){5-5}\cmidrule(lr){6-6}
\noalign{\vskip 3pt}
0 & \makebox[1.55cm][c]{61} & \makebox[1.80cm][c]{14.2\%} & \makebox[2.00cm][c]{0.0} & \makebox[2.05cm][c]{0.0} & \makebox[2.05cm][c]{} \\
1 & \makebox[1.55cm][c]{52} & \makebox[1.80cm][c]{12.1\%} & \makebox[2.00cm][c]{1.0} & \makebox[2.05cm][c]{1.1} & \makebox[2.05cm][c]{1.1} \\
2 & \makebox[1.55cm][c]{36} & \makebox[1.80cm][c]{8.4\%} & \makebox[2.00cm][c]{2.0} & \makebox[2.05cm][c]{2.2} & \makebox[2.05cm][c]{1.1} \\
3 & \makebox[1.55cm][c]{37} & \makebox[1.80cm][c]{8.6\%} & \makebox[2.00cm][c]{3.0} & \makebox[2.05cm][c]{3.3} & \makebox[2.05cm][c]{1.1} \\
4 & \makebox[1.55cm][c]{36} & \makebox[1.80cm][c]{8.4\%} & \makebox[2.00cm][c]{4.0} & \makebox[2.05cm][c]{4.6} & \makebox[2.05cm][c]{1.2} \\
5--9 & \makebox[1.55cm][c]{82} & \makebox[1.80cm][c]{19.1\%} & \makebox[2.00cm][c]{6.5} & \makebox[2.05cm][c]{8.1} & \makebox[2.05cm][c]{1.2} \\
10--19 & \makebox[1.55cm][c]{71} & \makebox[1.80cm][c]{16.6\%} & \makebox[2.00cm][c]{13.8} & \makebox[2.05cm][c]{20.1} & \makebox[2.05cm][c]{1.4} \\
20+ & \makebox[1.55cm][c]{54} & \makebox[1.80cm][c]{12.6\%} & \makebox[2.00cm][c]{36.7} & \makebox[2.05cm][c]{147.3} & \makebox[2.05cm][c]{3.3} \\
\hline
\noalign{\vskip 6pt}
\multicolumn{6}{l}{\textit{Panel B. Firm Characteristics by Public Recommendation Attention}} \\[-1pt]
\noalign{\hrule height 0.15pt}
\noalign{\vskip 2pt}
 & \makebox[1.55cm][c]{Low} & \makebox[1.80cm][c]{Middle} & \makebox[2.00cm][c]{High} & \makebox[2.05cm][c]{} & \makebox[2.05cm][c]{} \\
Variable & \makebox[1.55cm][c]{0--4} & \makebox[1.80cm][c]{5--19} & \makebox[2.00cm][c]{20+} & \makebox[2.05cm][c]{High $-$ Low} & \makebox[2.05cm][c]{$t$-stat} \\
\cmidrule(lr){1-1}\cmidrule(lr){2-2}\cmidrule(lr){3-3}\cmidrule(lr){4-4}\cmidrule(lr){5-5}\cmidrule(lr){6-6}
\noalign{\vskip 3pt}
Market cap (\$bn) & \makebox[1.55cm][c]{38.53} & \makebox[1.80cm][c]{87.28} & \makebox[2.00cm][c]{630.22} & \makebox[2.05cm][c]{591.70$^{***}$} & \makebox[2.05cm][c]{5.29} \\
Log market cap & \makebox[1.55cm][c]{10.29} & \makebox[1.80cm][c]{10.99} & \makebox[2.00cm][c]{12.62} & \makebox[2.05cm][c]{2.33$^{***}$} & \makebox[2.05cm][c]{13.03} \\
Listing age (years) & \makebox[1.55cm][c]{35.91} & \makebox[1.80cm][c]{35.14} & \makebox[2.00cm][c]{33.71} & \makebox[2.05cm][c]{-2.20} & \makebox[2.05cm][c]{-0.81} \\
Business equipment / tech (\%) & \makebox[1.55cm][c]{14.41} & \makebox[1.80cm][c]{19.74} & \makebox[2.00cm][c]{29.63} & \makebox[2.05cm][c]{15.22$^{**}$} & \makebox[2.05cm][c]{2.27} \\
Manufacturing (\%) & \makebox[1.55cm][c]{9.91} & \makebox[1.80cm][c]{9.21} & \makebox[2.00cm][c]{12.96} & \makebox[2.05cm][c]{3.05} & \makebox[2.05cm][c]{0.61} \\
Healthcare (\%) & \makebox[1.55cm][c]{8.56} & \makebox[1.80cm][c]{9.21} & \makebox[2.00cm][c]{3.70} & \makebox[2.05cm][c]{-4.85} & \makebox[2.05cm][c]{-1.51} \\
Other (\%) & \makebox[1.55cm][c]{17.57} & \makebox[1.80cm][c]{17.11} & \makebox[2.00cm][c]{22.22} & \makebox[2.05cm][c]{4.65} & \makebox[2.05cm][c]{0.74} \\
\hline\hline
\end{tabular}
}
\end{table}


The coverage gain from digital-twin interviews is therefore not just an increase in the sample size. The key gain is that the method makes the silent part of selective disclosure measurable. Public posts tell us what finfluencers choose to reveal. Digital-twin interviews ask the same public personas the same questions under a fixed protocol, including when no public recommendation appears. This means that the interview protocol can capture views about stocks that are underrepresented in public posts.



\hypertarget{Cross-Sectional Predictability as Evidence of Information Content}{%
\section{Cross-Sectional Predictability as Evidence of Information Content}\label{sec:xp}}

The central question in our paper is whether the digital twin interviews contain useful information about future stock outcomes. In this section, we show that they do: interview responses predict both future excess returns and future excess realized volatility in the cross-section of stocks.\footnote{The tests presented in this Section evaluate the informational content of twins' interviews. They do not evaluate the profitability of an implementable trading strategy. While we do form calendar-time portfolios, we do not study trading costs or liquidity. Our goal is to ask whether the interview variables contain economically meaningful information about future stock outcomes.}

The stock-pick interview branch produces three types of signals pertaining to future stock returns and volatility. \emph{Directional} signals measure whether the interview responses are buy-leaning or sell-leaning for a given stock on a given date. \emph{Disagreement} signals measure the extent to which low-speculation recommendations are split between buys and sells. And \emph{uncertainty} variables measure whether the panel of responses lacks conviction in either direction. We study these three signals in Tables \ref{tab:q1_stockpick_fmb_calendar_time_returns} through \ref{tab:stockpick_state_variables}.

\subsection{Directional Signals and Future Stock Returns}

We begin by studying the relation between directional signals obtained from digital twin interviews and the cross-section of future stock returns. Because event dates in our panel are spaced one day apart, return-measurement windows longer than one trading day necessarily overlap: adjacent $h$-day outcomes reuse many of the same daily returns. This dependence does not by itself imply biased coefficient estimates, but it induces serial correlation in the dependent variables and in the sequence of estimated cross-sectional slopes, so standard errors that ignore the overlap can overstate the level of statistical significance. 

We address this issue in the next two subsections. First, in Subsection \ref{sec:fm} we perform Fama--MacBeth regressions using Newey--West standard errors with lag length $h\mathbin{-}1$. We also perform calendar-time portfolio tests that aggregate all active event-date vintages into a daily long--short return series, so that our statistical inference is based on the time-series variation in returns. Second, in Subsection \ref{sec:sv}, we perform staggered-vintage regressions in which we space anchor dates $h$ trading days apart, eliminating overlap in future-return windows within each vintage. Together, these approaches assess whether the directional signal is robust to alternative treatments of overlapping outcomes.

\subsubsection{Fama--MacBeth Regressions and Calendar Time Portfolios}
\label{sec:fm}

Table \ref{tab:q1_stockpick_fmb_calendar_time_returns} reports the first stock-level return tests, which ask whether stocks with more buy-leaning interview responses earn higher future excess returns than other stocks. The four panels of Table \ref{tab:q1_stockpick_fmb_calendar_time_returns} (A-D) each use a different directional signal as independent variable (\emph{Tilt}, \emph{Net Buy Share}, \emph{Meaningful Tilt}, and \emph{Conviction-Weighted Meaningful Tilt}).

In Columns (1)--(3) of Table \ref{tab:q1_stockpick_fmb_calendar_time_returns} we estimate cross-sectional return predictability using Fama--MacBeth regressions \citep{FamaMacBeth1973}. Specifically, for each event date in the stock-pick branch, we estimate
\[
ExRet_{j,t}^{(h)}=a_{t,h}+b_{t,h}Direction_{j,t}+\varepsilon_{j,t}^{(h)}.
\]
Here, $ExRet_{j,t}^{(h)}$ is stock $j$'s future excess return over event horizon $h$, measured relative to the S\&P 500, and $Direction_{j,t}$ is one of four directional signals obtained from twin interview responses, measured before the return window begins. Regression coefficients $b_{t,h}$ are averaged across time ($t$), with standard errors computed from their intertemporal variation. These standard errors are Newey--West adjusted to account for serial correlation induced by overlapping return windows when the event horizon, $h$, exceeds one trading day \citep{NeweyWest1987}. 

We report the intertemporal averages, $\bar{b}_h$, of these regression coefficients for three event horizons: one, five, and ten trading days. The results show that the directional signals predict future excess returns, especially at the ten-day horizon. The clearest measure is \emph{Net Buy Share} (Panel B), the share of meaningful buy recommendations minus the share of meaningful sell recommendations for a stock-date. A ten-percentage-point increase in \emph{Net Buy Share} is associated with $40$ basis points higher excess return over the next ten trading days. The five-day estimate is also positive, though not statistically significant, while the one-day coefficient is close to zero. Thus, the relation is weak immediately after the interview, but stronger over the next several trading days.

The other directional variables provide similar results. \emph{Conviction-Weighted Meaningful Tilt} (Panel D), which multiplies the average meaningful recommendation tilt by the share of recommendations that are meaningful, also predicts higher future returns: a one-unit increase is associated with $16.9$ basis points higher excess return over the next ten trading days.\footnote{Since \emph{Conviction-Weighted Meaningful Tilt} has a standard deviation of $3.07$, a one-standard-deviation increase in \emph{Conviction-Weighted Meaningful Tilt} is associated with $51.9$ basis points higher excess return over the next ten trading days.} \emph{Tilt} and \emph{Meaningful Tilt} (Panels A and C) are also positive at the ten-day horizon. These measures are constructed differently, but they tell the same story. Stocks with more positive digital-twin interview responses tend to earn higher future excess returns.

\begin{table}[!htbp]
\caption{Digital-Twin Stock-Pick Direction and Future Excess Returns}
\label{tab:q1_stockpick_fmb_calendar_time_returns}
{\footnotesize{}This table combines daily Fama--MacBeth stock-pick return-predictability regressions with calendar-time long-short portfolio tests. Columns (1)--(3) report daily Fama--MacBeth cross-sectional regressions of cumulative future excess stock returns on one digital-twin direction signal measured before the return window. Excess returns are measured relative to the S\&P 500 and reported in percentage points over 1, 5, and 10 trading days. The table reports the time-series average of the daily slope estimates. Columns (4)--(6) report calendar-time portfolio tests using the same signals. On each event date, stocks are ranked by the indicated signal, the portfolio buys the top quintile and shorts the bottom quintile with equal weights within each leg, and each event-date vintage is held for the next 10 trading days, excluding the formation day. The daily calendar-time long-short return averages all active vintages on each holding date. H-L Mean and CAPM Alpha are daily percentage-point estimates while 10-day Alpha is 10 times the daily CAPM alpha. Market adjustment is estimated from a daily CAPM regression of the long-short return on the S\&P 500 return. Newey--West standard errors are in parentheses, using lag $h-1$ in columns (1)--(3) and lag 9 in columns (4)--(6) to allow for serial correlation from overlapping return windows and overlapping calendar-time vintages. Fama--MacBeth coefficients are scaled to a +10 recommendation-point move in Tilt or Meaningful Tilt, a +10 percentage-point move in Net Buy Share, or a +1-unit move in Conviction-Weighted Meaningful Tilt. N/days reports stock-event observations in columns (1)--(3) and calendar holding days in columns (4)--(6). Mean adjusted $R^2$ applies only to the Fama--MacBeth columns. The sample uses 429 stocks and 81 accounts over 50 event dates from December 15, 2025 to March 16, 2026; calendar-time holding dates run from December 16, 2025 to March 30, 2026. Indicators ***, **, * denote significance at the 1\%, 5\%, and 10\% level.}
{\footnotesize\par}
\vspace{5pt}
\centering{}
{\footnotesize
\setlength{\tabcolsep}{1.5pt}
\begin{tabular}{lcccccc}
\hline\hline
 & \makebox[1.88cm][c]{(1)} & \makebox[1.88cm][c]{(2)} & \makebox[1.88cm][c]{(3)} & \makebox[1.88cm][c]{(4)} & \makebox[1.88cm][c]{(5)} & \makebox[1.88cm][c]{(6)} \\
\noalign{\vskip 4pt}
 & \multicolumn{3}{c}{\makebox[5.60cm][c]{Dep. var.: future excess return (pp)}} & \multicolumn{3}{c}{\makebox[5.60cm][c]{Dep. var.: H-L portfolio return (pp)}} \\
\cmidrule(lr){2-4}\cmidrule(lr){5-7}
\noalign{\vskip 2pt}
 & \makebox[1.88cm][c]{1 day} & \makebox[1.88cm][c]{5 days} & \makebox[1.88cm][c]{10 days} & \makebox[1.88cm][c]{\shortstack{H-L\\Mean}} & \makebox[1.88cm][c]{\shortstack{CAPM\\Alpha}} & \makebox[1.88cm][c]{\shortstack{10-day\\Alpha}} \\
\cmidrule(lr){2-2}\cmidrule(lr){3-3}\cmidrule(lr){4-4}\cmidrule(lr){5-5}\cmidrule(lr){6-6}\cmidrule(lr){7-7}
\noalign{\vskip 4pt}
\multicolumn{7}{l}{\textit{Panel A. Tilt}} \\[-1pt]
\noalign{\hrule height 0.15pt}
\noalign{\vskip 2pt}
Direction signal & \makebox[1.88cm][c]{-0.010} & \makebox[1.88cm][c]{0.639} & \makebox[1.88cm][c]{1.594$^{***}$} & \makebox[1.88cm][c]{0.133$^{**}$} & \makebox[1.88cm][c]{0.120$^{**}$} & \makebox[1.88cm][c]{1.203$^{**}$} \\
 & \makebox[1.88cm][c]{(0.124)} & \makebox[1.88cm][c]{(0.424)} & \makebox[1.88cm][c]{(0.611)} & \makebox[1.88cm][c]{(0.061)} & \makebox[1.88cm][c]{(0.055)} & \makebox[1.88cm][c]{(0.549)} \\[1pt]
\noalign{\hrule height 0.15pt}
\noalign{\vskip 1pt}
N / days & \makebox[1.88cm][c]{21,410} & \makebox[1.88cm][c]{21,410} & \makebox[1.88cm][c]{21,410} & \makebox[1.88cm][c]{71} & \makebox[1.88cm][c]{71} & \makebox[1.88cm][c]{71} \\
Mean adj. $R^2$ & \makebox[1.88cm][c]{0.009} & \makebox[1.88cm][c]{0.013} & \makebox[1.88cm][c]{0.013} & \makebox[1.88cm][c]{} & \makebox[1.88cm][c]{} & \makebox[1.88cm][c]{} \\
\hline
\noalign{\vskip 4pt}
\multicolumn{7}{l}{\textit{Panel B. Net Buy Share}} \\[-1pt]
\noalign{\hrule height 0.15pt}
\noalign{\vskip 2pt}
Direction signal & \makebox[1.88cm][c]{-0.002} & \makebox[1.88cm][c]{0.166} & \makebox[1.88cm][c]{0.404$^{***}$} & \makebox[1.88cm][c]{0.120$^{*}$} & \makebox[1.88cm][c]{0.108$^{*}$} & \makebox[1.88cm][c]{1.082$^{*}$} \\
 & \makebox[1.88cm][c]{(0.030)} & \makebox[1.88cm][c]{(0.105)} & \makebox[1.88cm][c]{(0.152)} & \makebox[1.88cm][c]{(0.066)} & \makebox[1.88cm][c]{(0.061)} & \makebox[1.88cm][c]{(0.608)} \\[1pt]
\noalign{\hrule height 0.15pt}
\noalign{\vskip 1pt}
N / days & \makebox[1.88cm][c]{21,410} & \makebox[1.88cm][c]{21,410} & \makebox[1.88cm][c]{21,410} & \makebox[1.88cm][c]{71} & \makebox[1.88cm][c]{71} & \makebox[1.88cm][c]{71} \\
Mean adj. $R^2$ & \makebox[1.88cm][c]{0.010} & \makebox[1.88cm][c]{0.014} & \makebox[1.88cm][c]{0.016} & \makebox[1.88cm][c]{} & \makebox[1.88cm][c]{} & \makebox[1.88cm][c]{} \\
\hline
\noalign{\vskip 4pt}
\multicolumn{7}{l}{\textit{Panel C. Meaningful Tilt}} \\[-1pt]
\noalign{\hrule height 0.15pt}
\noalign{\vskip 2pt}
Direction signal & \makebox[1.88cm][c]{0.020} & \makebox[1.88cm][c]{0.112} & \makebox[1.88cm][c]{0.293$^{***}$} & \makebox[1.88cm][c]{0.109$^{**}$} & \makebox[1.88cm][c]{0.051} & \makebox[1.88cm][c]{0.514} \\
 & \makebox[1.88cm][c]{(0.034)} & \makebox[1.88cm][c]{(0.083)} & \makebox[1.88cm][c]{(0.109)} & \makebox[1.88cm][c]{(0.050)} & \makebox[1.88cm][c]{(0.041)} & \makebox[1.88cm][c]{(0.408)} \\[1pt]
\noalign{\hrule height 0.15pt}
\noalign{\vskip 1pt}
N / days & \makebox[1.88cm][c]{20,823} & \makebox[1.88cm][c]{20,823} & \makebox[1.88cm][c]{20,823} & \makebox[1.88cm][c]{71} & \makebox[1.88cm][c]{71} & \makebox[1.88cm][c]{71} \\
Mean adj. $R^2$ & \makebox[1.88cm][c]{0.010} & \makebox[1.88cm][c]{0.008} & \makebox[1.88cm][c]{0.004} & \makebox[1.88cm][c]{} & \makebox[1.88cm][c]{} & \makebox[1.88cm][c]{} \\
\hline
\noalign{\vskip 4pt}
\multicolumn{7}{l}{\textit{Panel D. Conviction-Weighted Meaningful Tilt}} \\[-1pt]
\noalign{\hrule height 0.15pt}
\noalign{\vskip 2pt}
Direction signal & \makebox[1.88cm][c]{0.000} & \makebox[1.88cm][c]{0.065} & \makebox[1.88cm][c]{0.169$^{**}$} & \makebox[1.88cm][c]{0.133$^{**}$} & \makebox[1.88cm][c]{0.117$^{**}$} & \makebox[1.88cm][c]{1.170$^{**}$} \\
 & \makebox[1.88cm][c]{(0.013)} & \makebox[1.88cm][c]{(0.048)} & \makebox[1.88cm][c]{(0.070)} & \makebox[1.88cm][c]{(0.057)} & \makebox[1.88cm][c]{(0.052)} & \makebox[1.88cm][c]{(0.519)} \\[1pt]
\noalign{\hrule height 0.15pt}
\noalign{\vskip 1pt}
N / days & \makebox[1.88cm][c]{20,823} & \makebox[1.88cm][c]{20,823} & \makebox[1.88cm][c]{20,823} & \makebox[1.88cm][c]{71} & \makebox[1.88cm][c]{71} & \makebox[1.88cm][c]{71} \\
Mean adj. $R^2$ & \makebox[1.88cm][c]{0.010} & \makebox[1.88cm][c]{0.013} & \makebox[1.88cm][c]{0.014} & \makebox[1.88cm][c]{} & \makebox[1.88cm][c]{} & \makebox[1.88cm][c]{} \\
\hline\hline
\end{tabular}
}
\end{table}


In Columns (4)--(6) of Table \ref{tab:q1_stockpick_fmb_calendar_time_returns} we use calendar-time portfolios as an alternative methodology to account for the overlapping return windows. Following \citet{MitchellStafford2000}, these tests aggregate the stock-level rankings into a daily high-minus-low portfolio return series and estimate performance from the time-series variation of that portfolio. The calendar-time test therefore asks a related portfolio question: do stocks ranked more favorably by the digital-twin panel outperform stocks ranked less favorably when the signal is implemented as a daily long-short portfolio?

On each event date, we rank stocks by a direction signal, buy the top quintile, short the bottom quintile, and hold that event-date portfolio for the next ten trading days. On each calendar day, we then average the returns of all active event-date vintages. Column (4) reports the average daily high-minus-low return, column (5) reports the CAPM alpha from a regression of the daily high-minus-low return on the S\&P 500 return, and column (6) reports the same alpha multiplied by ten to put it on a 10-trading-day scale.

The calendar-time results support the same interpretation as the event-time regressions. For \emph{Net Buy Share}, the high-minus-low portfolio earns an average daily return of $12.0$ basis points and a CAPM alpha of $10.8$ basis points per day, corresponding to a 10-day alpha of $108$ basis points. The estimates for \emph{Tilt} and \emph{Conviction-Weighted Meaningful Tilt} are similar and statistically significant. \emph{Meaningful Tilt} has a positive high-minus-low mean return, but its CAPM alpha is smaller and not statistically significant.\footnote{This weaker result is not surprising because \emph{Meaningful Tilt} conditions only on the subset of meaningful responses and does not account for how broadly that meaningful direction is supported across the interview panel. Once the same directional information is weighted by the share of meaningful recommendations (\emph{Conviction-Weighted Meaningful Tilt}), the calendar-time alpha is again positive and statistically significant.} Overall, the calendar-time evidence shows that the stock-pick signal also appears in a portfolio-level design, not only in event-time cross-sectional regressions.\footnote{Table \ref{tab:q1_stockpick_fmb_calendar_time_returns_value_weighted} in the Internet Appendix replicates Table \ref{tab:q1_stockpick_fmb_calendar_time_returns} using value weights both in the Fama--MacBeth regressions and in the calendar-time portfolios. The estimates in this table are much smaller and not statistically distinguishable from zero, indicating that the relation between interview signals and future returns is weaker among larger firms.}

\begin{figure}[!htbp]
\centering
\includegraphics[width=0.95\textwidth]{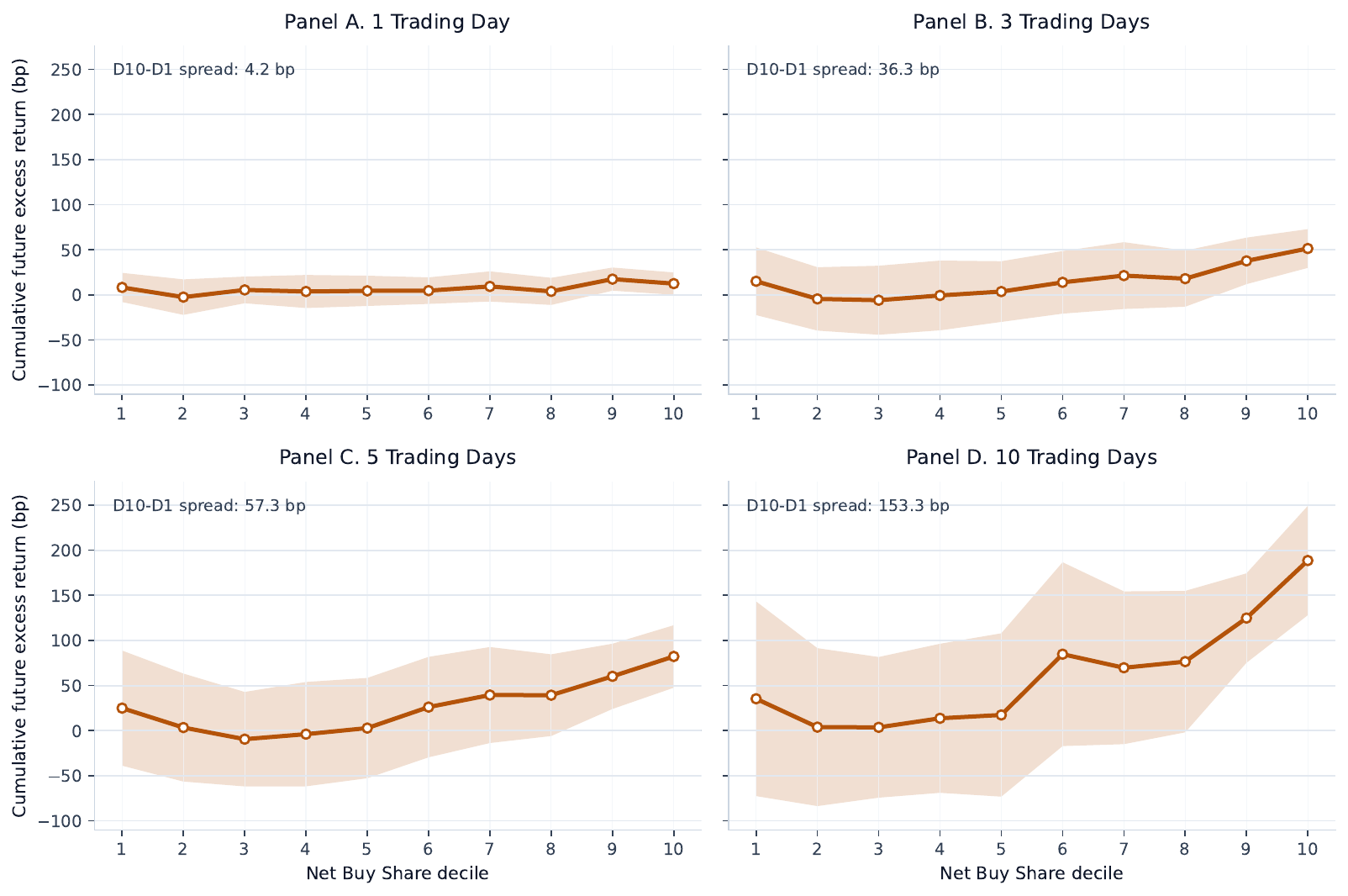}
\caption{Digital-Twin Stock-Pick Net Buy Share and Future Excess Returns}
\label{fig:stock_return_gradient}
\vspace{4pt}
\begin{minipage}{0.95\textwidth}
\footnotesize
This figure sorts stock-events into event-date deciles of Net Buy Share and plots equal-weight mean cumulative future excess returns. Returns are measured relative to the S\&P 500 and reported in basis points at the 1-, 3-, 5-, and 10-trading-day horizons. Deciles are formed within each event date. For each decile, the plotted value is the average return across event dates. D10-D1 is the top-minus-bottom decile spread. Shaded bands denote 90\% confidence intervals computed with Newey--West standard errors using lag length $h-1$ for horizon $h$. The sample uses 81 accounts, the 429-stock universe, and 50 stock-pick event dates from December 15, 2025 through March 16, 2026.
\end{minipage}
\end{figure}

Figure \ref{fig:stock_return_gradient} presents the same directional results visually. Stocks are sorted into event-day deciles based on \emph{Net Buy Share}, and the figure plots average future excess returns for each decile. The one-day panel is essentially flat, but the relation becomes positive at longer horizons. By the ten-day horizon, the top-minus-bottom decile spread is $153.3$ basis points. Figure \ref{fig:stock_return_gradient} corroborates the statistical result from Table \ref{tab:q1_stockpick_fmb_calendar_time_returns}: stocks that the digital twins rank more favorably tend to have higher subsequent excess returns, especially beyond the one-day horizon.

Table \ref{tab:q3_stockpick_returns_rv_controls} asks whether the results in Table \ref{tab:q1_stockpick_fmb_calendar_time_returns} survive controls for lagged stock-level excess returns and realized volatility. For each horizon \(h\) and event date \(t\), we estimate the cross-sectional regression
\[
ExRet_{j,t}^{(h)}
=
a_{t,h}
+b_{t,h} Direction_{j,t}
+\delta_{t,h} LagExRet_{j,t}^{(h)}
+\theta_{t,h} LagExVol_{j,t}^{(h)}
+\varepsilon_{j,t}^{(h)} .
\]
The controls are horizon-matched: \(LagExRet_{j,t}^{(h)}\) is the stock's cumulative excess return over the prior \(h\) trading days, and \(LagExVol_{j,t}^{(h)}\) is realized excess-return volatility over the same prior window. Regression coefficients ($b_{t,h}$, $\delta_{t,h}$, and $\theta_{t,h}$) are averaged across time ($t$), with Newey--West standard errors computed from the time series of daily coefficients using lag length $h-1$ \citep{NeweyWest1987}.

The controlled specifications support the results in Table \ref{tab:q1_stockpick_fmb_calendar_time_returns}. All four directional variables continue to predict the cross-section of stock returns at the ten-day horizon. For example, in Panel B, a ten-percentage-point increase in \emph{Net Buy Share} is associated with $38.5$ basis points higher future excess return at the ten-day horizon. Panel D gives the same message with \emph{Conviction-Weighted Meaningful Tilt}, where a one-point increase is associated with $15.8$ basis points higher ten-day excess return.\footnote{A one-standard-deviation increase in \emph{Conviction-Weighted Meaningful Tilt} is associated with $48.5$ basis points higher excess return over the next ten trading days.} Overall, this table shows that the directional signals are not simply restating a stock's recent excess return or realized volatility, but continue to predict future excess returns after those controls.

\begin{table}[!htbp]
\caption{Digital-Twin Stock-Pick Direction and Future Excess Returns with Controls}
\label{tab:q3_stockpick_returns_rv_controls}
{\footnotesize{}This table repeats the daily Fama--MacBeth specifications in columns (1)--(3) of Table \ref{tab:q1_stockpick_fmb_calendar_time_returns} after adding horizon-matched lagged excess return and lagged excess realized-volatility controls to each daily Fama--MacBeth cross-section. Return horizons, direction-signal scaling, Newey--West standard errors, N, and mean daily adjusted $R^2$ follow the Table \ref{tab:q1_stockpick_fmb_calendar_time_returns} conventions; N is the total number of usable stock-event observations across daily cross-sections. Control coefficients are percentage-point return changes for a one-percentage-point increase in the control. The sample uses 429 stocks, 81 accounts in the stock-pick interview branch, and 50 event dates from December 15, 2025 through March 16, 2026. Indicators ***, **, * denote statistical significance at the 1\%, 5\%, and 10\% level, respectively.}
{\footnotesize\par}
\vspace{5pt}
\centering{}
{\footnotesize
\setlength{\tabcolsep}{2.5pt}
\begin{tabular}{lccc}
\hline\hline
 & \makebox[2.75cm][c]{(1)} & \makebox[2.75cm][c]{(2)} & \makebox[2.75cm][c]{(3)} \\
\noalign{\vskip 4pt}
 & \multicolumn{3}{c}{\makebox[8.20cm][c]{Dependent variable: Cumulative future excess return (pp)}} \\
\cmidrule(lr){2-4}
\noalign{\vskip 2pt}
 & \makebox[2.75cm][c]{1 trading day} & \makebox[2.75cm][c]{5 trading days} & \makebox[2.75cm][c]{10 trading days} \\
\cmidrule(lr){2-2}\cmidrule(lr){3-3}\cmidrule(lr){4-4}
\noalign{\vskip 4pt}
\multicolumn{4}{l}{\textit{Panel A. Tilt}} \\[-1pt]
\noalign{\hrule height 0.15pt}
\noalign{\vskip 2pt}
Direction signal & \makebox[2.75cm][c]{0.035} & \makebox[2.75cm][c]{0.686} & \makebox[2.75cm][c]{1.515$^{*}$} \\
 & \makebox[2.75cm][c]{(0.125)} & \makebox[2.75cm][c]{(0.440)} & \makebox[2.75cm][c]{(0.805)} \\[1pt]
Lagged excess return & \makebox[2.75cm][c]{-0.029} & \makebox[2.75cm][c]{0.031} & \makebox[2.75cm][c]{0.061} \\
 & \makebox[2.75cm][c]{(0.023)} & \makebox[2.75cm][c]{(0.019)} & \makebox[2.75cm][c]{(0.066)} \\[1pt]
Lagged realized vol. & \makebox[2.75cm][c]{0.064$^{*}$} & \makebox[2.75cm][c]{0.118$^{**}$} & \makebox[2.75cm][c]{0.118} \\
 & \makebox[2.75cm][c]{(0.034)} & \makebox[2.75cm][c]{(0.055)} & \makebox[2.75cm][c]{(0.081)} \\[1pt]
\noalign{\hrule height 0.15pt}
\noalign{\vskip 1pt}
N & \makebox[2.75cm][c]{21,410} & \makebox[2.75cm][c]{21,410} & \makebox[2.75cm][c]{21,410} \\
Mean daily adj. $R^2$ & \makebox[2.75cm][c]{0.042} & \makebox[2.75cm][c]{0.035} & \makebox[2.75cm][c]{0.051} \\
\hline
\noalign{\vskip 4pt}
\multicolumn{4}{l}{\textit{Panel B. Net Buy Share}} \\[-1pt]
\noalign{\hrule height 0.15pt}
\noalign{\vskip 2pt}
Direction signal & \makebox[2.75cm][c]{0.008} & \makebox[2.75cm][c]{0.175} & \makebox[2.75cm][c]{0.385$^{**}$} \\
 & \makebox[2.75cm][c]{(0.030)} & \makebox[2.75cm][c]{(0.107)} & \makebox[2.75cm][c]{(0.195)} \\[1pt]
Lagged excess return & \makebox[2.75cm][c]{-0.029} & \makebox[2.75cm][c]{0.031} & \makebox[2.75cm][c]{0.060} \\
 & \makebox[2.75cm][c]{(0.023)} & \makebox[2.75cm][c]{(0.019)} & \makebox[2.75cm][c]{(0.066)} \\[1pt]
Lagged realized vol. & \makebox[2.75cm][c]{0.063$^{*}$} & \makebox[2.75cm][c]{0.118$^{**}$} & \makebox[2.75cm][c]{0.117} \\
 & \makebox[2.75cm][c]{(0.034)} & \makebox[2.75cm][c]{(0.055)} & \makebox[2.75cm][c]{(0.079)} \\[1pt]
\noalign{\hrule height 0.15pt}
\noalign{\vskip 1pt}
N & \makebox[2.75cm][c]{21,410} & \makebox[2.75cm][c]{21,410} & \makebox[2.75cm][c]{21,410} \\
Mean daily adj. $R^2$ & \makebox[2.75cm][c]{0.043} & \makebox[2.75cm][c]{0.036} & \makebox[2.75cm][c]{0.054} \\
\hline
\noalign{\vskip 4pt}
\multicolumn{4}{l}{\textit{Panel C. Meaningful Tilt}} \\[-1pt]
\noalign{\hrule height 0.15pt}
\noalign{\vskip 2pt}
Direction signal & \makebox[2.75cm][c]{0.034} & \makebox[2.75cm][c]{0.134} & \makebox[2.75cm][c]{0.306$^{*}$} \\
 & \makebox[2.75cm][c]{(0.033)} & \makebox[2.75cm][c]{(0.087)} & \makebox[2.75cm][c]{(0.172)} \\[1pt]
Lagged excess return & \makebox[2.75cm][c]{-0.030} & \makebox[2.75cm][c]{0.028} & \makebox[2.75cm][c]{0.060} \\
 & \makebox[2.75cm][c]{(0.023)} & \makebox[2.75cm][c]{(0.018)} & \makebox[2.75cm][c]{(0.063)} \\[1pt]
Lagged realized vol. & \makebox[2.75cm][c]{0.060$^{*}$} & \makebox[2.75cm][c]{0.119$^{**}$} & \makebox[2.75cm][c]{0.126} \\
 & \makebox[2.75cm][c]{(0.034)} & \makebox[2.75cm][c]{(0.056)} & \makebox[2.75cm][c]{(0.083)} \\[1pt]
\noalign{\hrule height 0.15pt}
\noalign{\vskip 1pt}
N & \makebox[2.75cm][c]{20,823} & \makebox[2.75cm][c]{20,823} & \makebox[2.75cm][c]{20,823} \\
Mean daily adj. $R^2$ & \makebox[2.75cm][c]{0.041} & \makebox[2.75cm][c]{0.030} & \makebox[2.75cm][c]{0.042} \\
\hline
\noalign{\vskip 4pt}
\multicolumn{4}{l}{\textit{Panel D. Conviction-Weighted Meaningful Tilt}} \\[-1pt]
\noalign{\hrule height 0.15pt}
\noalign{\vskip 2pt}
Direction signal & \makebox[2.75cm][c]{0.006} & \makebox[2.75cm][c]{0.070} & \makebox[2.75cm][c]{0.158$^{*}$} \\
 & \makebox[2.75cm][c]{(0.013)} & \makebox[2.75cm][c]{(0.049)} & \makebox[2.75cm][c]{(0.091)} \\[1pt]
Lagged excess return & \makebox[2.75cm][c]{-0.033} & \makebox[2.75cm][c]{0.031$^{*}$} & \makebox[2.75cm][c]{0.062} \\
 & \makebox[2.75cm][c]{(0.023)} & \makebox[2.75cm][c]{(0.019)} & \makebox[2.75cm][c]{(0.066)} \\[1pt]
Lagged realized vol. & \makebox[2.75cm][c]{0.058$^{*}$} & \makebox[2.75cm][c]{0.121$^{**}$} & \makebox[2.75cm][c]{0.126} \\
 & \makebox[2.75cm][c]{(0.034)} & \makebox[2.75cm][c]{(0.054)} & \makebox[2.75cm][c]{(0.080)} \\[1pt]
\noalign{\hrule height 0.15pt}
\noalign{\vskip 1pt}
N & \makebox[2.75cm][c]{20,823} & \makebox[2.75cm][c]{20,823} & \makebox[2.75cm][c]{20,823} \\
Mean daily adj. $R^2$ & \makebox[2.75cm][c]{0.042} & \makebox[2.75cm][c]{0.035} & \makebox[2.75cm][c]{0.052} \\
\hline\hline
\end{tabular}
}
\end{table}


\subsubsection{Staggered-Vintage Design}
\label{sec:sv}

The Fama--MacBeth and Calendar-Time methods reported above in Subsection \ref{sec:fm} provide two established ways to obtain statistical inference when multi-day return windows overlap. In our sample, however, both of these methods are likely to have low statistical power. In the Fama--MacBeth tests, standard errors are obtained from the intertemporal variation in  date-specific cross-sectional slope estimates. In the calendar-time tests, standard errors are obtained from the inter-temporal variation of daily long–short portfolio returns. Both methods, therefore, draw their statistical power primarily from the time-series dimension of the data, which is relatively short in our setting compared with the much larger cross-section of stocks observed on each event date.\footnote{See Subsection \ref{sec:limitations} for a discussion of this paper's limitations associated with the sample size.}

A pooled panel regression can potentially provide greater statistical power because it retains the full stock-level cross-section rather than reducing each date to a single slope estimate or portfolio return. The difficulty is that, at horizons longer than one trading day, adjacent dependent variables reuse many of the same daily returns. Pooling such observations without adequately addressing this mechanical overlap can understate standard errors and produce misleading inference. To address this problem, we implement a staggered-vintage design that combines the larger cross-sectional information set with non-overlapping future-return windows. Within each vintage, anchor dates are spaced $h$ trading days apart, ensuring that the corresponding $h$-day return windows do not overlap; shifting the initial anchor date across vintages allows us to examine every possible calendar alignment. This design thus provides a complementary, potentially higher-power test of the signal–return relation without mechanically overlapping dependent-variable windows. The results are presented in Table \ref{tab:staggered_vintage_stockpick_direction_returns_main}.

\begin{table}[!htbp]
\caption{Staggered-Vintage Stock-Pick Direction and Future Excess Returns}
\label{tab:staggered_vintage_stockpick_direction_returns_main}
{\footnotesize{}This table reports staggered-vintage tests of whether digital-twin stock-pick direction predicts future excess returns after controlling for lagged excess returns and realized volatility. Panel A uses five-day formation and return windows and five vintages; Panel B uses ten-day windows and ten vintages. Each vintage is a separate regression sample with a different starting date and non-overlapping future-return windows. Signals and controls are measured over the formation window ending on each anchor date, and the dependent variable is the stock's compounded excess return relative to the S\&P 500 over the following return window, in percentage points. Each vintage regression includes anchor-date fixed effects and standard errors clustered by stock. The table reports the mean and range of the vintage coefficients, the number that are positive, and the number that are positive and significant at the 10\% level. Mean $N$ and adjusted $R^2$ are averaged across vintages. Because the vintages share the same calendar, the counts are descriptive. Coefficients are scaled to a +10 recommendation-point increase in Tilt or Meaningful Tilt, a +10 percentage-point increase in Net Buy Share, or a +1-unit increase in Conviction-Weighted Meaningful Tilt. The final four rows show two-sided empirical $p$-values from 10,000 placebo repetitions; a small value means that the observed mean coefficient is unusual under that placebo. The unrestricted placebo reassigns complete four-signal histories across stocks. The AR simulation placebo creates artificial histories with similar distributions and persistence. The size and industry placebos reassign observed histories without self-matches within initial market-capitalization quartiles or Fama--French 12 industries. The industry design preserves within-industry comovement in signals and returns while breaking the link between a stock and its own signal history. The sample comprises 429 stocks and 81 stock-pick accounts, with anchor dates from December 15, 2025 through March 16, 2026.}
{\footnotesize\par}
\vspace{5pt}
\centering{}
{\footnotesize
\setlength{\tabcolsep}{2.8pt}
\begin{tabular}{lcccc}
\hline\hline
 & \makebox[2.55cm][c]{(1)} & \makebox[2.55cm][c]{(2)} & \makebox[2.55cm][c]{(3)} & \makebox[2.55cm][c]{(4)} \\
\noalign{\vskip 4pt}
 & \multicolumn{4}{c}{Direction Signal} \\
\cmidrule(lr){2-5}
\noalign{\vskip 2pt}
 & \makebox[2.55cm][c]{\shortstack{Tilt}} & \makebox[2.55cm][c]{\shortstack{Net\\Buy Share}} & \makebox[2.55cm][c]{\shortstack{Meaningful\\Tilt}} & \makebox[2.55cm][c]{\shortstack{Conv.-Wt.\\Meaningful Tilt}} \\
\cmidrule(lr){2-2}\cmidrule(lr){3-3}\cmidrule(lr){4-4}\cmidrule(lr){5-5}
\noalign{\vskip 2pt}
\multicolumn{5}{l}{\textit{Panel A. 5-Trading-Day Formation and Return Windows}} \\[-1pt]
\noalign{\hrule height 0.15pt}
\noalign{\vskip 2pt}
Mean coefficient & \makebox[2.55cm][c]{0.723} & \makebox[2.55cm][c]{0.190} & \makebox[2.55cm][c]{0.185} & \makebox[2.55cm][c]{0.080} \\
Vintage range & \makebox[2.55cm][c]{[0.470, 0.886]} & \makebox[2.55cm][c]{[0.131, 0.232]} & \makebox[2.55cm][c]{[0.097, 0.251]} & \makebox[2.55cm][c]{[0.051, 0.101]} \\
Positive vintages & \makebox[2.55cm][c]{5/5} & \makebox[2.55cm][c]{5/5} & \makebox[2.55cm][c]{5/5} & \makebox[2.55cm][c]{5/5} \\
Positive and significant & \makebox[2.55cm][c]{5/5} & \makebox[2.55cm][c]{5/5} & \makebox[2.55cm][c]{4/5} & \makebox[2.55cm][c]{5/5} \\
\noalign{\hrule height 0.15pt}
\noalign{\vskip 1pt}
Mean $N$ & \makebox[2.55cm][c]{5,320} & \makebox[2.55cm][c]{5,320} & \makebox[2.55cm][c]{5,298} & \makebox[2.55cm][c]{5,298} \\
Mean adj. $R^2$ & \makebox[2.55cm][c]{0.054} & \makebox[2.55cm][c]{0.055} & \makebox[2.55cm][c]{0.054} & \makebox[2.55cm][c]{0.054} \\
\noalign{\hrule height 0.15pt}
\noalign{\vskip 2pt}
\multicolumn{5}{l}{\textit{Placebo $p$-values}} \\
Unrestricted & \makebox[2.55cm][c]{0.0006} & \makebox[2.55cm][c]{0.0002} & \makebox[2.55cm][c]{0.0018} & \makebox[2.55cm][c]{0.0006} \\
AR simulation & \makebox[2.55cm][c]{0.0010} & \makebox[2.55cm][c]{0.0004} & \makebox[2.55cm][c]{0.0128} & \makebox[2.55cm][c]{0.0028} \\
Size quartiles & \makebox[2.55cm][c]{0.0002} & \makebox[2.55cm][c]{0.0002} & \makebox[2.55cm][c]{0.0016} & \makebox[2.55cm][c]{0.0002} \\
FF12 industries & \makebox[2.55cm][c]{0.1802} & \makebox[2.55cm][c]{0.0642} & \makebox[2.55cm][c]{0.7661} & \makebox[2.55cm][c]{0.1176} \\
\noalign{\vskip 4pt}
\hline
\noalign{\vskip 3pt}
\multicolumn{5}{l}{\textit{Panel B. 10-Trading-Day Formation and Return Windows}} \\[-1pt]
\noalign{\hrule height 0.15pt}
\noalign{\vskip 2pt}
Mean coefficient & \makebox[2.55cm][c]{1.687} & \makebox[2.55cm][c]{0.444} & \makebox[2.55cm][c]{0.434} & \makebox[2.55cm][c]{0.184} \\
Vintage range & \makebox[2.55cm][c]{[1.275, 2.133]} & \makebox[2.55cm][c]{[0.339, 0.546]} & \makebox[2.55cm][c]{[0.339, 0.495]} & \makebox[2.55cm][c]{[0.140, 0.229]} \\
Positive vintages & \makebox[2.55cm][c]{10/10} & \makebox[2.55cm][c]{10/10} & \makebox[2.55cm][c]{10/10} & \makebox[2.55cm][c]{10/10} \\
Positive and significant & \makebox[2.55cm][c]{10/10} & \makebox[2.55cm][c]{10/10} & \makebox[2.55cm][c]{10/10} & \makebox[2.55cm][c]{10/10} \\
\noalign{\hrule height 0.15pt}
\noalign{\vskip 1pt}
Mean $N$ & \makebox[2.55cm][c]{2,660} & \makebox[2.55cm][c]{2,660} & \makebox[2.55cm][c]{2,659} & \makebox[2.55cm][c]{2,659} \\
Mean adj. $R^2$ & \makebox[2.55cm][c]{0.042} & \makebox[2.55cm][c]{0.043} & \makebox[2.55cm][c]{0.041} & \makebox[2.55cm][c]{0.042} \\
\noalign{\hrule height 0.15pt}
\noalign{\vskip 2pt}
\multicolumn{5}{l}{\textit{Placebo $p$-values}} \\
Unrestricted & \makebox[2.55cm][c]{0.0004} & \makebox[2.55cm][c]{0.0002} & \makebox[2.55cm][c]{0.0010} & \makebox[2.55cm][c]{0.0006} \\
AR simulation & \makebox[2.55cm][c]{0.0014} & \makebox[2.55cm][c]{0.0016} & \makebox[2.55cm][c]{0.0210} & \makebox[2.55cm][c]{0.0042} \\
Size quartiles & \makebox[2.55cm][c]{0.0002} & \makebox[2.55cm][c]{0.0002} & \makebox[2.55cm][c]{0.0024} & \makebox[2.55cm][c]{0.0002} \\
FF12 industries & \makebox[2.55cm][c]{0.0966} & \makebox[2.55cm][c]{0.0268} & \makebox[2.55cm][c]{0.4748} & \makebox[2.55cm][c]{0.0858} \\
\hline\hline
\end{tabular}
}
\end{table}


The basic idea is to partition the data into several regression samples, which we call \emph{vintages}. The last day on which we measure the signal and controls is the \emph{anchor date}; the future-return window begins on the following trading day. Consider one five-day vintage whose first anchor is Friday, February 27, 2026. We measure signals and controls from February 23 through February 27 and predict returns from March 2 through March 6. The next anchor in that vintage is March 6, and its return window is March 9 through March 13. The two future-return windows are adjacent but do not overlap. Shifting the first anchor forward one trading day creates the next vintage. Repeating that shift produces five different starting dates at the five-day horizon and ten at the ten-day horizon.

Panel A of Table \ref{tab:staggered_vintage_stockpick_direction_returns_main} presents results for the five-day return measurement horizon. The construction of five-day vintages is summarized below using numbered trading days. Each row is a separate regression sample. Reading across a row shows that one return window ends immediately before the next begins. Reading down the rows shows how the starting date shifts.

\begin{center}
\setlength{\tabcolsep}{6pt}
\begin{tabular}{lcc}
\toprule
Vintage & Anchor positions & Future-return windows \\
\midrule
Vintage 1 & 1, 6, 11, \ldots  & 2--6, 7--11, 12--16, \ldots \\
Vintage 2 & 2, 7, 12, \ldots  & 3--7, 8--12, 13--17, \ldots \\
Vintage 3 & 3, 8, 13, \ldots  & 4--8, 9--13, 14--18, \ldots \\
Vintage 4 & 4, 9, 14, \ldots  & 5--9, 10--14, 15--19, \ldots \\
Vintage 5 & 5, 10, 15, \ldots & 6--10, 11--15, 16--20, \ldots \\
\bottomrule
\end{tabular}
\end{center}

Panel B of Table \ref{tab:staggered_vintage_stockpick_direction_returns_main} presents results for the ten-day return measurement horizon, with ten-day vintages constructed as follows: 

\begin{center}
\setlength{\tabcolsep}{6pt}
\begin{tabular}{lcc}
\toprule
Vintage & Anchor positions & Future-return windows \\
\midrule
Vintage 1  & 1, 11, 21, \ldots  & 2--11, 12--21, 22--31, \ldots \\
Vintage 2  & 2, 12, 22, \ldots  & 3--12, 13--22, 23--32, \ldots \\
\(\vdots\) & \(\vdots\)         & \(\vdots\) \\
Vintage 10 & 10, 20, 30, \ldots & 11--20, 21--30, 31--40, \ldots \\
\bottomrule
\end{tabular}
\end{center}

Throughout Table \ref{tab:staggered_vintage_stockpick_direction_returns_main}, the dependent variable is future excess return, the independent variable is one of the four $directional$ $signals$, and the controls are the same horizon-matched lagged excess-return and realized-volatility measures used in Table \ref{tab:q3_stockpick_returns_rv_controls}. Anchor-date fixed effects preserve the within-date comparison across stocks, and standard errors are clustered by stock. 

Columns (1)--(4) provide separate analyses for each of the four $directional$ $signals$. Within each column, we pool the stock-level observations assigned to a vintage and estimate one regression. The variables of interest are the coefficients $\widehat{b_v}$ of each $directional$ $signal$, for each vintage, $v$. This approach produces five coefficients at the five-day horizon (Panel A) and ten coefficients at the ten-day horizon (Panel B) for each signal. In the top half of each Table \ref{tab:staggered_vintage_stockpick_direction_returns_main} panel, we report the mean and range of these $\widehat{b_v}$ coefficients, the number that are positive, and the number that are positive and statistically significant at the $10\%$ level (using a two-tail test):
\[
\left.
\begin{array}{rcl}
\text{Vintage 1 sample}  & \longrightarrow & (\widehat b_{1},p_{1}) \\
\text{Vintage 2 sample}  & \longrightarrow & (\widehat b_{2},p_{2}) \\
                         & \vdots           &                         \\
\text{Vintage $n$ sample} & \longrightarrow & (\widehat b_{n},p_{n})
\end{array}
\right\}
\longrightarrow
\begin{array}{c}
\text{mean and range of }\widehat b_v \\
\text{number of positive }\widehat b_v \\
\text{number that are positive and significant}
\end{array}
\]

For these conventional regression tests, the null hypothesis is \(H_0: b_v = 0\). Therefore, zero is the reference value for the vintage-specific t-statistics and for the significance counts reported in the top half of each panel.

The coefficient $\widehat{b_v}$ is positive in every vintage and statistically significant at the 10\% level in nearly every one. All $20$ five-day coefficients are positive, and $19$ of them are positive and significant at the $10\%$ level (see top half of Panel A). All $40$ ten-day coefficients are positive and significant at the 10\% level, and their magnitudes are also similar to the controlled Fama--MacBeth estimates (see top half of Panel B). A ten-percentage-point increase in \emph{Net Buy Share} (see Column (2)) predicts an average of $19.0$ basis points higher five-day excess return across vintages, with estimates ranging from $13.1$ to $23.2$ basis points. The corresponding ten-day average is $44.4$ basis points, with a range of $33.9$ to $54.6$ basis points. Table \ref{tab:q3_stockpick_returns_rv_controls} reports comparable estimates of $17.5$ and $38.5$ basis points.\footnote{\emph{Conviction-Weighted Meaningful Tilt} gives the same message. A one-unit increase predicts an average of $8.0$ basis points at five days and $18.4$ basis points at ten days, with vintage ranges of $5.1$--$10.1$ and $14.0$--$22.9$ basis points. The corresponding Table \ref{tab:q3_stockpick_returns_rv_controls} estimates are $7.0$ and $15.8$ basis points.} Thus, our results survive non-overlapping return windows and every possible starting date.

The vintages represent alternative calendar alignments of the same finite sample and are not independent; the observed signals also exhibit time-series persistence. To account for these features, we use simulated placebo distributions to assess whether the staggered-vintage procedure, when applied to signals with similar empirical properties but no stock-specific link to future returns, could generate an average coefficient as large as the one observed. We construct four placebo benchmarks that preserve different features of the observed data. Three reassign complete observed signal histories among stocks, while the fourth generates artificial histories with similar distributions and persistence. For each placebo draw, we rerun the full staggered-vintage analysis and calculate the mean coefficient across vintages. Repeating this process 10,000 times produces a reference distribution against which to compare the observed mean of the $\widehat{b_v}$ coefficients. 

The placebo tests use a different inferential benchmark. In the bottom half of each Table \ref{tab:staggered_vintage_stockpick_direction_returns_main} panel, we compare the observed mean coefficient across vintages with the empirical distribution of mean coefficients generated by the 10,000 random draws under each placebo procedure. The reported two-sided empirical $p$-value measures how unusual the observed mean coefficient is relative to the corresponding placebo distribution, rather than whether it differs from zero.\footnote{The empirical mean of the placebo distribution summarizes the center of this benchmark distribution. Thus, zero and the placebo benchmark coincide approximately when the simulated distribution is centered near zero, but they need not coincide, for example, when industry-level comovement shifts the placebo distribution to the right of zero. We examine these industry comobements in our last placebo test below.}

The \emph{unrestricted} placebo (top line in each panel's bottom half) reassigns each stock's complete four-signal history across the $429$-stock universe. This preserves the observed time pattern and the relationships among the four direction measures but breaks the link between a stock's signal history and its returns. Under this empirical distribution, the p-values of the mean $\widehat{b_v}$ coefficients are $0.0018$ or smaller, suggesting that the observed directional signals do not produce similar coefficients when they are assigned arbitrarily across stocks. 

\begin{sidewaysfigure}[p]
\global\pdfpageattr{/Rotate 90}
\afterpage{\global\pdfpageattr{}}
\centering
\begin{minipage}[t]{0.455\linewidth}
\centering
Panel A. Unrestricted Placebo

\includegraphics[width=\linewidth]{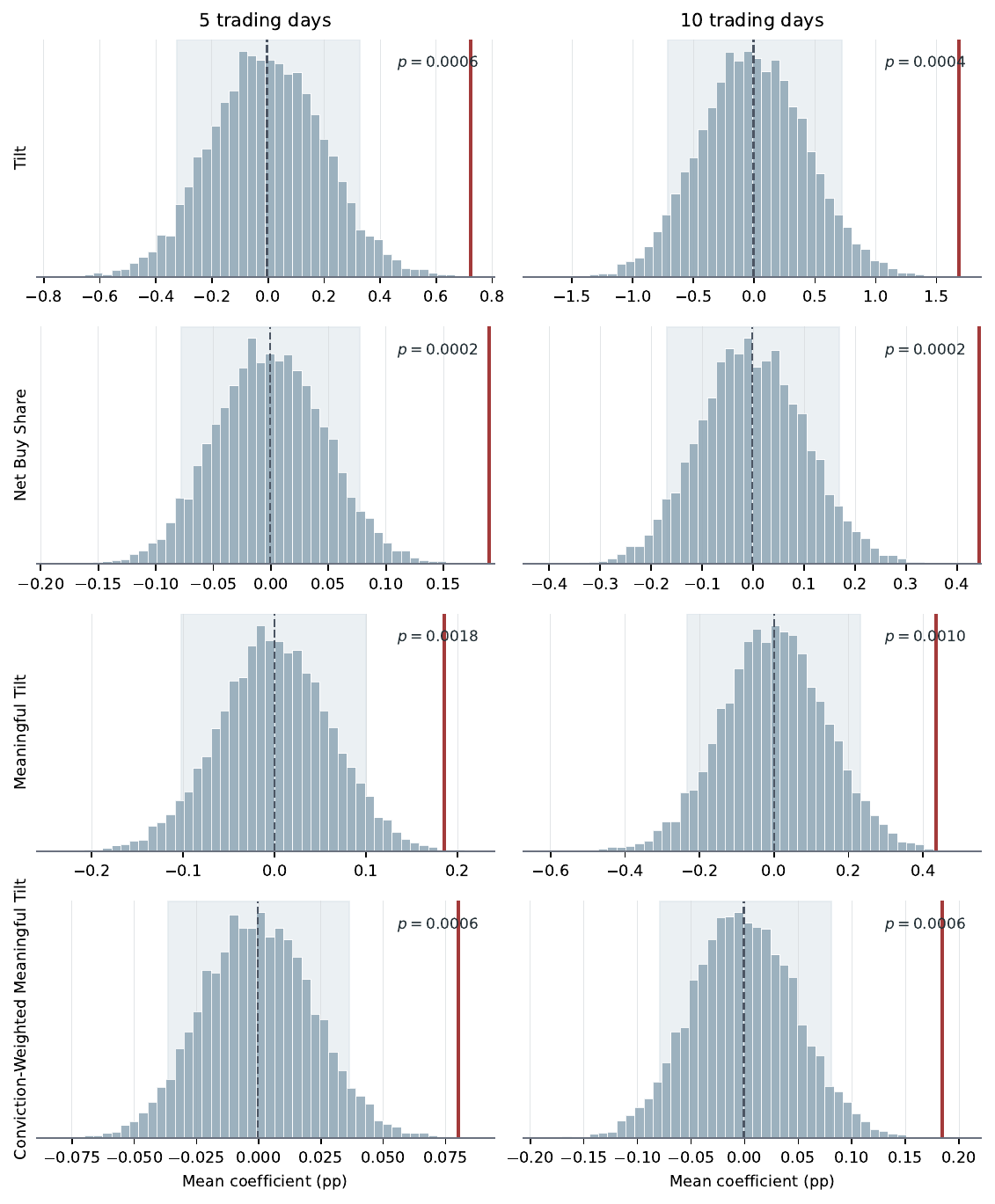}
\end{minipage}\hfill
\begin{minipage}[t]{0.455\linewidth}
\centering
Panel B. Industry Placebo

\includegraphics[width=\linewidth]{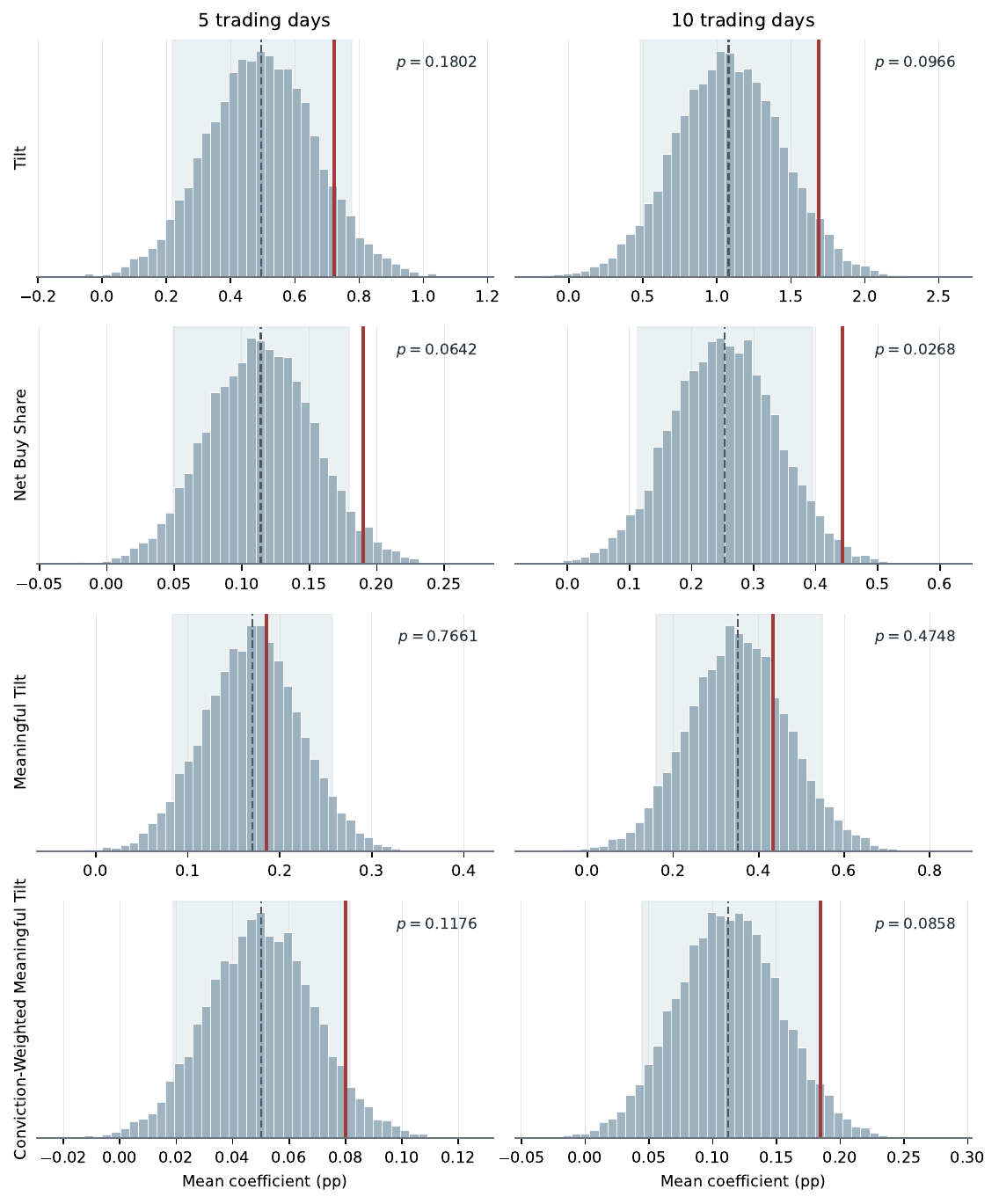}
\end{minipage}
\caption{Placebo Distributions for Staggered-Vintage Return Coefficients}
\label{fig:staggered_vintage_placebo_histograms}
\vspace{3pt}
\begin{minipage}{0.98\linewidth}
\footnotesize
This figure compares the mean staggered-vintage coefficients in the actual data with distributions produced by 10,000 placebo reassignments. Panel A randomly reassigns each stock's complete four-signal history across the 429-stock universe. Panel B reassigns complete histories without self-matches among stocks in the same Fama--French 12 industry. Each reassignment moves the four direction measures together and is used at both return horizons. Rows correspond to the four direction measures, and columns correspond to the five- and ten-day horizons. In each histogram, gray bars show the placebo mean coefficients, the shaded area contains the middle 90\% of them, the dashed line marks their mean, and the solid red line marks the mean coefficient in the actual data. Reported $p$-values are two-sided empirical comparisons between the observed coefficients and their placebo distributions. Coefficient scaling and the sample follow Table \ref{tab:staggered_vintage_stockpick_direction_returns_main}.
\end{minipage}
\end{sidewaysfigure}

This result is also illustrated visually in Panel A of Figure \ref{fig:staggered_vintage_placebo_histograms}, which shows eight histograms comparing the mean $\widehat{b_v}$ coefficients in the actual data with those produced by the unrestricted placebo simulations. In each histogram, the gray bars show the empirical distribution of the $10{,}000$ placebo mean coefficients, the shaded area contains the $90\%$ confidence interval, the dashed line shows the average of the placebo coefficients, and the solid line to the right is the mean $\widehat{b_v}$ coefficient in the actual data. Across all four measures and two event horizons, the unrestricted distributions are centered near zero, while the solid lines lie far to the right, outside the $90\%$ confidence interval. This visual separation corroborates the $p$-values shown in the table.

\providecommand{\staggeredVintageFigurePath}{live/01_exhibits/figures}


The \emph{AR simulation} placebo (second line in each panel's bottom half) asks whether persistent signals in a short sample could mechanically produce similar coefficients. It generates artificial histories with the same overall signal distributions and approximately the same adjacent-date persistence as the observed histories.\footnote{For \emph{Net Buy Share}, the observed mean and standard deviation are $0.1078$ and $0.1423$, compared with averages of $0.1078$ and $0.1421$ across the artificial panels. The average adjacent-date correlation is $0.361$ in the data and $0.359$ in the simulations. For the other direction measures, observed average correlations range from $0.261$ to $0.461$, compared with $0.262$ to $0.465$ in the simulations. The four measures are generated separately, so the exercise reproduces each measure's behavior rather than the exact mathematical links among them.} The \emph{AR simulation} $p$-values range from $0.0004$ to $0.0210$, indicating that persistence and the short sample do not by themselves explain the observed positive $\widehat{b_v}$ coefficients.

The \emph{Size-quartile} placebo (third line in each panel's bottom half) restricts the reassignment to stocks in the same beginning-of-sample market-capitalization quartile. This preserves broad size differences while breaking the stock--directional-signal match. Every  $p$-value obtained under the \emph{Size-quartile} distribution is $0.0024$ or smaller, suggesting that broad differences in firm size do not explain the positive  $\widehat{b_v}$ coefficients either.

Finally, in the \emph{FF12 industries} placebo (bottom line of each panel) we reassign directional signals to other stocks using each signal's observed history, but restrict the reassignment more sharply. We move a stock's complete history to another stock in the same Fama--French 12-industry classification, using the four-digit SIC-code mapping provided in the Kenneth R. French Data Library,\footnote{See https://mba.tuck.dartmouth.edu/pages/faculty/ken.french/data\_library/det\_12\_ind\_port.html.} and no stock keeps its own history. The test therefore asks whether the actual directional signals produced by digital twins remain informative after keeping the shared industry component in place.

Because signals and returns are both correlated within industries, these placebo distributions are shifted to the right of zero, as shown in Panel B of Figure \ref{fig:staggered_vintage_placebo_histograms}. This shift is expected and makes the \emph{FF12 industries} placebo a more stringent benchmark than the preceding placebos: the observed mean coefficient must be large relative to a null distribution that already preserves industry-level comovement in both signals and returns.\footnote{Digital-twin signal histories are correlated within industries because the interviews can express common sector and thematic views across related stocks. Returns of stocks in the same industry also move together. By reassigning histories only within industries, the placebo keeps both sources of industry comovement in place while breaking the match between a particular stock and a particular signal history.} Nonetheless, even relative to this more stringent benchmark, the observed mean $\widehat{b_v}$ coefficients remain statistically significant at the 10\% level for the ten-day horizon in the case of three of our four directional signals: \emph{Tilt}, \emph{Net Buy Share}, and \emph{Conviction-Weighted Meaningful Tilt} (Columns (1), (2), and (4) of Panel B). The mean $\widehat{b_v}$ coefficient obtained with \emph{Net Buy Share} also preserves its significance at the 10\% level for the five-day horizon (Column (2) of Panel A). For example, for a ten-percentage-point increase in \emph{Net Buy Share}, excess returns obtained under the industry-placebo distribution are $11.4$ basis points at the five-day horizon and $25.3$ basis points at ten-day horizon, compared with observed coefficients of $19.0$ and $44.4$ basis points. The corresponding $p$-values are $0.0642$ and $0.0268$.

Taken together, the placebo tests conducted with the \emph{Unrestricted}, \emph{AR simulation}, and \emph{Size-quartiles} methods show that the statistical significance of $\widehat{b_v}$ estimates is not readily explained by arbitrary signal-to-stock matching, broad size differences, or persistent signals in a short sample. The \emph{FF 12 industries} placebo tests show that some predictability is shared within industries, as expected when the interviews elicit sector and thematic views. Yet, after preserving within-industry correlation in both signals and returns, \emph{Net Buy Share} remains statistically significant at both five- and ten-day horizons, as do ten-day \emph{Tilt} and \emph{Conviction-Weighted Meaningful Tilt}. Thus, while industry comovement appears to contribute to the signal-return relation, it does not fully account for it.

\subsection{The Silent Region Return Signal}
\label{sec:silent}

We now return to the silent region, which consists of stock-date pairs that are not specifically covered by human finfluencers. We ask whether cross-sectional return predictability in the silent region is different from predictability resulting from actual posts. We show that return predictability is much stronger in the silent region. Recommendations made by digital twins that are associated with public posts have weaker predictability. 

Table \ref{tab:q7_stockpick_public_post_comparison} compares public recommendations with the additional coverage created by the interview protocol. The table follows the same controlled Fama--MacBeth design as Table \ref{tab:q3_stockpick_returns_rv_controls}, and each panel again uses one directional signal measure. The difference is that each panel now separates three sources of information. The first row uses the public-post signal on stock-dates that have both a public post and an interview signal for the same stock on the same date. The second row uses the digital-twin interview signal on those same overlap stock-dates. The third row uses the digital-twin interview signal on stock-dates with no same-date public post. Because the signal definitions are fixed across public posts and interviews, differences across rows reflect differences in the information used to build the signal: observed public recommendations, interview responses on stock-dates that overlap with public posts, or interview responses in the no-post region.\footnote{The four direction measures are constructed the same way for public posts and interviews. Recommendations are mapped to the same 0--100 scale, neutral is 50, buy and sell indicators use the same 65/35 cutoffs, and the same formulas produce \emph{Tilt}, \emph{Net Buy Share}, \emph{Meaningful Tilt}, and \emph{Conviction-Weighted Meaningful Tilt}. The difference is support. Public-post signals use only observed public recommendations, so an overlap stock-date is often based on a single public account-stock-date event; in the overlap sample, the median public stock-date has one such event. As a result, public \emph{Net Buy Share} is often a discrete value such as $-1$, $0$, or $1$, whereas the interview signal averages across the digital-twin panel responses for the same stock-date.}

The three rows ask three separate questions. The top row (public-signal-overlap) asks whether the visible recommendation predicts returns. The middle row (interview-signal-overlap) holds fixed the publicly visible stock-date and asks whether the broader interview panel adds information on that same stock-date. The bottom row (interview-signal-no-post) asks whether the interview protocol recovers directional information in the silent region.

The results show two clear patterns. First, public recommendations do not predict future excess returns in the overlap sample. Across all four signal definitions, the public-signal coefficients are small and statistically insignificant. Second, the interview signal is strongest in the silent region. For stock-dates with no same-date public post, the interview coefficients are positive across three of the four direction measures and become larger at longer horizons. A ten-percentage-point increase in interview \emph{Net Buy Share} is associated with $24$ basis points higher five-day excess return and $50$ basis points higher ten-day excess return. The \emph{Tilt} and \emph{Conviction-Weighted Meaningful Tilt} rows again show the same broad pattern, while \emph{Meaningful Tilt} is statistically insignificant. Together, the results show that the silent region does more than increase coverage. It is where the interview design recovers its strongest directional return signal.

The interview-signal estimates on the overlap row answer a narrower question. When a stock is already publicly discussed, does the broader interview panel add anything beyond the public post? On overlap dates, the public channel reveals that at least one account chose to post about the stock, while the interview protocol adds contemporaneous views from many accounts that did not post. Most interview-overlap coefficients are not statistically distinguishable from zero, so this evidence should not be overstated. The main exceptions are \emph{Net Buy Share}---where a ten-percentage-point increase in interview \emph{Net Buy Share} is associated with $31$ basis points higher ten-day excess return---and \emph{Tilt}. This suggests that broader interview coverage may add some information even on stock-dates with public posts, but the evidence is much weaker than in the no-post region.

\begin{table}[!htbp]
\caption{Coverage Gain, Public Recommendations, and Interview Signals}
\label{tab:q7_stockpick_public_post_comparison}
{\footnotesize{}This table compares observed public recommendations with the additional stock coverage created by the interview protocol. The dependent variable is cumulative future excess stock return relative to the S\&P 500, in percentage points. Each panel reports controlled daily Fama--MacBeth regressions for one signal definition in three samples: public-post signals on exact public-post/interview stock-date overlaps, interview signals on those same overlaps, and interview signals on stock-dates with no same-date public post. Public and interview signals use the same 0--100 recommendation-score scale, neutral score of 50, buy/sell cutoffs of 65/35, meaningful cutoff of 15 points from neutral, and aggregation formula. Public-signal rows are observed-disclosure aggregates and may be based on few posts; interview-signal rows are digital-twin panel aggregates. All rows use 429 stocks and 81 accounts (65 have public stock-recommendation posts). The interview panel contains 21,410 stock-events: 3,251 (15.2\%) coincide with a same-stock, same-date public recommendation post, and 18,159 (84.8\%) have no same-date public post. Return columns report 1-, 5-, and 10-trading-day horizons. N reports usable stock-event observations after requiring the signal, future return, and horizon-matched lagged excess-return and realized-volatility controls. Coefficient scales are +10 percentage points for Net Buy Share, +10 recommendation points for Tilt and Meaningful Tilt, and +1 unit for Conviction-Weighted Meaningful Tilt. Newey--West standard errors with lag $h-1$ are shown beneath estimates. Indicators ***, **, * denote significance at the 1\%, 5\%, and 10\% level.}
{\footnotesize\par}
\vspace{5pt}
\centering{}
{\footnotesize
\setlength{\tabcolsep}{2.5pt}
\begin{tabular}{lcccc}
\hline\hline
 & \makebox[1.30cm][c]{(1)} & \makebox[2.75cm][c]{(2)} & \makebox[2.75cm][c]{(3)} & \makebox[2.75cm][c]{(4)} \\
\noalign{\vskip 4pt}
 & \makebox[1.30cm][c]{} & \multicolumn{3}{c}{\makebox[8.20cm][c]{Dependent variable: Cumulative future excess return (pp)}} \\
\cmidrule(lr){3-5}
\noalign{\vskip 2pt}
 & \makebox[1.30cm][c]{N} & \makebox[2.75cm][c]{1 trading day} & \makebox[2.75cm][c]{5 trading days} & \makebox[2.75cm][c]{10 trading days} \\
\cmidrule(lr){2-2}\cmidrule(lr){3-3}\cmidrule(lr){4-4}\cmidrule(lr){5-5}
\noalign{\vskip 3pt}
\noalign{\vskip 4pt}
\multicolumn{5}{l}{\textit{Panel A. Tilt}} \\[-1pt]
\noalign{\hrule height 0.15pt}
\noalign{\vskip 2pt}
Public Signal (Overlap) & \makebox[1.30cm][c]{3,251} & \makebox[2.75cm][c]{-0.027} & \makebox[2.75cm][c]{0.008} & \makebox[2.75cm][c]{0.077} \\
 & \makebox[1.30cm][c]{} & \makebox[2.75cm][c]{(0.063)} & \makebox[2.75cm][c]{(0.111)} & \makebox[2.75cm][c]{(0.170)} \\[1pt]
Interview Signal (Overlap) & \makebox[1.30cm][c]{3,251} & \makebox[2.75cm][c]{-0.043} & \makebox[2.75cm][c]{0.411} & \makebox[2.75cm][c]{1.072$^{*}$} \\
 & \makebox[1.30cm][c]{} & \makebox[2.75cm][c]{(0.170)} & \makebox[2.75cm][c]{(0.440)} & \makebox[2.75cm][c]{(0.605)} \\[1pt]
Interview Signal (No Post) & \makebox[1.30cm][c]{18,159} & \makebox[2.75cm][c]{0.190} & \makebox[2.75cm][c]{1.017$^{**}$} & \makebox[2.75cm][c]{2.087$^{**}$} \\
 & \makebox[1.30cm][c]{} & \makebox[2.75cm][c]{(0.147)} & \makebox[2.75cm][c]{(0.455)} & \makebox[2.75cm][c]{(0.906)} \\[1pt]
\hline
\noalign{\vskip 4pt}
\multicolumn{5}{l}{\textit{Panel B. Net Buy Share}} \\[-1pt]
\noalign{\hrule height 0.15pt}
\noalign{\vskip 2pt}
Public Signal (Overlap) & \makebox[1.30cm][c]{3,251} & \makebox[2.75cm][c]{-0.008} & \makebox[2.75cm][c]{-0.003} & \makebox[2.75cm][c]{0.003} \\
 & \makebox[1.30cm][c]{} & \makebox[2.75cm][c]{(0.011)} & \makebox[2.75cm][c]{(0.025)} & \makebox[2.75cm][c]{(0.016)} \\[1pt]
Interview Signal (Overlap) & \makebox[1.30cm][c]{3,251} & \makebox[2.75cm][c]{-0.008} & \makebox[2.75cm][c]{0.134} & \makebox[2.75cm][c]{0.312$^{**}$} \\
 & \makebox[1.30cm][c]{} & \makebox[2.75cm][c]{(0.041)} & \makebox[2.75cm][c]{(0.100)} & \makebox[2.75cm][c]{(0.132)} \\[1pt]
Interview Signal (No Post) & \makebox[1.30cm][c]{18,159} & \makebox[2.75cm][c]{0.044} & \makebox[2.75cm][c]{0.242$^{**}$} & \makebox[2.75cm][c]{0.503$^{**}$} \\
 & \makebox[1.30cm][c]{} & \makebox[2.75cm][c]{(0.035)} & \makebox[2.75cm][c]{(0.114)} & \makebox[2.75cm][c]{(0.218)} \\[1pt]
\hline
\noalign{\vskip 4pt}
\multicolumn{5}{l}{\textit{Panel C. Meaningful Tilt}} \\[-1pt]
\noalign{\hrule height 0.15pt}
\noalign{\vskip 2pt}
Public Signal (Overlap) & \makebox[1.30cm][c]{2,476} & \makebox[2.75cm][c]{0.014} & \makebox[2.75cm][c]{0.030} & \makebox[2.75cm][c]{-0.113} \\
 & \makebox[1.30cm][c]{} & \makebox[2.75cm][c]{(0.056)} & \makebox[2.75cm][c]{(0.134)} & \makebox[2.75cm][c]{(0.185)} \\[1pt]
Interview Signal (Overlap) & \makebox[1.30cm][c]{3,225} & \makebox[2.75cm][c]{0.129} & \makebox[2.75cm][c]{0.021} & \makebox[2.75cm][c]{0.238} \\
 & \makebox[1.30cm][c]{} & \makebox[2.75cm][c]{(0.092)} & \makebox[2.75cm][c]{(0.166)} & \makebox[2.75cm][c]{(0.187)} \\[1pt]
Interview Signal (No Post) & \makebox[1.30cm][c]{17,589} & \makebox[2.75cm][c]{0.025} & \makebox[2.75cm][c]{0.149} & \makebox[2.75cm][c]{0.310} \\
 & \makebox[1.30cm][c]{} & \makebox[2.75cm][c]{(0.032)} & \makebox[2.75cm][c]{(0.092)} & \makebox[2.75cm][c]{(0.202)} \\[1pt]
\hline
\noalign{\vskip 4pt}
\multicolumn{5}{l}{\textit{Panel D. Conviction-Weighted Meaningful Tilt}} \\[-1pt]
\noalign{\hrule height 0.15pt}
\noalign{\vskip 2pt}
Public Signal (Overlap) & \makebox[1.30cm][c]{2,476} & \makebox[2.75cm][c]{0.001} & \makebox[2.75cm][c]{0.005} & \makebox[2.75cm][c]{-0.003} \\
 & \makebox[1.30cm][c]{} & \makebox[2.75cm][c]{(0.006)} & \makebox[2.75cm][c]{(0.015)} & \makebox[2.75cm][c]{(0.019)} \\[1pt]
Interview Signal (Overlap) & \makebox[1.30cm][c]{3,225} & \makebox[2.75cm][c]{-0.001} & \makebox[2.75cm][c]{0.033} & \makebox[2.75cm][c]{0.074} \\
 & \makebox[1.30cm][c]{} & \makebox[2.75cm][c]{(0.017)} & \makebox[2.75cm][c]{(0.043)} & \makebox[2.75cm][c]{(0.046)} \\[1pt]
Interview Signal (No Post) & \makebox[1.30cm][c]{17,589} & \makebox[2.75cm][c]{0.024} & \makebox[2.75cm][c]{0.109$^{**}$} & \makebox[2.75cm][c]{0.239$^{**}$} \\
 & \makebox[1.30cm][c]{} & \makebox[2.75cm][c]{(0.017)} & \makebox[2.75cm][c]{(0.056)} & \makebox[2.75cm][c]{(0.109)} \\[1pt]
\hline\hline
\end{tabular}
}
\end{table}


\subsection{Disagreement, Uncertainty, and Future Volatility}

We now turn to evaluating \emph{disagreement} and \emph{uncertainty} signals obtained from our digital twin interviews and ask whether they have different implications for future returns and volatility. This question is important for the paper's contribution. In public posts, silence is hard to interpret. A stock may be missing because views are split, because conviction is low, or because no one chose to post. Digital-twin interviews reduce this problem in two ways. They ask the same accounts about the same stock universe on each event date, and they ask each account to report both its directional view and its uncertainty about that view. That design lets us separate clear disagreement from broad uncertainty. Unlike the preceding subsection (\ref{sec:silent}), the tests here use all stock-dates in the interview panel, not only the silent region, and the state variables are constructed solely from interview responses.

Our analysis is summarized in Table \ref{tab:stockpick_state_variables}, which uses the same controlled Fama--MacBeth design as the return-predictability tables above. For each outcome \(Y\), horizon \(h\), and event date \(t\), we estimate the cross-sectional regression
\[
Y_{j,t}^{(h)}
=
a_{t,h}^{Y}
+b_{t,h}^{Y} State_{j,t}
+\delta_{t,h}^{Y} LagExRet_{j,t}^{(h)}
+\theta_{t,h}^{Y} LagExVol_{j,t}^{(h)}
+\varepsilon_{j,t}^{Y,(h)}.
\]
Here, \(Y_{j,t}^{(h)}\) is either stock \(j\)'s future excess return or future excess realized volatility over horizon \(h\). The state variable \(State_{j,t}\) is either \emph{Informative Polarization} or \emph{Uncertainty Score}, measured from the digital-twin stock-pick interviews before the outcome window begins. The controls, \(LagExRet_{j,t}^{(h)}\) and \(LagExVol_{j,t}^{(h)}\), are measured over the matching pre-event horizon. The table reports the average of the daily slope estimate, $b_{t,h}$, with Newey--West standard errors computed from the time series of daily coefficients using lag length $h-1$ \citep{NeweyWest1987}.

\emph{Informative Polarization} is a proxy for \emph{disagreement}. This variable measures whether digital twins with clear, low-speculation views about a stock disagree with each other. \emph{Informative Polarization} is high when these informative interview recommendations are split between buys and sells, and it is zero when they all point in the same direction. Thus, the variable captures disagreement among informed, non-neutral views, not general uncertainty. 

\emph{Uncertainty Score} is a proxy for the broader concept of \emph{uncertainty}. This variable is high when recommendations are mostly neutral, when non-neutral recommendations are speculative, or when meaningful recommendations lack supporting information. It is low when many digital twins give clear buy or sell recommendations that are grounded in the supplied account-specific material.

Both state variables are normalized to a scale from zero to one, and the table reports coefficients for a $0.10$-unit increase in these state variables.\footnote{Section \ref{appx: Data Appendix} of the Internet Appendix provides additional details on how these two state variables were constructed and normalized to the range from zero to one.}

Panel A of Table \ref{tab:stockpick_state_variables} shows the results using \emph{Future Excess Returns} as the dependent variable. \emph{Informative Polarization} is negatively related to future returns across the reported horizons. The ten-day coefficient implies that a ten-percentage-point increase predicts 27 basis points lower future excess return. This negative relation is broadly consistent with theoretical and empirical research linking disagreement to lower subsequent returns. In \citet{miller1977risk}, investors' inability to register pessimistic views causes prices to place greater weight on relatively optimistic valuations, implying lower subsequent returns. \citet{DietherMalloyScherbina2002} document a similar empirical relation using analyst forecast dispersion. Because our measure captures disagreement among digital-twin interview responses rather than analyst forecasts, we view this comparison as suggestive rather than direct.

The evidence is weaker for \emph{Uncertainty Score}. Its ten-day coefficient is also negative, but it is not statistically significant. This is not too surprising because uncertainty captures the speculative or noncommittal nature of responses, rather than disagreement between informative bullish and bearish views.

Panel B of Table \ref{tab:stockpick_state_variables} shows the results using \emph{Future Excess Volatility} as dependent variable. This Panel provides the most relevant results from Table \ref{tab:stockpick_state_variables}. \emph{Informative Polarization} predicts \emph{higher} future volatility at every horizon. A ten-percentage-point increase predicts $12$ basis points higher excess realized volatility at the ten-day horizon. This is what we would expect from a disagreement variable. When clear, low-speculation recommendations are split between the buy and sell sides, the stock's future price path becomes more volatile. In contrast, \emph{Uncertainty Score} has the opposite pattern. A ten-percentage-point increase predicts $27$ basis points \emph{lower} excess realized volatility at the ten-day horizon, with the same negative and significant sign at the shorter horizons. 

The contrast between these two state variables is a key contribution of our paper. The positive sign on \emph{Informative Polarization} shows that our digital twin interviews recover not only \emph{directional} informational content but also content about future \emph{volatility}. When twins have strong, diverging opinions about a given stock, the stock's returns are more volatile in the future. This result further corroborates the validity of our experimental design in that it provides a different setting in which interviews obtained from digital twins reveal value-relevant information.

\emph{Uncertainty Score}, on the other hand, captures weak convictions among digital twins. A stock-date pair with a high \emph{Uncertainty Score} means that the digital twins' directional views are mostly neutral, speculative, or unsupported. That corresponds to a low-attention state. A low-attention stock should have fewer opinion-driven trades, thus lower future realized volatility. This interpretation is consistent with models in which investor attention increases volatility by increasing the amount of information processed into prices. In \cite{AndreiHasler2015}, time-varying attention to past dividends generates time-varying stock-market volatility. Our \emph{Uncertainty Score} appears to capture the opposite state: a lack of grounded attention or conviction in the interview panel, which is followed by less volatile returns in the future.

\begin{table}[!htbp]
\caption{Digital-Twin Stock-Pick Disagreement, Uncertainty, and Future Outcomes}
\label{tab:stockpick_state_variables}
{\footnotesize{}This table reports daily Fama--MacBeth regressions of future outcomes on digital-twin stock-pick responses before the outcome window. Panel A uses cumulative future excess returns, measured relative to the S\&P 500. Panel B uses future excess realized volatility. Outcomes are reported in percentage points over 1, 5, and 10 trading days. Coefficients are percentage-point changes in the outcome for a +0.10 increase in the independent variable. Informative Polarization captures disagreement among meaningful low-speculation recommendations. Uncertainty Score is one minus average certainty coverage, where meaningful recommendations receive certainty weight $1-\mathrm{speculation}/100$ and non-meaningful responses receive zero weight. All rows use 429 stocks and 81 accounts. Each daily cross-section includes horizon-matched controls for lagged excess return and lagged excess realized volatility. Newey--West standard errors with lag $h-1$ are shown beneath estimates. Informative Polarization requires at least three meaningful low-speculation recommendations, so those rows use 10,536 stock-event observations over 50 event days; Uncertainty Score rows use 21,410 stock-event observations over 50 event days. All rows also require nonmissing future outcomes and matching lagged controls. The sample runs from December 15, 2025 through March 16, 2026. Indicators ***, **, * denote significance at the 1\%, 5\%, and 10\% level.}
{\footnotesize\par}
\vspace{7pt}
\centering{}
{\footnotesize
\setlength{\tabcolsep}{2.5pt}
\begin{tabular}{lccc}
\hline\hline
 & \makebox[2.75cm][c]{(1)} & \makebox[2.75cm][c]{(2)} & \makebox[2.75cm][c]{(3)} \\
\noalign{\vskip 6pt}
\multicolumn{4}{l}{\textit{Panel A. Cumulative Future Excess Returns}} \\[-1pt]
\noalign{\hrule height 0.15pt}
\noalign{\vskip 2pt}
 & \multicolumn{3}{c}{\makebox[8.20cm][c]{Dependent variable: Cumulative future excess return (pp)}} \\
\cmidrule(lr){2-4}
\noalign{\vskip 2pt}
 & \makebox[2.75cm][c]{1 trading day} & \makebox[2.75cm][c]{5 trading days} & \makebox[2.75cm][c]{10 trading days} \\
\cmidrule(lr){2-2}\cmidrule(lr){3-3}\cmidrule(lr){4-4}
\noalign{\vskip 3pt}
Informative Polarization & \makebox[2.75cm][c]{-0.038$^{**}$} & \makebox[2.75cm][c]{-0.093$^{**}$} & \makebox[2.75cm][c]{-0.270$^{***}$} \\
 & \makebox[2.75cm][c]{(0.017)} & \makebox[2.75cm][c]{(0.045)} & \makebox[2.75cm][c]{(0.063)} \\[2pt]
Uncertainty Score & \makebox[2.75cm][c]{0.007} & \makebox[2.75cm][c]{-0.198} & \makebox[2.75cm][c]{-0.378} \\
 & \makebox[2.75cm][c]{(0.071)} & \makebox[2.75cm][c]{(0.259)} & \makebox[2.75cm][c]{(0.419)} \\[2pt]
\hline
\noalign{\vskip 6pt}
\multicolumn{4}{l}{\textit{Panel B. Future Excess Realized Volatility}} \\[-1pt]
\noalign{\hrule height 0.15pt}
\noalign{\vskip 2pt}
 & \multicolumn{3}{c}{\makebox[8.20cm][c]{Dependent variable: Future excess realized volatility (pp)}} \\
\cmidrule(lr){2-4}
\noalign{\vskip 2pt}
 & \makebox[2.75cm][c]{1 trading day} & \makebox[2.75cm][c]{5 trading days} & \makebox[2.75cm][c]{10 trading days} \\
\cmidrule(lr){2-2}\cmidrule(lr){3-3}\cmidrule(lr){4-4}
\noalign{\vskip 3pt}
Informative Polarization & \makebox[2.75cm][c]{0.038$^{***}$} & \makebox[2.75cm][c]{0.100$^{***}$} & \makebox[2.75cm][c]{0.117$^{***}$} \\
 & \makebox[2.75cm][c]{(0.010)} & \makebox[2.75cm][c]{(0.020)} & \makebox[2.75cm][c]{(0.034)} \\[2pt]
Uncertainty Score & \makebox[2.75cm][c]{-0.105$^{***}$} & \makebox[2.75cm][c]{-0.198$^{***}$} & \makebox[2.75cm][c]{-0.266$^{***}$} \\
 & \makebox[2.75cm][c]{(0.034)} & \makebox[2.75cm][c]{(0.068)} & \makebox[2.75cm][c]{(0.058)} \\[2pt]
\hline\hline
\end{tabular}
}
\end{table}


\subsection{Informational Content of Digital Twin Interviews}

Taken together, the results in Section \ref{sec:xp} show that the stock-pick branch of the digital-twin interviews recovers informative signals about future stock returns and volatility. \emph{Directional} signals, such as \emph{Net Buy Share}, predict the cross-section of stock returns, especially in the silent region where the human finfluencer does not make a public recommendation. \emph{Disagreement} among digital twins predicts lower future returns and higher future volatility, consistent with the idea in \cite{miller1977risk} and the related analyst disagreement literature that followed. \emph{Uncertainty} behaves differently. Our \emph{Uncertainty Score} is high when the panel gives mostly neutral, speculative, or weakly supported recommendations. It therefore captures the \emph{absence of attention} or \emph{absence of grounded conviction}, not fundamental uncertainty in the usual asset-pricing sense.\footnote{Our use of the term \emph{uncertainty} differs from the traditional distinction between risk and uncertainty in economics. Since \citet{Knight1921}, risk typically refers to situations in which the relevant probability distribution is known or can be measured, while uncertainty refers to situations in which the probability distribution itself is unknown. \citet{Ellsberg1961} provides an illustration of this distinction, showing that individuals treat known probabilities differently from unknown or ambiguous probabilities. Our \emph{Uncertainty Score} is different. It does not measure uncertainty about the true probability distribution of stock returns. It measures the absence of grounded conviction, or the lack of attention, in the digital-twin panel: recommendations are neutral, speculative, or weakly supported by account-specific material.} The fact that high uncertainty predicts lower future volatility is consistent with the attention-based interpretation: stocks that draw little attention from the digital twins have lower future volatility, consistent with the predictions of \cite{AndreiHasler2015}.

These results further support the validity of our research design. Digital twins do not merely produce generic market commentary. Their responses contain stock-level information that maps into different future outcomes: direction predicts returns, disagreement predicts volatility, and uncertainty predicts stability. In this sense, the twins behave like account-conditioned representations of finfluencers' public personas. And, their interviews produce structured responses that are useful precisely because they are tied to the public-facing accounts we study.

This Section's main contribution, therefore, is about measuring the silent region under selective disclosure. Our predictability tests are not designed to prove that finfluencers are good stock pickers or that an investor could earn positive abnormal returns by following the twins' recommendations. Additional evidence, including out-of-sample validation, would be required to make that claim. The result here is narrower, but central to the paper. Digital-twin interviews produce stock-level public-persona belief proxies with relevant economic content, particularly in cases where public disclosures do not exist. They identify which stocks the panel favors, which stocks divide the panel, and which stocks draw weak conviction. This is the kind of information our research design was intended to recover.



\hypertarget{Macro Beliefs and Future Market Returns}{%
\section{Macro Beliefs and Future Market Returns}\label{sec: Macro Beliefs and Future Market Returns}}

Section \ref{sec:xp} studies the stock-level information contained in the stock-pick branch of the digital-twin interviews. We now turn to the macro branch, which asks the same digital twins about broad market conditions, including recession risk, business conditions, investor sentiment, market direction, and interest rates. These questions and their responses let us ask whether the interviews also recover market-level belief proxies.

We estimate predictive regressions of the form
\[
Ret_{t}^{M,(h)}=\alpha_h+\beta_h MacroSignal_t+\Gamma_h Controls_t^{(h)}+\varepsilon_{t}^{(h)},
\]
where $Ret_{t}^{M,(h)}$ is the future cumulative S\&P 500 return over horizon $h$, and $MacroSignal_t$ is a macro belief proxy, either \emph{Sentiment} or \emph{Disagreement}, measured before the return window begins. The controls are horizon-matched lagged S\&P 500 returns and lagged VIX returns. Standard errors are Newey--West adjusted with lag length $h-1$.

This test is distinct from the stock-level tests in Section \ref{sec:xp}. There, the tests ask whether stocks favored by the digital-twin panel outperform other stocks on the same date. Here, the analysis asks whether market-level beliefs on a given date predict future broad-market returns. Thus, the two branches of the interview protocol speak to different objects, with the stock-pick branch measuring relative stock preference and the macro branch measuring aggregate market beliefs.

Table \ref{tab:q1_q3_combined_sentiment_returns} first asks whether average macro \emph{Sentiment} predicts future market returns. The table reports three sentiment measures. Panel A uses \emph{Net Sentiment 3 of 7}, defined as the share of interviewed digital twins classified as bullish minus the share classified as bearish. A twin is classified as bullish when at least three of the seven core macro responses are strongly bullish and low-speculation, and bearish when at least three are strongly bearish and low-speculation. Panel B uses the analogous measure based on six macro questions, excluding the interest-rate question. Panel C uses a continuous sentiment score that averages the bullishness of the macro responses.\footnote{Section \ref{sec: The Interview Protocol as a Research Instrument} describes these variables. Section \ref{appx: Data Appendix} of the Internet Appendix gives the exact construction of all sentiment variables.}

\begin{table}[p]
\caption{Digital-Twin Macro Sentiment and Future Market Returns}
\label{tab:q1_q3_combined_sentiment_returns}
{\footnotesize{}This table reports controlled predictive regressions of future cumulative S\&P 500 returns on macro sentiment from the macro interview branch. Returns are measured in percentage points over 1, 5, and 10 trading days. Each specification includes horizon-matched lagged S\&P 500 returns and lagged VIX returns as controls. Panels A--C identify the sentiment measure. Coefficients are percentage-point changes in future S\&P 500 returns for a +1-unit increase in the sentiment measure. Newey--West standard errors are shown beneath each estimate, with lag length $h-1$ for horizon $h$. N reports trading-day observations. The sample uses the 81-account macro interview branch and contains 53 trading days from December 15, 2025 through March 16, 2026. Indicators ***, **, * denote statistical significance at the 1\%, 5\%, and 10\% level, respectively.}
{\footnotesize\par}
\vspace{10pt}
\centering{}
{\footnotesize
\renewcommand{\arraystretch}{0.96}
\setlength{\tabcolsep}{2.5pt}
\begin{tabular}{lccc}
\hline\hline
 & \makebox[2.75cm][c]{(1)} & \makebox[2.75cm][c]{(2)} & \makebox[2.75cm][c]{(3)} \\
\noalign{\vskip 4pt}
 & \multicolumn{3}{c}{\makebox[8.20cm][c]{Dependent variable: Future S\&P 500 return (pp)}} \\
\cmidrule(lr){2-4}
\noalign{\vskip 2pt}
 & \makebox[2.75cm][c]{1 trading day} & \makebox[2.75cm][c]{5 trading days} & \makebox[2.75cm][c]{10 trading days} \\
\cmidrule(lr){2-2}\cmidrule(lr){3-3}\cmidrule(lr){4-4}
\noalign{\vskip 7pt}
\multicolumn{4}{l}{\textit{Panel A. Net Sentiment (3 of 7)}} \\[-1pt]
\noalign{\hrule height 0.15pt}
\noalign{\vskip 2pt}
Sentiment & \makebox[2.75cm][c]{0.991} & \makebox[2.75cm][c]{3.090} & \makebox[2.75cm][c]{3.330} \\
 & \makebox[2.75cm][c]{(1.247)} & \makebox[2.75cm][c]{(3.067)} & \makebox[2.75cm][c]{(3.030)} \\[1pt]
Lagged return & \makebox[2.75cm][c]{-0.233} & \makebox[2.75cm][c]{-0.588$^{***}$} & \makebox[2.75cm][c]{-0.280} \\
 & \makebox[2.75cm][c]{(0.194)} & \makebox[2.75cm][c]{(0.177)} & \makebox[2.75cm][c]{(0.363)} \\[1pt]
Lagged VIX return & \makebox[2.75cm][c]{-0.013} & \makebox[2.75cm][c]{-0.116$^{***}$} & \makebox[2.75cm][c]{-0.140$^{***}$} \\
 & \makebox[2.75cm][c]{(0.036)} & \makebox[2.75cm][c]{(0.034)} & \makebox[2.75cm][c]{(0.049)} \\[1pt]
\noalign{\hrule height 0.15pt}
\noalign{\vskip 1pt}
N & \makebox[2.75cm][c]{53} & \makebox[2.75cm][c]{53} & \makebox[2.75cm][c]{53} \\
Adj. $R^2$ & \makebox[2.75cm][c]{-0.025} & \makebox[2.75cm][c]{0.245} & \makebox[2.75cm][c]{0.459} \\
\hline
\noalign{\vskip 7pt}
\multicolumn{4}{l}{\textit{Panel B. Net Sentiment (3 of 6)}} \\[-1pt]
\noalign{\hrule height 0.15pt}
\noalign{\vskip 2pt}
Sentiment & \makebox[2.75cm][c]{1.385} & \makebox[2.75cm][c]{3.442} & \makebox[2.75cm][c]{4.276} \\
 & \makebox[2.75cm][c]{(1.248)} & \makebox[2.75cm][c]{(2.938)} & \makebox[2.75cm][c]{(2.689)} \\[1pt]
Lagged return & \makebox[2.75cm][c]{-0.247} & \makebox[2.75cm][c]{-0.589$^{***}$} & \makebox[2.75cm][c]{-0.246} \\
 & \makebox[2.75cm][c]{(0.195)} & \makebox[2.75cm][c]{(0.175)} & \makebox[2.75cm][c]{(0.353)} \\[1pt]
Lagged VIX return & \makebox[2.75cm][c]{-0.015} & \makebox[2.75cm][c]{-0.114$^{***}$} & \makebox[2.75cm][c]{-0.131$^{***}$} \\
 & \makebox[2.75cm][c]{(0.036)} & \makebox[2.75cm][c]{(0.033)} & \makebox[2.75cm][c]{(0.046)} \\[1pt]
\noalign{\hrule height 0.15pt}
\noalign{\vskip 1pt}
N & \makebox[2.75cm][c]{53} & \makebox[2.75cm][c]{53} & \makebox[2.75cm][c]{53} \\
Adj. $R^2$ & \makebox[2.75cm][c]{-0.014} & \makebox[2.75cm][c]{0.251} & \makebox[2.75cm][c]{0.468} \\
\hline
\noalign{\vskip 7pt}
\multicolumn{4}{l}{\textit{Panel C. Continuous Sentiment (Quadratic)}} \\[-1pt]
\noalign{\hrule height 0.15pt}
\noalign{\vskip 2pt}
Sentiment & \makebox[2.75cm][c]{-1.910} & \makebox[2.75cm][c]{15.707} & \makebox[2.75cm][c]{27.572} \\
 & \makebox[2.75cm][c]{(13.033)} & \makebox[2.75cm][c]{(17.919)} & \makebox[2.75cm][c]{(31.567)} \\[1pt]
Sentiment$^2$ & \makebox[2.75cm][c]{6.007} & \makebox[2.75cm][c]{-7.785} & \makebox[2.75cm][c]{-24.224} \\
 & \makebox[2.75cm][c]{(25.109)} & \makebox[2.75cm][c]{(36.885)} & \makebox[2.75cm][c]{(68.233)} \\[1pt]
Lagged return & \makebox[2.75cm][c]{-0.201} & \makebox[2.75cm][c]{-0.619$^{***}$} & \makebox[2.75cm][c]{-0.369} \\
 & \makebox[2.75cm][c]{(0.228)} & \makebox[2.75cm][c]{(0.133)} & \makebox[2.75cm][c]{(0.286)} \\[1pt]
Lagged VIX return & \makebox[2.75cm][c]{-0.007} & \makebox[2.75cm][c]{-0.071$^{***}$} & \makebox[2.75cm][c]{-0.089$^{*}$} \\
 & \makebox[2.75cm][c]{(0.043)} & \makebox[2.75cm][c]{(0.026)} & \makebox[2.75cm][c]{(0.050)} \\[1pt]
\noalign{\hrule height 0.15pt}
\noalign{\vskip 1pt}
N & \makebox[2.75cm][c]{53} & \makebox[2.75cm][c]{53} & \makebox[2.75cm][c]{53} \\
Adj. $R^2$ & \makebox[2.75cm][c]{-0.050} & \makebox[2.75cm][c]{0.388} & \makebox[2.75cm][c]{0.549} \\
\hline\hline
\end{tabular}
}
\end{table}


Across all three measures of \emph{Sentiment} and three investment horizons, the \emph{Sentiment} coefficients are not statistically distinguishable from zero. Thus, in this sample, average macro sentiment from the digital-twin panel does not reliably predict future broad-market returns.

Table \ref{tab:q4_disagreement_main} next asks whether \emph{Disagreement} among the digital twins' macro views predicts future market returns. This parallels the stock-pick analysis in Table \ref{tab:stockpick_state_variables}, where \emph{Informative Polarization} measures disagreement across stock-level recommendations and predicts lower future stock returns. The macro setting asks the analogous market-level question. Two days can have similar average sentiment, but on one day the twins may largely agree, while on another day their views may be far apart. We measure aggregate \emph{Disagreement} using \emph{Disagreement IQR}, the interquartile range of account-level bull scores.\footnote{Bull scores translate each macro response to a common $0$--$100$ scale, oriented so that higher values are more bullish for the stock market. The account-level composite bull score averages those responses within an account-day. Section \ref{appx: Data Appendix} of the Internet Appendix gives the exact construction of this variable.}

Panel A reports controlled specifications with \emph{Disagreement} as the main independent variable. Panel B adds \emph{Net Sentiment 3 of 7} as a second independent variable, to ask whether \emph{Disagreement} contains information beyond the average bullishness of the panel. In both panels, the regressions include horizon-matched lagged S\&P 500 returns and lagged VIX returns as controls.

The results show a negative relation between disagreement and future market returns. In Panel A, a $10$-point increase in the bull-score \emph{Disagreement} measure predicts $77.1$ basis points lower S\&P 500 returns over the next five trading days and $86.4$ basis points lower returns over the next ten trading days. The ten-day coefficient remains negative and statistically significant after adding \emph{Net Sentiment 3 of 7} as a control in Panel B. Thus, the macro branch does not show reliable return predictability from average sentiment, but it does show that greater dispersion in macro beliefs is followed by weaker broad-market returns.

\begin{table}[p]
\caption{Digital-Twin Macro Disagreement and Future Market Returns}
\label{tab:q4_disagreement_main}
{\footnotesize{}This table reports controlled predictive regressions of future cumulative S\&P 500 returns on macro disagreement from the macro interview branch. Returns are measured in percentage points over 1, 5, and 10 trading days. The disagreement measure is Disagreement IQR (3 of 7). Each specification includes horizon-matched lagged S\&P 500 returns and lagged VIX returns as controls. Panel A reports disagreement-only specifications. Panel B adds Net Sentiment (3 of 7) to the same controlled specification. Disagreement coefficients and standard errors are scaled to a 10 bull-score-point increase in the disagreement measure. Sentiment coefficients are percentage-point changes in future S\&P 500 returns for a +1-unit increase in sentiment. Newey--West standard errors are shown beneath each estimate, with lag length $h-1$ for horizon $h$. $N$ reports trading-day observations. The sample uses the 81-account macro interview branch and contains 53 trading days from December 15, 2025 through March 16, 2026. Indicators ***, **, * denote statistical significance at the 1\%, 5\%, and 10\% level, respectively.}
{\footnotesize\par}
\vspace{10pt}
\centering{}
{\footnotesize
\renewcommand{\arraystretch}{0.96}
\setlength{\tabcolsep}{2.5pt}
\begin{tabular}{lccc}
\hline\hline
 & \makebox[2.75cm][c]{(1)} & \makebox[2.75cm][c]{(2)} & \makebox[2.75cm][c]{(3)} \\
\noalign{\vskip 4pt}
 & \multicolumn{3}{c}{\makebox[8.20cm][c]{Dependent variable: Future S\&P 500 return (pp)}} \\
\cline{2-4}
\noalign{\vskip 2pt}
 & \makebox[2.75cm][c]{1 trading day} & \makebox[2.75cm][c]{5 trading days} & \makebox[2.75cm][c]{10 trading days} \\
\cmidrule(lr){2-2}\cmidrule(lr){3-3}\cmidrule(lr){4-4}
\noalign{\vskip 7pt}
\multicolumn{4}{l}{\textit{Panel A. Disagreement with Lagged Market Controls}} \\[-1pt]
\noalign{\hrule height 0.15pt}
\noalign{\vskip 2pt}
Disagreement & \makebox[2.75cm][c]{-0.069} & \makebox[2.75cm][c]{-0.771$^{*}$} & \makebox[2.75cm][c]{-0.864$^{*}$} \\
 & \makebox[2.75cm][c]{(0.212)} & \makebox[2.75cm][c]{(0.452)} & \makebox[2.75cm][c]{(0.510)} \\[1pt]
Lagged return & \makebox[2.75cm][c]{-0.212} & \makebox[2.75cm][c]{-0.558$^{***}$} & \makebox[2.75cm][c]{-0.362} \\
 & \makebox[2.75cm][c]{(0.197)} & \makebox[2.75cm][c]{(0.157)} & \makebox[2.75cm][c]{(0.265)} \\[1pt]
Lagged VIX return & \makebox[2.75cm][c]{-0.012} & \makebox[2.75cm][c]{-0.116$^{***}$} & \makebox[2.75cm][c]{-0.146$^{***}$} \\
 & \makebox[2.75cm][c]{(0.039)} & \makebox[2.75cm][c]{(0.026)} & \makebox[2.75cm][c]{(0.039)} \\[1pt]
\noalign{\hrule height 0.15pt}
\noalign{\vskip 1pt}
N & \makebox[2.75cm][c]{53} & \makebox[2.75cm][c]{53} & \makebox[2.75cm][c]{53} \\
Adj. $R^2$ & \makebox[2.75cm][c]{-0.036} & \makebox[2.75cm][c]{0.281} & \makebox[2.75cm][c]{0.482} \\
\hline
\noalign{\vskip 7pt}
\multicolumn{4}{l}{\textit{Panel B. Disagreement and Sentiment with Lagged Market Controls}} \\[-1pt]
\noalign{\hrule height 0.15pt}
\noalign{\vskip 2pt}
Disagreement & \makebox[2.75cm][c]{0.136} & \makebox[2.75cm][c]{-0.761} & \makebox[2.75cm][c]{-0.863$^{*}$} \\
 & \makebox[2.75cm][c]{(0.356)} & \makebox[2.75cm][c]{(0.512)} & \makebox[2.75cm][c]{(0.444)} \\[1pt]
Sentiment & \makebox[2.75cm][c]{1.502} & \makebox[2.75cm][c]{0.089} & \makebox[2.75cm][c]{0.013} \\
 & \makebox[2.75cm][c]{(2.026)} & \makebox[2.75cm][c]{(2.786)} & \makebox[2.75cm][c]{(2.651)} \\[1pt]
Lagged return & \makebox[2.75cm][c]{-0.229} & \makebox[2.75cm][c]{-0.559$^{***}$} & \makebox[2.75cm][c]{-0.362} \\
 & \makebox[2.75cm][c]{(0.195)} & \makebox[2.75cm][c]{(0.165)} & \makebox[2.75cm][c]{(0.310)} \\[1pt]
Lagged VIX return & \makebox[2.75cm][c]{-0.011} & \makebox[2.75cm][c]{-0.115$^{***}$} & \makebox[2.75cm][c]{-0.146$^{***}$} \\
 & \makebox[2.75cm][c]{(0.037)} & \makebox[2.75cm][c]{(0.026)} & \makebox[2.75cm][c]{(0.050)} \\[1pt]
\noalign{\hrule height 0.15pt}
\noalign{\vskip 1pt}
N & \makebox[2.75cm][c]{53} & \makebox[2.75cm][c]{53} & \makebox[2.75cm][c]{53} \\
Adj. $R^2$ & \makebox[2.75cm][c]{-0.043} & \makebox[2.75cm][c]{0.267} & \makebox[2.75cm][c]{0.471} \\
\hline\hline
\end{tabular}
}
\end{table}

\begin{figure}[!htbp]
\centering
\includegraphics[width=0.95\textwidth]{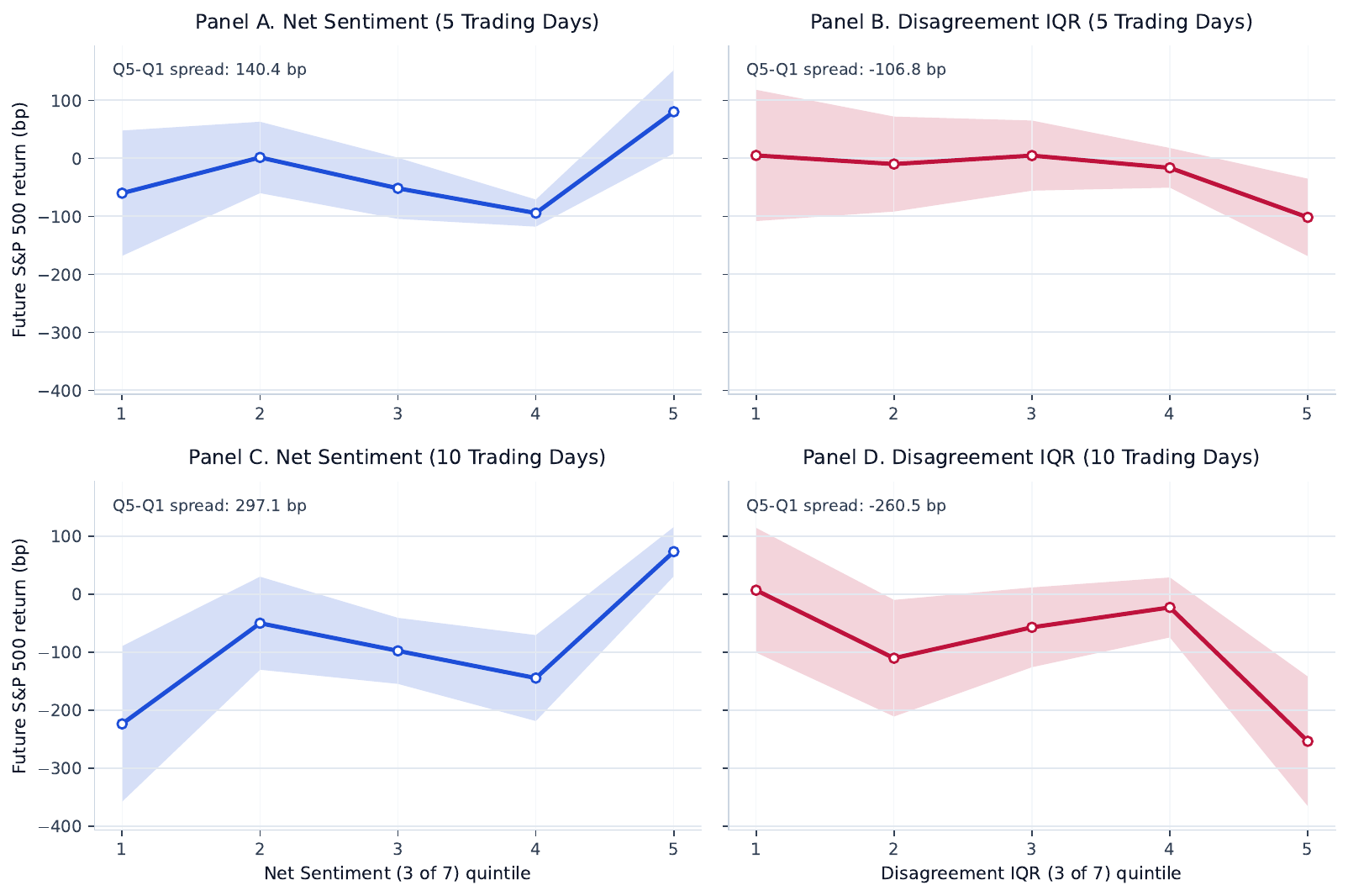}
\caption{Digital-Twin Macro Sentiment, Disagreement, and Future Market Returns}
\label{fig:macro_binned_relation}
\vspace{4pt}
\begin{minipage}{0.95\textwidth}
\footnotesize
This figure sorts trading days into quintiles of the indicated macro belief proxy and plots average future S\&P 500 returns, reported in basis points. Panels A and C use Net Sentiment (3 of 7). Panels B and D use Disagreement IQR (3 of 7). The top row reports 5-trading-day future returns, and the bottom row reports 10-trading-day future returns. The sample uses 81 accounts and 53 trading days from December 15, 2025 through March 16, 2026. Q5-Q1 is the top-minus-bottom quintile spread. Shaded bands denote 90\% confidence intervals for quintile means, computed with Newey--West standard errors using lag length $h-1$ for horizon $h$.
\end{minipage}
\end{figure}

Figure \ref{fig:macro_binned_relation} shows the macro predictability patterns visually. Panels A and C of Figure \ref{fig:macro_binned_relation} sort days into quintiles of \emph{Net Sentiment}. The highest-sentiment quintile beats the lowest-sentiment quintile by $140.4$ basis points over five days (Panel A) and $297.1$ basis points over ten days (Panel C). However, as shown above, this pattern is not statistically significant in a more formal setting. Panels B and D of Figure \ref{fig:macro_binned_relation} sort days into quintiles of \emph{Disagreement IQR}. The highest-disagreement quintile underperforms the lowest-disagreement quintile by $106.8$ basis points over five days (Panel B) and by $260.5$ basis points over ten days (Panel D), a relation that is statistically significant in the more formal setting.

Taken together, these results show that the interviews in the macro branch are informative, but not because average sentiment forecasts market returns. The level of bullishness in the digital-twin panel is not reliably related to future S\&P 500 returns. Instead, the stronger signal is disagreement across twins. Days with more dispersed macro beliefs are followed by weaker broad-market performance, even after controlling for average sentiment, lagged market returns, and lagged VIX returns. This pattern complements the stock-pick evidence: the interviews recover useful belief proxies not only about which stocks are relatively favored, but also about when market-level views are more contested. In this sample, belief dispersion, rather than average optimism, is the macro belief proxy with predictive content.

The results in this section are consistent with the broader asset-pricing evidence that market timing is difficult, even when cross-sectional return predictability is present. Prior work documents predictable cross-sectional return patterns, including size, value, and momentum \citep{FamaFrench1992,JegadeeshTitman1993}, while the market-timing literature emphasizes the difficulty of reliably forecasting broad market movements \citep{GoyalWelch2008, GoyalWelchZafirov2024}. The stronger contribution of our paper remains the stock-level measurement result: the digital-twin interviews recover economically meaningful variation across stocks, especially in the silent region where public recommendations are absent.



\hypertarget{Discussion and Applications Beyond Financial Social Media}{%
\section{Discussion}\label{sec: Discussion and Applications Beyond Financial Social Media}}

\subsection{Applications Beyond Financial Social Media}
\label{sec:applications}

The selective-disclosure problem is not unique to financial influencers on social media. Many social-science settings involve agents who choose what to reveal, when to reveal it, how strongly to express it, and when to remain silent. The design in this paper offers a way to study that problem. Researchers can build digital twins of human agents from public statements, ask fixed questions in real time, validate the output where public statements provide a benchmark, and then conduct experiments using the added coverage from the silent region. The goal is not to replace public statements but to measure beliefs alongside them. The main benefit of this research design is that the resulting panel includes both disclosed views and silent-region belief proxies, so it can be used to study a broader range of views.

In the finance literature, sell-side analysts are one natural extension of our research. These analysts disclose ratings and price targets, and their public personas can be observed through report language, conference-call questions, and public commentary. Prior work shows that analyst recommendations and revisions can affect prices and predict subsequent stock returns \citep{womack1996brokerage, barber2001can, jegadeesh2006value, sorescu2006cross}. Still, analysts choose which firms to emphasize, which risks to highlight, and when to update a recommendation. A digital-twin design could help researchers measure the views implied by an analyst's public-report persona across a wider range of covered firms and risks, especially in cases where the public reports are silent. The same logic applies to executives, fund managers, financial journalists, market commentators, and financial advisers.

Beyond finance, the same measurement problem appears in polling, political communication, labor-market disclosure, journalism, central banking, and other public-facing settings.\footnote{For related work on survey nonresponse, response bias, and digital-twin polling, see \cite{keeter2017low}, \cite{kennedy2019response}, \cite{groves2008impact}, \cite{brick2013explaining}, \cite{mercer2023comparing}, \cite{claassen2025biased}, and \cite{cerina20252024}. For strategic issue emphasis in politics, see \cite{petrocik1996issue}, \cite{sniderman2004structure}, and \cite{Riker1996}. For labor-market signaling, disclosure, and audit-study designs, see \cite{Spence1973}, \cite{pallais2014inefficient}, \cite{kessler2019incentivized}, \cite{kuhn2013gender}, and \cite{bertrand2004emily}.} The details differ, but the common problem is that public records reveal, in part, an agent's decision to share a particular view, not the full set of views held by that agent. A digital-twin design could help researchers compare what public agents say voluntarily with what these same agents would say when questioned through their corresponding digital twins. Then, researchers could study both what is voluntarily disclosed and what becomes visible only because of the digital twin interviews.

\subsection{Scope, Limitations, and Future Research}
\label{sec:limitations}

The prospective nature of our research design imposes a practical constraint on scale. Although the \emph{stock-pick} panel is substantial in the cross-sectional dimension—containing 21,410 stock-date observations across 429 stocks—it is generated from 81 finfluencer personas over 50 event dates, while the \emph{macro} panel spans 53 event dates. Building this panel and conducting repeated daily interviews of the digital twins—particularly the account-by-stock interviews—required substantial API usage, token consumption, and associated expenditures. These resource demands constrained the duration of the data-collection period.

At the same time, our main stock-level directional findings are robust across different empirical approaches: Fama--MacBeth regressions with Newey–West inference, Calendar-Time portfolios, and staggered-vintage regressions with non-overlapping return windows and placebo benchmarks. These complementary designs do not substitute for a longer sample, but they reduce the concern that the findings reflect one particular estimator or treatment of overlapping outcomes. Future research can extend the design across longer periods, additional personas and platforms, broader stock universes, and different market regimes. We hope that such extensions will corroborate and refine our findings while further establishing digital-twin interviews as a measurement instrument for studying selective disclosure.

Importantly, digital-twin interviews are measurement instruments, not claims about hidden views. They require transparent conditioning material, fixed protocols, real-time archives, and validation benchmarks. Researchers should be especially careful when the underlying material is private, sensitive, or not clearly part of a public persona. The goal is not to infer private beliefs of human agents. It is to create comparable records in settings where public disclosure is incomplete. Used carefully, this approach can help social scientists study a common problem: important information is often unobserved not because it does not exist, but because someone chose not to disclose it.



\hypertarget{Conclusion}{%
\section{Conclusion}\label{sec: Conclusion}}

This paper starts from a simple measurement problem. Public financial posts in social media are selective disclosures. They reveal what finfluencers choose to say, not the full set of views that these finfluencers could have expressed. Silence, therefore, is ambiguous. A research design based on actual public posts can measure disclosed content, but it cannot observe the silent region where no post appears.

We address that problem with repeated, real-time digital-twin interviews of finfluencers. The interviews use a fixed protocol, are timestamped before the relevant return windows, and produce public-persona belief proxies for the overall market and for a broad set of stocks. We do not claim that our method recovers the private beliefs of these finfluencers. Our claim is narrower, but important. Standardized responses from digital-twin interviews let us study both the views that appear in public posts and the belief proxies that become visible only because the interview protocol asks.

The evidence supports the premise of our research design. Digital-twin interview responses align with public recommendations, lean toward later public recommendations even before they appear, and retain account-specific structure after common market conditions are removed. The silent region is large, accounting for $84.8\%$ of the stock-pick panel, and the stock-level interview responses carry valuable economic content. Stocks with more buy-leaning digital-twin responses earn higher future excess returns. Stocks with greater disagreement have higher future volatility, while stocks that lack attention have more stable price patterns in the future. At the aggregate level, average macro sentiment does not reliably predict broad-market returns, but disagreement in macro beliefs predicts weaker future market performance.

Our main contribution is measurement under selective disclosure. The paper is not a claim that digital twins reveal private beliefs, and it is not primarily a claim that finfluencers can pick stocks. Instead, this paper shows that standardized, real-time digital-twin interviews can make the silent region of a public persona visible. In settings where public communication is selective, we show that what is not said can also be measured.


\newpage
\bibliographystyle{jf}
\bibliography{references}

\clearpage
\newpage
\processdelayedfloats
\clearpage
\newpage
\hypersetup{pageanchor=false}
\setcounter{section}{0}
\setcounter{subsection}{0}
\setcounter{subsubsection}{0}
\setcounter{table}{0}
\setcounter{figure}{0}
\setcounter{postfigure}{0}
\setcounter{posttable}{0}
\renewcommand{\thesection}{A}
\renewcommand{\thesubsection}{\Alph{subsection}}
\renewcommand{\thetable}{A\arabic{table}}
\renewcommand{\thefigure}{A\arabic{figure}}
\renewcommand{\theposttable}{A\arabic{posttable}}
\renewcommand{\thepostfigure}{A\arabic{postfigure}}
\renewcommand{\thesubsubsection}{\thesubsection.\arabic{subsubsection}}

\setcounter{page}{1}

\section*{Internet Appendix\\ \\``Talking to Digital Twins: Selective Disclosure\\and Belief Measurement in Financial Social Media''}
\label{sec:internet_appendix}

\subsection{Data Appendix}\label{appx: Data Appendix}

This section of the Internet Appendix describes the data-generating process outlined in Section \ref{sec: The Interview Protocol as a Research Instrument}. It covers the selection of the finfluencer panel, the daily digital-twin interviews, both the stock-pick and macro interview branches, the public-post comparison panel, the timing rules, and the variable dictionary used in the empirical tests.

\subsubsection{Finfluencer Panel Selection}

This project is designed to study public personas on X that post finance-related content. We do not attempt to construct a census of all finance accounts on X. Instead, we construct a fixed panel of accounts that regularly discuss markets and stocks, have a meaningful public audience, and provide enough observable content to support repeated digital-twin interviews. The panel was fixed before the interviews began, so interview responses could not affect which accounts entered the sample.

Our initial approach to building the panel was to identify finfluencers algorithmically, by combining a finance-specific word list with a list of exchange-listed ticker symbols (cashtags). An account was flagged as a candidate finfluencer when its recent posts contained a sufficiently high density of finance vocabulary---drawn from the \citet{LoughranMcDonald2011} financial dictionaries---together with frequent references to tickers. The appeal of this approach was scale and objectivity: in principle it could screen a very large number of accounts with little manual judgment and could be reproduced by other researchers. In practice, however, the lexicon-plus-ticker screen produced too many false positives to support a credible persona panel. Two sources of noise were especially severe. First, ticker symbols frequently coincide with common English words and abbreviations, so ordinary posts were misclassified as stock commentary. Cashtags such as \$IT, \$ALL, \$ON, \$A, \$CAR, \$NOW, \$REAL, and \$GOOD match everyday language at least as often as they identify the underlying equities (Gartner, Allstate, ON Semiconductor, Agilent, Avis, ServiceNow, The RealReal, and Gladstone Commercial, respectively), so casual posts were repeatedly flagged as ticker references. Second, finance vocabulary is shared with many non-investing domains: words such as ``share,'' ``interest,'' ``credit,'' ``bond,'' ``security,'' ``margin,'' ``stock,'' and ``options'' appear routinely in posts about retail inventory, personal relationships, online gaming, sports betting, and promotional giveaways. As a result, the automated screen swept in news bots and headline aggregators, corporate and investor-relations handles, cryptocurrency and giveaway-promotion accounts, and general-interest accounts that merely used finance-sounding language---none of which represent the kind of market-commentary persona our design requires. 

Because the method selected accounts on the basis of token co-occurrence rather than on whether an account maintains a recognizable, finance-oriented public persona, its precision was low, and the flagged set required so much manual pruning that the automation offered little practical advantage. We therefore pivoted to curated, human-compiled lists of finance-focused accounts. 

The sample was assembled in three stages. First, we compiled candidate accounts from third-party online directories and public lists, including The CFO Club, Investopedia, Investing.io, and Yahoo Finance.\footnote{The source pages were consulted between April 2025 and June 2025. The full list, with hyperlinks, is as follows: \href{https://thecfoclub.com/leadership/finance-twitter-accounts/}{The CFO Club}; \href{https://www.investopedia.com/financial-edge/0712/10-twitter-feeds-investors-should-follow.aspx}{Investopedia}; \href{https://x.feedspot.com/investing_twitter_influencers/}{Feedspot, Investing X Influencers}; \href{https://x.feedspot.com/trading_twitter_influencers/}{Feedspot, Trading X Influencers}; \href{https://www.lendinvest.com/blog/financial-advice-from-twitter/}{LendInvest}; \href{https://www.agilitypr.com/resources/top-influencers/top-10-uk-social-media-influencers-finance/}{Agility PR}; \href{https://investing.io/investor-twitter-accounts/}{Investing.io}; \href{https://tradersunion.com/interesting-articles/day-trading-what-is-day-trading/best-day-traders-to-follow-on-twitter/}{Traders Union}; \href{https://bullishbears.com/best-twitter-traders/}{Bullish Bears}; \href{https://www.asktraders.com/learn-to-trade/trading-guide/top-10-trader-twitter-to-follow/}{Asktraders}; \href{https://www.securities.io/top-10-x-twitter-accounts-to-help-you-learn-chart-analysis/}{Securities.io}; \href{https://optionshawk.com/best-traders-to-follow-on-financial-twitter/}{OptionsHawk}; \href{https://realtrading.com/trading-blog/7-best-day-traders-follow-twitter/}{Real Trading}; \href{https://finance.yahoo.com/news/30-best-investors-twitter-social-171430248.html}{Yahoo Finance}; and \href{https://www.reddit.com/r/Trading/comments/1jh0z5i/any_good_trading_accounts_to_follow_on_twitter/}{r/Trading}.} The consulted sources span trading generalists, long-horizon investors, wealth managers, technical analysts, day traders, market-sentiment commentators, options and derivatives traders, self-identified bullish and bearish investors, and chart-based technical analysts.

In the second stage, we manually screened each candidate account and its post history from December 2024 through June 2025 against a topical inclusion rule. To be retained, an account had to post primarily about tradable financial instruments---individual stocks, equity indices, options, futures, or other derivatives---with content that took the form of recommendations, directional calls, or technical analysis of specific tickers. We excluded accounts whose principal content was real estate, personal finance advice, pure macroeconomic commentary without tradable recommendations, generic financial-news aggregation, aggressive ticker promotion, paid-signal advertising, course or lecture sales, mentoring offers, or similar commercial content.

In the third stage, we applied an additional manual screen to the remaining candidates. Accounts were retained if they had at least $5{,}000$ followers, a sustained posting history of at least $10$ posts since profile creation so there is sufficient information to support repeated digital-twin interviews, and language and topical content consistent with a financial persona rather than a general social-media personality.

The final panel contains $81$ finfluencer accounts on X that were observed in both interview branches of our study. The panel was finalized in June 2025, before any interviews were conducted.\footnote{The main sample contains the $81$ X accounts observed in both interview branches. The macro interview branch contains one additional account, for a total of $82$ accounts, while the stock-pick branch contains the $81$ accounts used in the main analysis. Because the stock-level analysis requires accounts to appear in the stock-pick branch, and to keep the sample consistent across the paper, the main analysis uses the $81$ accounts observed in both branches. All tests that can be estimated with all $82$ accounts are quantitatively similar when the \nth{82} account is retained.}

The X handles of the $81$ accounts used in our study are listed below. These are the public personas that supply the account-conditioned respondents (digital twins) for the macro and stock-pick interview panels.

\begin{center}
\par\medskip
{\footnotesize
\setlength{\tabcolsep}{3pt}
\begin{tabular}{lllll}
\hline\hline
1OptionsTrading & abnormalreturns & AdeptMarket & anymantrading & AswathDamodaran \\
BezosCrypto & Blackopstocks & BreakoutStocks & BrianStutland & ChandlerTrading \\
Chariot\_Invest & ContrarianShort & CrossStocks & Darkminer71 & DataDInvesting \\
davidmoadel & daytradesignals & DCDOWORK & DevotedDividend & dirtcheapstocks \\
Dividend\_Dr & DKellerCMT & dmdsplyinvestor & DV\_Situations & ElliottForecast \\
GlobalStockPick & gurgavin & hiddensmallcaps & I\_Am\_The\_ICT & ideahive \\
InvestorsLive & ivanhoff2 & Jake\_\_Wujastyk & LindaRaschke & LizAnnSonders \\
MarketMovesMatt & MasteredTrader & MichaelGoodwell & OMillionaires & OnlyOTrades \\
OptionAlpha & optionscjp & OptionsDepth & OptionsHawk & OptiontradinIQ \\
profitly & RealDayTrading & ReturnsJourney & SeekingAlpha & Selling4Premium \\
sentimentrader & seth\_fin & SethCL & StockMKTNewz & StockOptionCole \\
StocksOnSpaces & StocksToTrade & Stocktwits & SuburbanDrone & SwaggyStocks \\
TAftermath2020 & TheAlphaThought & TheStalwart & TheTradingTank & timothysykes \\
traderstewie & TradesTrey & tradewithprof & Trading0secrets & TradingwithFun1 \\
TrendSpider & TSXtrad3r & ukarlewitz & value\_invest12 & ValueStockGeek \\
ValueWolf & VertiCallAlgo & VROStocks & vsourbh & ZacksResearch \\
zerohedge & & & & \\
\hline\hline
\end{tabular}
}
\end{center}

\subsubsection{Daily Digital-Twin Interviews}

For each X account in the panel, we create an account-conditioned digital twin. The digital twin is an LLM respondent conditioned on the account's public persona.\footnote{\label{fn:gptnote}We retain the model identifier for every digital-twin interview response. In the retained analysis samples (after timing and deduplication), nearly all responses were generated with \texttt{gpt-5-mini-2025-08-07}; \texttt{gpt-5-nano-2025-08-07} was used only during a brief interval, corresponding to eight event dates in each interview branch. Model choice reflected API availability and cost considerations, not any review or analysis of the interview responses. All queries were issued programmatically through OpenAI's API rather than the consumer web (ChatGPT) interface, allowing us to automate and reproduce the prompts at scale. API access is a paid service, and the volume of queries required for this study entailed substantial usage charges to OpenAI.} The conditioning material includes profile metadata and recent posts. As the account produces new public content and as market conditions change, the conditioning material is updated. The interview questions remain fixed for the core measures. This design gives us a repeated panel of public-persona belief proxies.

Before launching the main interview protocol, we conducted two daily pilot studies, from August 1, 2025 through August 31, 2025, and again from September 29, 2025 to December 14, 2025. During these pilot periods, we queried the digital twins about aggregate market conditions and about content appearing in the finfluencers' contemporaneous public posts. The pilots were designed to evaluate feasibility rather than to produce the main empirical results. They allowed us to estimate the cost of repeated real-time interviews, identify operational constraints, and refine the workflow used in the main experiment.

We began our main experiment on December 15, 2025. The interviews were conducted almost daily through March 14, 2026.\footnote{A few days are missing from this period because of ordinary operational issues such as backlogs. The alignment rules described below in Subsection \ref{sec:allign} explain how we handled these irregularities.} The interview protocol has two layers of instructions. The first level gives the model the account profile, recent posts, and permission to retrieve clearly relevant market context through web search. When using web search, the prompt tells the model to prioritize reputable finance-focused sources and to treat social-media or forum content mainly as sentiment. The second level of the prompt then asks for one structured answer to each question. Three excerpts from the protocol capture these instructions:

\begin{quote}
    \small
    ``Select the most likely response based strictly on the provided profile data. The chosen response must be the most accurate representation of the profile.''
    
    \medskip
    ``You may also use web search to retrieve relevant, up-to-date information about the specific tickers, sectors, asset classes, strategies, and macro themes that this profile is likely to care about.''
    
    \medskip
    ``You must give an answer for every question while maintaining the persona and perspective of the financial influencer profile provided.''
\end{quote}

Together, these instructions bound the interview responses. The model can use outside market context when it is clearly relevant, but the selected response must still be supported by the account material or by clearly relevant retrieved information. If the sources are uncertain or conflicting, the model is told to acknowledge that uncertainty.

Within the \emph{macro} branch, questions are structured with forced responses: each response includes a substantive answer, a selected symbol/category, an explanation, and a speculation score. The speculation score is not a confidence score about the future state of the market. It records how much useful account information supports the selected response.\footnote{The prompt defines speculation on a $0$ to $100$ scale: $0$--$20$ is low speculation, $21$--$40$ is moderate-low, $41$--$60$ is moderate, $61$--$80$ is moderate-high, and $81$--$100$ is high. This scale matters in the analysis because the headline macro signal treats answers with speculation scores of $40$ or below as sufficiently grounded in the account's public persona to contribute to strong bullish or bearish classifications.} The prompt also asks the model to preserve a structured response format so the answers can be analyzed consistently. The first macro question illustrates the structure:

\begin{quote}
    \small
    \textbf{Question:} What is your best estimate of the probability that the U.S. economy will enter a recession in the next 12 months?
    
    \medskip
    A1) 0 to 20\% (Very unlikely)\\
    A2) 21 to 40\% (Unlikely)\\
    A3) 41 to 60\% (About as likely as not)\\
    A4) 61 to 80\% (Likely)\\
    A5) 81 to 100\% (Very likely)
    
    \medskip
    \textbf{Required response fields:}\\
    \textbf{question:} [Question text]\\
    \textbf{explanation:} [Features of the supplied information that support the selected response and speculation score]\\
    \textbf{symbol:} [Selected symbol, such as A1]\\
    \textbf{category:} [Full text of selected response category]\\
    \textbf{speculation:} [Speculation score from 0 to 100]
\end{quote}

The raw files produced from every daily interview retain the explanation and category fields. In our macro analysis, we use the selected symbol and speculation score for the seven core questions listed below. We then translate the selected symbols into question-specific bull scores, oriented so that higher values are more bullish. For example, on the recession question above, an answer of A4, corresponding to a $61$ to $80\%$ recession probability, is translated into a bull score of $30$.

\begin{quote}
    \small
    \begin{enumerate}[leftmargin=*,itemsep=2pt,topsep=2pt]
        \item What is your best estimate of the probability that the U.S. economy will enter a recession in the next 12 months?
        \item Considering business conditions in the country as a whole, do you think that during the next 12 months we will have good times financially, bad times, or something in between?
        \item In the next six months, do you expect business conditions to be better, worse, or the same?
        \item In your view, is overall investor sentiment currently bearish, neutral, or bullish?
        \item Do you expect the direction of the stock market over the next six months to be up, unchanged, or down?
        \item In the next six months, do you expect U.S. stock market indices to rise, stay about the same, or fall?
        \item In the next 12 months, do you expect interest rates on U.S. bonds to rise, stay about the same, or fall?
    \end{enumerate}
\end{quote}

Within the \emph{stock-pick} branch we used the same account-conditioning protocol as we did in the \emph{macro} branch, but asked a much larger set of questions. Instead of asking each digital twin for a broad market view, we asked them for a separate view on each stock in the covered universe. As a result, this interview branch was significantly more expensive to run.\footnote{We did not begin the \emph{Stock-Pick} branch until after the conclusion of the two pilot studies because we required additional time to define the scope of our research (studying the finfluencers' silent region), obtain cost estimates for accessing OpenAI's API at a much larger scale, and identify funding sources.} After aligning trading days to interview timestamps, the cleaned stock-pick files run from December 15, 2025 through March 16, 2026.\footnote{The last interview was on March 14, 2026, corresponding to the trading day on March 16, 2026.}

The first stock-pick runs were intentionally broad: we asked each digital twin for reflections on every ticker in the Russell $4{,}000$. That breadth was useful for learning what the process could support, but it was too costly to run as a daily panel. We therefore shifted the interviews to a narrower set of $429$ large-cap stocks that are a subset of the Russell $4{,}000$ index. In all of the analyses in the paper, we use this smaller, recurring universe of stocks rather than the broad Russell $4{,}000$ universe we interviewed for a small number of days. The resulting stock-pick sample contains $429$ recurring tickers, $21{,}410$ stock-by-trading-day observations, and $50$ trading days.

The complete list of the 429 tickers follows.

\newpage

\begingroup
\footnotesize
\setlength{\tabcolsep}{3pt}
\setlength{\LTleft}{0pt plus 1fill}
\setlength{\LTright}{0pt plus 1fill}

\begin{longtable}{llllllllllll}
\hline\hline
\endfirsthead

\hline\hline
\endhead

\hline\hline
\endfoot

\hline\hline
\endlastfoot

A & AAPL & ABBV & ABNB & ABT & ADBE & ADI & ADM & ADP & ADSK & AEE & AEP \\
AES & AFL & AIG & AIZ & AJG & AKAM & ALB & ALGN & ALL & AMAT & AMD & AME \\
AMGN & AMP & AMZN & ANET & AOS & APA & APD & APH & APO & APP & ATO & AVGO \\
AVY & AWK & AXON & AXP & AZO & BA & BAC & BALL & BAX & BBY & BDX & BEN \\
BIIB & BK & BKNG & BKR & BLDR & BLK & BMY & BR & BRO & BSX & BX & C \\
CAG & CAH & CARR & CAT & CBRE & CDNS & CDW & CEG & CF & CFG & CHD & CHRW \\
CHTR & CI & CINF & CL & CLX & CMCSA & CME & CMG & CMI & CMS & CNC & CNP \\
COF & COIN & COO & COP & COST & CPB & CPRT & CRL & CRM & CRWD & CSCO & CSGP \\
CSX & CTAS & CTRA & CTSH & CTVA & CVS & CVX & D & DAL & DASH & DD & DDOG \\
DE & DECK & DELL & DG & DGX & DHI & DHR & DIS & DLTR & DOV & DOW & DPZ \\
DRI & DTE & DUK & DVA & DVN & DXCM & EA & EBAY & ECL & ED & EFX & EIX \\
EL & ELV & EME & EMR & EOG & EPAM & EQT & ERIE & ES & ETR & EVRG & EW \\
EXC & EXPD & EXPE & F & FANG & FAST & FCX & FDS & FDX & FE & FFIV & FICO \\
FIS & FISV & FITB & FOXA & FSLR & FTNT & FTV & GD & GDDY & GE & GEHC & GEN \\
GILD & GIS & GL & GLW & GM & GNRC & GOOG & GOOGL & GPC & GPN & GS & GWW \\
HAL & HAS & HBAN & HCA & HD & HIG & HII & HLT & HOLX & HON & HOOD & HPE \\
HPQ & HRL & HSIC & HSY & HUBB & HUM & HWM & IBKR & IBM & ICE & IDXX & IEX \\
IFF & INCY & INTC & INTU & IP & IQV & IR & ISRG & IT & ITW & J & JBHT \\
JBL & JKHY & JNJ & JPM & K & KDP & KEY & KEYS & KHC & KKR & KLAC & KMB \\
KMI & KO & KR & KVUE & L & LDOS & LEN & LH & LHX & LII & LKQ & LLY \\
LMT & LNT & LOW & LRCX & LULU & LUV & LVS & LW & LYV & MA & MAR & MAS \\
MCD & MCHP & MCK & MCO & MDLZ & MET & META & MGM & MHK & MKC & MLM & MMC \\
MMM & MNST & MO & MOH & MOS & MPC & MPWR & MRK & MRNA & MS & MSCI & MSFT \\
MSI & MTB & MTCH & MTD & MU & NDAQ & NDSN & NEE & NEM & NFLX & NI & NKE \\
NOC & NOW & NRG & NSC & NTAP & NTRS & NUE & NVDA & NVR & NWS & NWSA & ODFL \\
OKE & OMC & ON & ORCL & ORLY & OTIS & OXY & PANW & PAYC & PAYX & PCAR & PCG \\
PEG & PEP & PFE & PFG & PG & PGR & PH & PHM & PKG & PLTR & PM & PNC \\
PNW & PODD & POOL & PPG & PPL & PRU & PSX & PTC & PWR & PYPL & QCOM & REGN \\
RF & RJF & RL & RMD & ROK & ROL & ROP & ROST & RSG & RTX & SBUX & SCHW \\
SHW & SJM & SMCI & SNA & SNPS & SO & SPGI & SRE & STLD & STT & STZ & SWK \\
SWKS & SYF & SYK & SYY & T & TAP & TDG & TDY & TECH & TER & TFC & TGT \\
TJX & TMO & TMUS & TPL & TPR & TRGP & TRMB & TROW & TRV & TSCO & TSLA & TSN \\
TTD & TTWO & TXN & TXT & TYL & UAL & UBER & UHS & ULTA & UNH & UNP & UPS \\
URI & USB & V & VLO & VLTO & VMC & VRSK & VRSN & VRTX & VST & VTRS & VZ \\
WAB & WAT & WBD & WDAY & WDC & WEC & WFC & WM & WMB & WMT & WRB & WSM \\
WST & WYNN & XEL & XOM & XYL & YUM & ZBH & ZBRA & ZTS &  &  &  \\

\end{longtable}
\endgroup

The stock-pick prompt includes stock-by-stock questions. A representative stock-level question and response template is:

\begin{quote}
    \small
    \textbf{Question:} Based on your general knowledge of market conditions, indicate whether you would BUY, HOLD, or SELL (SHORT SELL) the stock ticker: [COMPANY NAME (TICKER)].
    
    \medskip
    \textbf{Required response fields:}\\
    \textbf{question:} [COMPANY NAME (TICKER)]\\
    \textbf{explanation:} [Detailed explanation for recommendation]\\
    \textbf{recommendation:} [Value selected]\\
    \textbf{confidence:} [Value selected]\\
    \textbf{speculation:} [Speculation score selected]\\
    \textbf{expected holding period:} [Number of days selected]\\
    \textbf{primary catalyst type:} [Earnings/Macro/Product/Regulatory/Flow-Technical/Other]
\end{quote}

The prompt defines the recommendation field on a $0$--$100$ scale, where $0$ is a very strong sell recommendation and $100$ is a very strong buy recommendation. It describes $40$--$60$ as hold, $60$--$80$ as moderate buy, and $80$ or above as strong buy; values below $40$ are sell-leaning. Confidence is also reported on a $0$--$100$ scale. Speculation follows the same interpretation as in the macro interview: higher values mean the supplied account material contains less useful information for the stock-specific answer.

The raw daily stock-pick files are extremely wide because each account-day contains fields for many tickers. We reshape those files into a long panel with one row per account, interview, and ticker. The compact cleaned panel keeps the recommendation and speculation fields; the long explanations remain in the raw files and are not used in the main stock-pick analysis. When the model supplies a text recommendation instead of a number, we map SELL or SHORT to $25$, HOLD to $50$, and BUY to $75$ before any stock-date measures are built.

For each stock and event date, we then aggregate across the interviewed digital twins. The basic direction measure is the average recommendation score centered at $50$. A recommendation is treated as meaningful only if it is at least $15$ points away from neutral, so scores of $65$ or above are meaningful buys and scores of $35$ or below are meaningful sells. \emph{Informative Polarization} uses the subset of meaningful recommendations with speculation scores of $40$ or below, and we require at least three such recommendations before reporting it. \emph{Uncertainty Score} is high when a stock-event has little directional conviction: for example, when many responses are near neutral, speculative, or both.

\subsubsection{Public-Post Comparison Panel}

We also construct a public-post comparison panel. This panel does not include digital twins but rather the $81$ human finfluencers who make actual public posts. We identify posts that mention specific stocks and classify the direction of the disclosed recommendation. The panel gives us an observable benchmark for digital twin interview data. When a human finfluencer publicly recommends a ticker on a given trading date, we can compare that recommendation with the digital-twin interview response for the same finfluencer, ticker, and date. When no matching public stock-recommendation post exists for a stock-event, the digital twin interview panel measures the silent region: the part of the belief space that cannot be observed from public recommendations alone. We use the public-post panel for validation and for the paper's selective-disclosure comparisons.

\subsubsection{Timing Alignment and Sample Filters}\label{sec:allign}

The purpose of our empirical tests is to ask whether information contained in the digital-twin interviews predicts later market outcomes. We therefore impose a strict, conservative timing rule to align interviews in market-time. As such, the return window used to evaluate interview responses begins only after the interview has been assigned to an effective trading date. This prevents an interview from being evaluated using returns that were already realized, or partly realized, when the interview response was generated.

We align interviews to market time using a 3:30 p.m. U.S. Eastern Time (ET) cutoff; a notably conservative cutoff considering that trading continues until 4:00 p.m. each day. If an interview occurs on a trading day before 3:30 p.m. ET, we keep its calendar date and map that date to the corresponding trading date (e.g., an interview completed on a Tuesday before 3:30 p.m. ET is mapped to that day, Tuesday). If an interview occurs on a trading day at or after 3:30 p.m. ET, we first shift it forward by one calendar day and then map the resulting date to the next trading date (e.g., an interview completed on a Tuesday after 3:30 p.m. ET is mapped to the next day, Wednesday). Interviews that occur on non-trading days are mapped to the next trading date. The mapped trading date is the event date used in the analysis.\footnote{We use 3:30 p.m. ET, rather than the market close, as a conservative buffer against very late interviews being paired with same-day closing prices or end-of-day information. Relative to a 4:00 p.m. cutoff, this choice has little effect. In the macro branch, $38$ of $7,107$ interview records before de-duplication map differently. The final daily macro dataset has the same $53$ event dates under either cutoff, and only one daily signal row changes. On that date, January 23, 2026, continuous sentiment changes from $0.327$ to $0.323$. In the stock-pick branch, $90$ of $7,431$ interview runs map differently before de-duplication. After keeping one run per account-event date, the only substantive difference is that a 4:00 p.m. cutoff would add one March 5, 2026 stock-pick event-date cross-section based on only nine interviews completed between 3:57 p.m. and 3:59 p.m. ET. Existing stock-event signal rows are unchanged across the two timing conventions.}

Returns are then measured over close-to-close trading-day windows following the event date. If an interview is assigned to event date $t$, the one-day return is measured from the close of trading on $t$ to the close of trading on the next S\&P 500 trading day. More generally, the $h$-day return is the cumulative close-to-close return over the next $h$ trading days. Thus, an interview conducted at 2:00 p.m. ET on a Tuesday is assigned to Tuesday, and its one-day return is measured from Tuesday's close to Wednesday's close. An interview conducted at 3:45 p.m. ET on that same Tuesday is shifted forward, assigned to Wednesday, and evaluated using returns beginning after Wednesday's close. A Sunday interview is assigned to the next trading day, typically Monday, and its one-day return is measured from Monday's close to Tuesday's close.

When more than one interview response is available for the same account and event date, we retain only the latest response that maps to that event date. In the stock-pick branch, this de-duplication is applied at the account-event-date level before forming stock-level recommendation measures, so a single account does not receive multiple weights on the same event date. In the macro branch, the same logic retains one account-level macro response per event date before forming daily macro belief measures.\footnote{Here, ``latest'' means the latest available run after sorting by the source run date and the observed interview timestamp, not the latest intended target date. A small number of interview runs were completed in catch-up batches because computational-resource and processing constraints prevented every scheduled run from being executed on its target calendar night. We apply the event-date rule to the observed interview timestamp. When multiple runs for the same account map to the same event date, the latest observed run is retained.}

\begin{figure}[!htbp]
\centering
\includegraphics[width=0.92\textwidth]{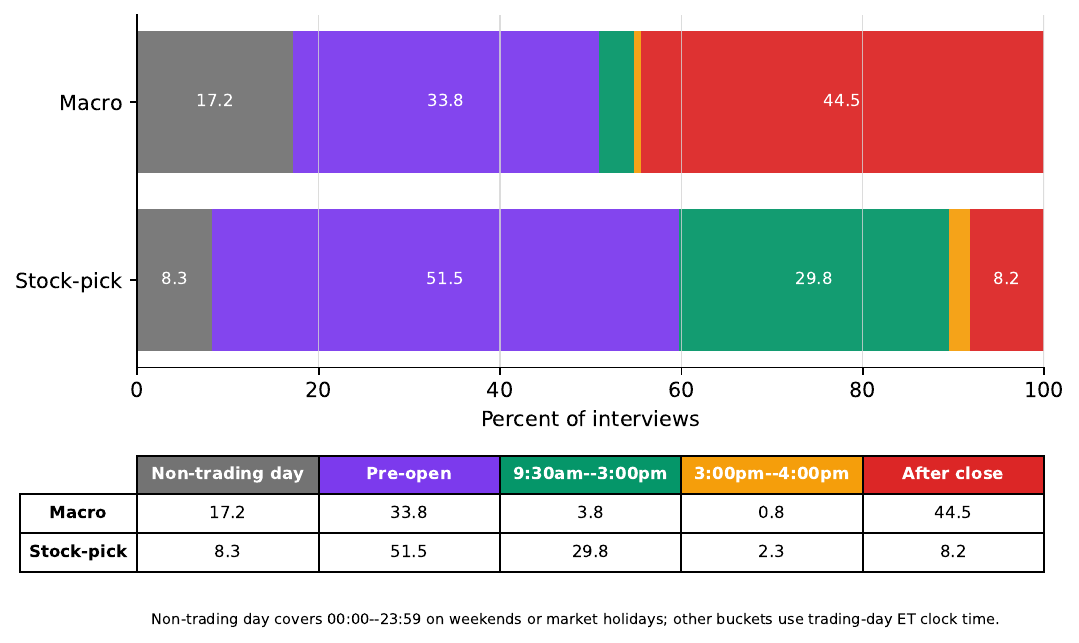}
\caption{Timing of Digital-Twin Interviews}
\label{fig:interview_timing}
\vspace{4pt}
\begin{minipage}{0.95\textwidth}
\footnotesize
The figure reports the timing distribution for macro and stock-pick interviews during the shared sample period. Non-trading day covers interviews occurring between 00:00 and 23:59 ET on weekends or S\&P 500 market holidays. The remaining categories use actual ET clock time on S\&P 500 trading days: pre-open is before 9:30am, regular hours are 9:30am--3:00pm, the late-market bucket is 3:00pm--4:00pm, and after close is 4:00pm or later.
\end{minipage}
\end{figure}

Figure \ref{fig:interview_timing} summarizes the timing of the interviews that enter the analysis after applying the timing rule and account-date de-duplication. The figure shows that very few interviews occur near the market close. In the macro branch, only $0.8\%$ of interviews occur between 3:00 p.m. and 4:00 p.m. ET on trading days. In the stock-pick branch, $2.3\%$ occur between 3:00 p.m. and 4:00 p.m. The broader timing pattern reflects the sequencing of the interview pipeline. Macro interviews are typically launched after the market close and completed overnight, so most macro interviews occur after close or before the next open. The stock-pick branch is run after the macro branch, so fewer stock-pick interviews complete during the evening; instead, many complete before the open or during regular trading hours on the next trading day.

We apply the same market-time convention to public recommendations, using the public-post timestamp. A public recommendation posted at or before 3:30 p.m. ET keeps its calendar date before being mapped to a trading date. A public recommendation posted after 3:30 p.m. ET is first shifted forward by one calendar day and then mapped to the next S\&P 500 trading date. If multiple public recommendations by the same account about the same stock map to the same public event date, we collapse them to one account-stock-event-date public signal. The collapsed observation averages the public recommendation scores and related post-level measures, while retaining the number of posts and the first and last mention times. This keeps the interview and public-post samples aligned under the same market-time convention and the same account-stock-event-date unit of observation.

The remaining filters are simple sample-construction choices. First, the paper uses the $81$ accounts observed in both the macro and stock-pick branches. Second, an account-day observation in the macro branch must contain enough usable core-question information to support the daily belief measures. In practice, it must have at least four observed speculation scores among the seven core questions, and a macro trading day must have at least five contributing finfluencers. Third, the stock-pick analysis uses the recurring 429-ticker universe described above.

\subsubsection{Variable Definitions}
The tables and tests in the paper use variables built from the digital-twin interviews and public-post data. The variable dictionary below records the construction, unit, and observation unit for the main variables used in the paper, including the macro sentiment and disagreement measures, stock-pick direction and uncertainty measures, public-post comparison variables, and market return and volatility outcomes.\footnote{\label{fn:event}An event date is the trading date assigned to an interview response or public post after applying the market-time alignment rule. It may differ from the calendar date on which the interview or post occurred. Throughout the paper, event date refers to this aligned trading date. A stock-event is a stock-by-event-date observation.}

\begingroup
\setstretch{1.0}
\footnotesize
\setlength{\tabcolsep}{3pt}
\renewcommand{\arraystretch}{1.08}
\begin{longtable}{@{}P{0.17\textwidth}P{0.55\textwidth}P{0.11\textwidth}P{0.12\textwidth}@{}}
\multicolumn{4}{@{}l}{\textbf{Data Appendix Variable Definitions}}\\
\addlinespace[4pt]
\toprule
Variable & Construction and interpretation & Unit & \shortstack[l]{Sample\\unit} \\
\midrule
\endfirsthead
\multicolumn{4}{@{}l}{\textbf{Data Appendix Variable Definitions} \textit{(continued)}}\\
\addlinespace[4pt]
\toprule
Variable & Construction and interpretation & Unit & \shortstack[l]{Sample\\unit} \\
\midrule
\endhead
\midrule
\multicolumn{4}{r}{\textit{Continued on next page}} \\
\endfoot
\bottomrule
\endlastfoot

\multicolumn{4}{@{}l}{\textit{Stock-pick interview panel}} \\
\midrule
Recommendation score & Parsed 0--100 stock-pick recommendation score for an account-stock interview response. The neutral value is 50. Higher values are more buy-leaning and lower values are more sell-leaning. When a response is available only as text, we map sell or short to 25, hold to 50, and buy to 75 before aggregation. Used to construct Tilt, Net Buy Share, Meaningful Tilt, Conviction-Weighted Meaningful Tilt, Informative Polarization, Uncertainty Score, and the public-overlap validation measures. & Score points & Account $\times$ stock $\times$ event date \\
\addlinespace[3pt]
Stock-pick speculation score & Parsed 0--100 speculation score for a stock-pick response. Higher values mean the stock-level answer is less grounded in the supplied account material. A meaningful recommendation is informative when this score is observed and at most 40. Used to construct Informative Polarization, Uncertainty Score, and the low-speculation validation filters. & Score points & Account $\times$ stock $\times$ event date \\
\addlinespace[3pt]
Recommendation Count & Number of parsed account-stock recommendations contributing to a stock-event observation. This count is reported in the summary statistics and is the denominator for Meaningful Recommendation Share, Net Buy Share, and Uncertainty Score. & Count & Stock $\times$ event date \\
\addlinespace[3pt]
Meaningful Recommendation Share & Share of parsed recommendations with $|\text{rec\_score}-50| \ge 15$ for the stock-event. This records how much of the recommendation set moves materially away from neutral and is used to construct Conviction-Weighted Meaningful Tilt. & Share in $[0,1]$ & Stock $\times$ event date \\
\addlinespace[3pt]
Tilt & Average recommendation score for the stock-event, centered at 50: $\overline{\text{rec\_score}}_{it}-50$. Positive values mean the recommendation set leans buy-side. & Rec. points & Stock $\times$ event date \\
\addlinespace[3pt]
Net Buy Share & Meaningful buy share minus meaningful sell share for the stock-event. Meaningful buys have recommendation scores at least 65, and meaningful sells have scores at most 35. Both shares are divided by the total number of parsed recommendations for the stock-event, not only by meaningful recommendations. & Net share in $[-1,1]$ & Stock $\times$ event date \\
\addlinespace[3pt]
Meaningful Tilt & Average recommendation score centered at 50, using only recommendations with $|\text{rec\_score}-50| \ge 15$. This removes the neutral center of the raw recommendation distribution. Used to construct Conviction-Weighted Meaningful Tilt. & Rec. points & Stock $\times$ event date \\
\addlinespace[3pt]
Conviction-Weighted Meaningful Tilt & Meaningful Tilt multiplied by Meaningful Recommendation Share. The measure is defined when the stock-event has at least one meaningful recommendation. High values require both directional strength and many recommendations away from neutral. Tables sometimes shorten the label to Conviction-Weighted Tilt. & Index & Stock $\times$ event date \\
\addlinespace[3pt]
Informative Polarization & $4 \times \text{buy share}^{inf}_{it}\times \text{sell share}^{inf}_{it}$ for the stock-event, where the buy and sell shares are measured within the informative subset. That subset requires both meaningful direction and a speculation score at most 40. The factor of 4 rescales the index so it reaches one when informative buys and sells are evenly split. The measure is zero when informative recommendations are all on one side and rises toward one when informative recommendations are sharply split. The paper requires at least three informative recommendations. & Index in $[0,1]$ & Stock $\times$ event date \\
\addlinespace[3pt]
Uncertainty Score & Stock-event index that is low when many responses give clear, grounded buy or sell views and high when responses are mostly neutral, weak, or speculative. Each meaningful recommendation receives a grounding weight of $1-\text{speculation}/100$; non-meaningful recommendations and meaningful recommendations without a parsed speculation score receive weight zero. The score is one minus the average grounding weight across all parsed recommendations. & Index in $[0,1]$ & Stock $\times$ event date \\
\addlinespace[3pt]
Market capitalization & Market capitalization for stock $i$, matched to the event date. Used for value-weighted stock-pick estimates and for the public-attention firm-characteristic comparison, where it is reported in billions of dollars. & Dollars & Stock $\times$ event date or firm \\
\addlinespace[3pt]
log(Market Cap) & Natural logarithm of event-date market capitalization for stock $i$. Reported in the summary statistics and used as a firm-size characteristic in the public-attention comparison. & Log dollars & Stock $\times$ event date or firm \\
\addlinespace[3pt]
Excess Return ($h$ days) & Cumulative excess stock return relative to the S\&P 500 over the next $h \in \{1,5,10\}$ trading days, excluding the event date. The measure sums daily stock-minus-S\&P 500 log returns over the window and converts the cumulative excess log return to simple-return form. & Percentage points & Stock $\times$ event date \\
\addlinespace[3pt]
Excess Realized Volatility ($h$ days) & Square root of the sum of squared daily stock-minus-S\&P 500 log returns over the next $h \in \{1,5,10\}$ trading days, excluding the event date. The reported measure is multiplied by 100. & Percentage points & Stock $\times$ event date \\
\addlinespace[3pt]

\midrule
\multicolumn{4}{@{}l}{\textit{Macro interview panel}} \\
\midrule
Bull score & A 0--100 coding of each macro answer, with higher values always meaning a more bullish market view and 50 meaning neutral. We map the ordered response categories to scores separately for each question. When a category gives a range, we use its midpoint before orienting the score. For example, a 61--80\% recession-probability answer is coded as 30 because high recession risk is bearish for stocks. Used to construct Net Sentiment, Continuous Sentiment, Disagreement IQR, and the macro validation diagnostics. & Score points & Account $\times$ question $\times$ day \\
\addlinespace[3pt]
Speculation score & A 0--100 score recorded with each macro answer. Higher values mean the answer is less supported by the account material supplied to the digital twin. The low-speculation screen used for Net Sentiment and Continuous Sentiment keeps answers with scores at most 40; High-Speculation Share counts answers with scores at least 60. The availability of these scores is also used to retain macro account-days. & Score points & Account $\times$ question $\times$ day \\
\addlinespace[3pt]
Net Sentiment \quad (3 of 7) & Daily bullish share minus daily bearish share across retained account-days, using Q1--Q7. An account-day is bullish if at least three answers have Bull score at least 70 and Speculation score at most 40. It is bearish if at least three answers have Bull score at most 30 and Speculation score at most 40. Mixed account-days stay in the denominator but are not counted as bullish or bearish. & Net share in $[-1,1]$ & Trading day \\
\addlinespace[3pt]
Net Sentiment \quad (3 of 6) & Same construction as Net Sentiment (3 of 7), but using Q1--Q6 only. This six-question version drops Q7, the question about the expected direction of interest rates on U.S. bonds over the next 12 months. & Net share in $[-1,1]$ & Trading day \\
\addlinespace[3pt]
Continuous Sentiment & Daily average of account-level normalized Bull scores across Q1--Q7. Each answer with an observed Bull score and Speculation score at most 40 is transformed to $(\text{Bull score}-50)/50$, so the scale runs from bearish to bullish. The account-day score requires at least three valid core questions, and the daily variable averages those account-day scores. Some regressions also include the square of this variable to allow a curved sentiment-return relation. & Index in $[-1,1]$ & Trading day \\
\addlinespace[3pt]
Disagreement IQR (3 of 7) & Interquartile range, across retained accounts on a trading day, of account-level composite Bull scores. The account-level composite averages available Q1--Q7 Bull scores for that account-day. This spread is computed from Bull-score levels, not from the bullish and bearish classification flags used in Net Sentiment. Higher values mean macro views are more dispersed across accounts. & Bull-score points & Trading day \\
\addlinespace[3pt]
High-Speculation Share, Q1--Q7 & Daily average, across retained account-days, of the share of Q1--Q7 Speculation scores that are at least 60. This variable is reported in the summary statistics. & Share in $[0,1]$ & Trading day \\
\addlinespace[3pt]
Contributing Influencers per Day & Number of distinct panel accounts retained on a trading day after the macro interview is aligned to the trading calendar, de-duplicated, and required to have at least four observed speculation scores among Q1--Q7. This count can be below 81 because some account-days are missing or filtered. & Count & Trading day \\
\addlinespace[3pt]
S\&P 500 Return ($h$ days) & Cumulative S\&P 500 simple return over horizon $h$, built by summing daily S\&P 500 log returns and converting the sum back to a simple return. Forward returns use the next $h$ trading days after the aligned interview date and exclude the event day. Lagged returns use the previous $h$ trading days. The main horizons are $h \in \{1,5,10\}$. & Percentage points & Trading day \\
\addlinespace[3pt]
S\&P 500 Realized Volatility ($h$ days) & Square root of the sum of squared S\&P 500 daily log returns over horizon $h$. The macro realized-volatility controls use lagged windows. For the one-day horizon, the control is the absolute value of the previous day's S\&P 500 log return; for the 5- and 10-day horizons, it uses the prior $h$ trading days. The reported measure is multiplied by 100. & Percentage points & Trading day \\
\addlinespace[3pt]
VIX Return ($h$ days) & Cumulative VIX simple return over horizon $h$, built by summing daily VIX log returns and converting the sum back to a simple return. The macro return regressions use horizon-matched lagged VIX returns for $h \in \{1,5,10\}$ as the VIX-based control. & Percentage points & Trading day \\
\addlinespace[3pt]

\midrule
\multicolumn{4}{@{}l}{\textit{Public-post comparison variables}} \\
\midrule
Public Net Buy Share & Public-post version of Net Buy Share. Public stock-recommendation posts are assigned to an effective trading date from the mention timestamp, using the 3:30 p.m. Eastern Time cutoff. When an account has multiple posts about the same stock on the same effective date, those posts are collapsed to one account-stock-event-date public signal by averaging the public recommendation scores and related post-level measures. Public Net Buy Share is meaningful public buys minus meaningful public sells, divided by total account-level public recommendations for the stock-event. & Net share in $[-1,1]$ & Stock $\times$ event date \\
\addlinespace[3pt]
Digital-Twin Interview Net Buy Share & Net Buy Share from the stock-pick interview branch, used alongside Public Net Buy Share in the exact-overlap and no-public-post comparisons. & Net share in $[-1,1]$ & Stock $\times$ event date \\
\addlinespace[3pt]
Public direction & Buy or sell direction of an observed public recommendation. Public buys have recommendation scores at least 65, and public sells have scores at most 35. Neutral or conflicting public directions are excluded when a validation test needs a buy/sell benchmark. Used in the public-overlap sign-alignment, AUC, and signed-tilt validation tests. & Buy/sell sign & Account $\times$ stock $\times$ event date \\
\addlinespace[3pt]
Public-post days & Number of distinct effective public-post dates for a firm during the stock-pick interview window from December 15, 2025 through March 16, 2026, using public recommendations from the stock-pick interview account set. Used to form the public-attention distribution. & Count & Firm \\
\addlinespace[3pt]
Public posts & Number of observed public stock-recommendation posts for a firm during the stock-pick interview window, using the same effective-date public-post sample as Public-post days. Public-attention distribution rows report per-firm means within public-post-day bins. & Count & Firm \\
\addlinespace[3pt]
Posts per public day & Firm-level ratio of public stock-recommendation posts to public-post days, computed for firms with at least one public-post day. & Ratio & Firm \\
\addlinespace[3pt]
Public attention bins & Firm groups based on Public-post days during the stock-pick interview window. The distribution bins are 0, 1, 2, 3, 4, 5--9, 10--19, and 20+ public-post days; the characteristic comparison groups are low attention (0--4), middle attention (5--19), and high attention (20+). & Category & Firm \\
\addlinespace[3pt]
Firm characteristics & Firm-level variables used in the public-attention comparison. Table \ref{tab:stockpick_public_recommendation_attention} reports market capitalization, log market capitalization, CRSP listing age, and selected industry indicators. Market capitalization is matched from the stock-level market data used in the return-linked stock-pick panel. Listing age and industry classifications are built from CRSP monthly data as of December 31, 2024. Continuous characteristics are winsorized at the pooled 1st and 99th percentiles. & Mixed & Firm \\
\addlinespace[3pt]
\end{longtable}
\endgroup


\setcounter{table}{0}
\setcounter{posttable}{0}


\subsubsection{Capital IQ and CRSP Data}

This section lists the market-data variables used to construct the return, volatility, size, and firm-characteristic variables in the paper.

From S\&P Capital IQ, we use adjusted close price levels, $P_{i,d}$, for U.S. equities and selected market series, where $i$ represents the specific stock or macro-level series, and $d$ is the trading day.\footnote{The exact function used is CIQ(.,IQ\_CLOSEPRICE\_ADJ,.). This function generates daily prices that are adjusted for dividends and stock splits.} The equity file contains a Capital IQ security identifier, company name, exchange-ticker identifier, exchange, ticker label, and a panel of daily adjusted close prices. We convert these adjusted close prices to daily log returns (log denotes the natural logarithm),
\[
r_{i,d}=\log(P_{i,d}/P_{i,d-1}),
\]
setting returns to missing when either the current or prior adjusted close price is missing, zero, or negative. These daily log returns are used to construct future stock returns, stock excess returns relative to the S\&P 500, and realized-volatility measures. We also use Capital IQ to obtain closing prices for the S\&P 500 and daily values of the volatility index, VIX. The S\&P 500 series provides the trading calendar, market returns, and the benchmark return used to compute stock excess returns. The VIX series provides the VIX-return controls used in the macro predictive regressions. Finally, we use stock-level market capitalization from the same market-data files to construct event-date market capitalization, log market capitalization, and value-weighted robustness tests.

From CRSP\footnote{CRSP refers to Morningstar's \emph{Center for Research in Security Prices}, formerly a product of the University of Chicago.}, we use monthly security data to construct the firm characteristics in Table \ref{tab:stockpick_public_recommendation_attention}. We match tickers to CRSP securities, use the first observed CRSP date to compute listing age as of December 31, 2024, and use SIC codes to form industry indicators based on the Fama--French 12-industry classification. The table reports listing age and selected industry shares, along with market capitalization and log market capitalization from the return-linked stock-level market data. Continuous firm characteristics are winsorized at the pooled 1st and 99th percentiles.

\subsection{Validation and Public-Disclosure Appendix}\label{app:validation_public_disclosure}

In this section of the Internet Appendix we present, in Table \ref{tab:validation_public_overlap_appendix}, additional tests to assess the robustness of the results presented in Table \ref{tab:validation_main} (Section \ref{sec: Validation: Does the Instrument Work?}) of the main paper. Recall that Table \ref{tab:validation_main} shows that the digital-twin interview outputs look like account-conditioned public-persona responses, not like unconditional, generic market commentary. Table \ref{tab:validation_public_overlap_appendix} extends the public-recommendation overlap test beyond that result.

Specifically, in Table \ref{tab:validation_main} of the main paper we use exact same-account, same-stock, same-date overlaps. Here, in Table \ref{tab:validation_public_overlap_appendix}, we ask whether the result also holds when public recommendations are allowed to fall on nearby trading dates. It does. In the exact-date sample, meaningful low-speculation interview directions align with public recommendations $91.5\%$ of the time. In the narrow one-trading-day window, alignment is still $90.5\%$, and the area-under-the-curve (AUC) is $0.742$. These nearby-window checks support the same conclusion as Table \ref{tab:validation_main}: when public recommendations provide an observable benchmark, the digital-twin interview responses usually point in the same direction and rank public buys above public sells.

\begin{table}[!htbp]
\caption{Additional Validation of Digital-Twin Interview Belief Proxies}
\label{tab:validation_public_overlap_appendix}
{\footnotesize{}This table provides additional detail for the public-recommendation overlap validation in Table \ref{tab:validation_main}. It compares digital-twin stock-pick interview responses with public recommendations made by the same account about the same stock in the core 429-stock universe. Exact-date matches require the interview and public recommendation to have the same effective trading date. Narrow-window matches also allow the public recommendation to fall one trading day before or after the interview date. Forward-window matches allow the public recommendation to fall on the same trading date or one trading day after the interview date. Sign alignment and AUC are the same validation measures used in Table \ref{tab:validation_main}. Sign alignment uses the meaningful low-speculation interview subset, and AUC is the probability that the interview score ranks a public buy above a public sell. Benchmark is the null or placebo benchmark used in Table \ref{tab:validation_main}. Percentage rows are in percentage points, AUC is unitless, and N is row-specific. The $p$-value column reports the sign-test or AUC randomization p-value for the Actual-Benchmark difference. The public-overlap validation branch uses the shared 81-account analysis sample.}
{\footnotesize\par}
\vspace{7pt}
\centering{}
{\footnotesize
\begin{tabular}{p{6.4cm}ccccc}
\hline\hline
 & (1) & (2) & (3) & (4) & (5) \\
\noalign{\vskip 2pt}
Measure & Actual & Benchmark & Difference & $p$-value & N \\
\cmidrule(lr){1-1}\cmidrule(lr){2-2}\cmidrule(lr){3-3}\cmidrule(lr){4-4}\cmidrule(lr){5-5}\cmidrule(lr){6-6}
Exact-date matches: Sign alignment & 91.5\% & 50.0\% & 41.5 pp & $<0.001$ & 2,439 \\
Exact-date matches: AUC & 0.776 & 0.522 & 0.255 & $<0.001$ & 4,063 \\
Narrow-window matches: Sign alignment & 90.5\% & 50.0\% & 40.5 pp & $<0.001$ & 5,385 \\
Narrow-window matches: AUC & 0.742 & 0.511 & 0.231 & $<0.001$ & 9,797 \\
Forward-window matches: Sign alignment & 90.1\% & 50.0\% & 40.1 pp & $<0.001$ & 3,825 \\
Forward-window matches: AUC & 0.725 & 0.509 & 0.215 & $<0.001$ & 7,056 \\
\hline\hline
\end{tabular}
}
\end{table}


\subsection{Stock-Pick Results Appendix}\label{app:stockpick_results}

In this section of the Internet Appendix we report two robustness checks for the \emph{stock-pick} analysis presented in Section \ref{sec:xp} of the main paper. The main text shows that stocks with more favorable digital-twin stock-pick signals earn higher future excess returns in equal-weighted daily Fama--MacBeth regressions, Calendar-Time portfolios, and Staggered-Vintage regressions (with appropriate controls in each case). Here, we ask whether this result is driven by larger firms or by the repeated use of overlapping return windows.

Table \ref{tab:q1_stockpick_fmb_calendar_time_returns_value_weighted} repeats the analysis in Table \ref{tab:q1_stockpick_fmb_calendar_time_returns} using event-date market capitalization weights. Columns (1)--(3) estimate the daily Fama--MacBeth regressions by weighted least squares, and columns (4)--(6) value-weight stocks within the high- and low-signal legs of the calendar-time portfolios. The value-weighted estimates are much smaller than the equal-weighted estimates and are not statistically distinguishable from zero. The main stock-pick result is therefore not a large-cap effect. It is stronger in the equal-weighted cross-section, where smaller and larger stock-events receive comparable weight.

\begin{table}[!htbp]
\caption{Value-Weighted Digital-Twin Stock-Pick Direction and Future Excess Returns}
\label{tab:q1_stockpick_fmb_calendar_time_returns_value_weighted}
{\footnotesize{}This table is the value-weighted analogue of Table \ref{tab:q1_stockpick_fmb_calendar_time_returns}. Columns (1)--(3) report daily Fama--MacBeth cross-sectional WLS regressions of cumulative future excess stock returns on one digital-twin direction signal measured before the return window. Stock-events are weighted by event-date market capitalization. Excess returns are measured relative to the S\&P 500 and reported in percentage points over 1, 5, and 10 trading days. The table reports the time-series average of the daily slope estimates. Columns (4)--(6) report calendar-time portfolio tests using the same signals. On each event date, stocks are ranked by the indicated signal, the portfolio buys the top quintile and shorts the bottom quintile, stocks are value-weighted within each leg using event-date market capitalization, and each event-date vintage is held for the next 10 trading days, excluding the formation day. The daily calendar-time long-short return averages all active vintages on each holding date. H-L Mean and CAPM Alpha are daily percentage-point estimates while 10-day Alpha is 10 times the daily CAPM alpha. Market adjustment is estimated from a daily CAPM regression of the long-short return on the S\&P 500 return. Newey--West standard errors are in parentheses, using lag $h-1$ in columns (1)--(3) and lag 9 in columns (4)--(6). Coefficients are scaled to a +10 recommendation-point move in Tilt or Meaningful Tilt, a +10 percentage-point move in Net Buy Share, or a +1-unit move in Conviction-Weighted Meaningful Tilt. N/days reports stock-event observations in columns (1)--(3) and calendar holding days in columns (4)--(6). Mean weighted adjusted $R^2$ applies only to the Fama--MacBeth columns. The sample uses 429 stocks and 81 accounts over 50 event dates from December 15, 2025 to March 16, 2026; calendar-time holding dates run from December 16, 2025 to March 30, 2026. Indicators ***, **, * denote significance at the 1\%, 5\%, and 10\% level.}
{\footnotesize\par}
\vspace{5pt}
\centering{}
{\footnotesize
\setlength{\tabcolsep}{1.5pt}
\begin{tabular}{lcccccc}
\hline\hline
 & \makebox[1.88cm][c]{(1)} & \makebox[1.88cm][c]{(2)} & \makebox[1.88cm][c]{(3)} & \makebox[1.88cm][c]{(4)} & \makebox[1.88cm][c]{(5)} & \makebox[1.88cm][c]{(6)} \\
\noalign{\vskip 4pt}
 & \multicolumn{3}{c}{\makebox[5.60cm][c]{Dep. var.: future excess return (pp)}} & \multicolumn{3}{c}{\makebox[5.60cm][c]{Dep. var.: H-L portfolio return (pp)}} \\
\cmidrule(lr){2-4}\cmidrule(lr){5-7}
\noalign{\vskip 2pt}
 & \makebox[1.88cm][c]{1 day} & \makebox[1.88cm][c]{5 days} & \makebox[1.88cm][c]{10 days} & \makebox[1.88cm][c]{\shortstack{H-L\\Mean}} & \makebox[1.88cm][c]{\shortstack{CAPM\\Alpha}} & \makebox[1.88cm][c]{\shortstack{10-day\\Alpha}} \\
\cmidrule(lr){2-2}\cmidrule(lr){3-3}\cmidrule(lr){4-4}\cmidrule(lr){5-5}\cmidrule(lr){6-6}\cmidrule(lr){7-7}
\noalign{\vskip 4pt}
\multicolumn{7}{l}{\textit{Panel A. Tilt}} \\[-1pt]
\noalign{\hrule height 0.15pt}
\noalign{\vskip 2pt}
Direction signal & \makebox[1.88cm][c]{-0.181} & \makebox[1.88cm][c]{-0.197} & \makebox[1.88cm][c]{-0.478} & \makebox[1.88cm][c]{0.016} & \makebox[1.88cm][c]{0.011} & \makebox[1.88cm][c]{0.107} \\
 & \makebox[1.88cm][c]{(0.140)} & \makebox[1.88cm][c]{(0.488)} & \makebox[1.88cm][c]{(0.667)} & \makebox[1.88cm][c]{(0.092)} & \makebox[1.88cm][c]{(0.097)} & \makebox[1.88cm][c]{(0.968)} \\[1pt]
\noalign{\hrule height 0.15pt}
\noalign{\vskip 1pt}
N / days & \makebox[1.88cm][c]{21,410} & \makebox[1.88cm][c]{21,410} & \makebox[1.88cm][c]{21,410} & \makebox[1.88cm][c]{71} & \makebox[1.88cm][c]{71} & \makebox[1.88cm][c]{71} \\
Mean wtd. adj. $R^2$ & \makebox[1.88cm][c]{0.051} & \makebox[1.88cm][c]{0.053} & \makebox[1.88cm][c]{0.042} & \makebox[1.88cm][c]{} & \makebox[1.88cm][c]{} & \makebox[1.88cm][c]{} \\
\hline
\noalign{\vskip 4pt}
\multicolumn{7}{l}{\textit{Panel B. Net Buy Share}} \\[-1pt]
\noalign{\hrule height 0.15pt}
\noalign{\vskip 2pt}
Direction signal & \makebox[1.88cm][c]{-0.041} & \makebox[1.88cm][c]{-0.032} & \makebox[1.88cm][c]{-0.108} & \makebox[1.88cm][c]{0.012} & \makebox[1.88cm][c]{0.008} & \makebox[1.88cm][c]{0.078} \\
 & \makebox[1.88cm][c]{(0.033)} & \makebox[1.88cm][c]{(0.120)} & \makebox[1.88cm][c]{(0.187)} & \makebox[1.88cm][c]{(0.094)} & \makebox[1.88cm][c]{(0.099)} & \makebox[1.88cm][c]{(0.986)} \\[1pt]
\noalign{\hrule height 0.15pt}
\noalign{\vskip 1pt}
N / days & \makebox[1.88cm][c]{21,410} & \makebox[1.88cm][c]{21,410} & \makebox[1.88cm][c]{21,410} & \makebox[1.88cm][c]{71} & \makebox[1.88cm][c]{71} & \makebox[1.88cm][c]{71} \\
Mean wtd. adj. $R^2$ & \makebox[1.88cm][c]{0.042} & \makebox[1.88cm][c]{0.047} & \makebox[1.88cm][c]{0.041} & \makebox[1.88cm][c]{} & \makebox[1.88cm][c]{} & \makebox[1.88cm][c]{} \\
\hline
\noalign{\vskip 4pt}
\multicolumn{7}{l}{\textit{Panel C. Meaningful Tilt}} \\[-1pt]
\noalign{\hrule height 0.15pt}
\noalign{\vskip 2pt}
Direction signal & \makebox[1.88cm][c]{-0.003} & \makebox[1.88cm][c]{-0.027} & \makebox[1.88cm][c]{0.043} & \makebox[1.88cm][c]{-0.003} & \makebox[1.88cm][c]{-0.038} & \makebox[1.88cm][c]{-0.378} \\
 & \makebox[1.88cm][c]{(0.059)} & \makebox[1.88cm][c]{(0.177)} & \makebox[1.88cm][c]{(0.232)} & \makebox[1.88cm][c]{(0.081)} & \makebox[1.88cm][c]{(0.086)} & \makebox[1.88cm][c]{(0.859)} \\[1pt]
\noalign{\hrule height 0.15pt}
\noalign{\vskip 1pt}
N / days & \makebox[1.88cm][c]{20,823} & \makebox[1.88cm][c]{20,823} & \makebox[1.88cm][c]{20,823} & \makebox[1.88cm][c]{71} & \makebox[1.88cm][c]{71} & \makebox[1.88cm][c]{71} \\
Mean wtd. adj. $R^2$ & \makebox[1.88cm][c]{0.025} & \makebox[1.88cm][c]{0.025} & \makebox[1.88cm][c]{0.016} & \makebox[1.88cm][c]{} & \makebox[1.88cm][c]{} & \makebox[1.88cm][c]{} \\
\hline
\noalign{\vskip 4pt}
\multicolumn{7}{l}{\textit{Panel D. Conviction-Weighted Meaningful Tilt}} \\[-1pt]
\noalign{\hrule height 0.15pt}
\noalign{\vskip 2pt}
Direction signal & \makebox[1.88cm][c]{-0.018} & \makebox[1.88cm][c]{-0.025} & \makebox[1.88cm][c]{-0.054} & \makebox[1.88cm][c]{0.020} & \makebox[1.88cm][c]{0.012} & \makebox[1.88cm][c]{0.118} \\
 & \makebox[1.88cm][c]{(0.014)} & \makebox[1.88cm][c]{(0.049)} & \makebox[1.88cm][c]{(0.067)} & \makebox[1.88cm][c]{(0.084)} & \makebox[1.88cm][c]{(0.088)} & \makebox[1.88cm][c]{(0.879)} \\[1pt]
\noalign{\hrule height 0.15pt}
\noalign{\vskip 1pt}
N / days & \makebox[1.88cm][c]{20,823} & \makebox[1.88cm][c]{20,823} & \makebox[1.88cm][c]{20,823} & \makebox[1.88cm][c]{71} & \makebox[1.88cm][c]{71} & \makebox[1.88cm][c]{71} \\
Mean wtd. adj. $R^2$ & \makebox[1.88cm][c]{0.055} & \makebox[1.88cm][c]{0.059} & \makebox[1.88cm][c]{0.045} & \makebox[1.88cm][c]{} & \makebox[1.88cm][c]{} & \makebox[1.88cm][c]{} \\
\hline\hline
\end{tabular}
}
\end{table}


Table \ref{tab:nonoverlap_stockpick_direction_returns} provides an alternative way to address correlation among dependent-variable observations arising from overlapping return windows. In the main text we use Calendar-Time portfolios and daily Fama--MacBeth regressions with Newey-West standard errors to determine whether digital twins' recommendations can predict returns at the ten-day investment horizon. We also use the staggered-vintage design to exclude overlapping return windows. Here, in Table \ref{tab:nonoverlap_stockpick_direction_returns}, we also address the overlapping-window problem by partitioning trading days into non-overlapping blocks, where a block is a pre-defined period of time ranging from two to five trading days. Within each block, we average a stock's directional signal and compute the lagged return and volatility controls. We then use those block variables to predict the same stock's excess return in the next block. This gives a block-level Fama--MacBeth test in which each cross-section uses only information from the prior block and adjacent dependent variables do not reuse the same return days.

\begin{table}[!htbp]
\caption{Non-Overlapping Stock-Pick Direction and Future Excess Returns with Controls}
\label{tab:nonoverlap_stockpick_direction_returns}
{\footnotesize{}This table is a non-overlapping-period robustness companion to Table \ref{tab:q3_stockpick_returns_rv_controls}. Each period-specific cross-section regresses next-period stock excess returns on the prior-period direction signal and two controls: the stock's formation-period excess return and formation-period excess realized volatility. The controls are computed over the same non-overlapping formation block as the signal and are included in the regressions but not reported. The signal is averaged within stock and formation period, then used to predict the same stock's cumulative excess return over the next non-overlapping return period. Unlike Table \ref{tab:q3_stockpick_returns_rv_controls}, whose forward return windows overlap and therefore use Newey--West lags of $h-1$, this design partitions trading dates into non-overlapping blocks; standard errors are therefore Newey--West with lag 0 on the period-specific Fama--MacBeth slope series. Two-day blocks are Mon--Tue, Wed--Thu, and Friday-only blocks with Friday excess log returns multiplied by 2; three-day blocks are Mon--Wed blocks scaled by $2/3$ and Thu--Fri blocks unscaled; five-day blocks run Wednesday through the following Tuesday. Coefficients are percentage-point changes in next-period excess returns for a +10 recommendation-point increase in Tilt or Meaningful Tilt, a +10 percentage-point increase in Net Buy Share, or a +1-unit increase in Conviction-Weighted Meaningful Tilt. Stars use a $T-1$ t-reference, where $T$ is the number of formation-period slopes. N is the number of stock-period observations. The sample uses 429 stocks and 81 stock-pick accounts, with formation periods from December 15, 2025 through March 13, 2026. Indicators ***, **, * denote significance at the 1\%, 5\%, and 10\% level.}
{\footnotesize\par}
\vspace{5pt}
\centering{}
{\footnotesize
\setlength{\tabcolsep}{2.5pt}
\begin{tabular}{lccc}
\hline\hline
 & \makebox[2.85cm][c]{(1)} & \makebox[2.85cm][c]{(2)} & \makebox[2.85cm][c]{(3)} \\
\noalign{\vskip 3pt}
 & \multicolumn{3}{c}{\makebox[8.55cm][c]{Dep. var.: next non-overlapping excess return (pp)}} \\
\cmidrule(lr){2-4}
\noalign{\vskip 2pt}
 & \makebox[2.85cm][c]{\shortstack{2-day\\blocks}} & \makebox[2.85cm][c]{\shortstack{3-day\\blocks}} & \makebox[2.85cm][c]{\shortstack{5-day\\Wed--Tue}} \\
\cmidrule(lr){2-2}\cmidrule(lr){3-3}\cmidrule(lr){4-4}
\noalign{\vskip 4pt}
\multicolumn{4}{l}{\textit{Panel A. Tilt}} \\[-1pt]
\noalign{\hrule height 0.15pt}
\noalign{\vskip 2pt}
Direction signal & \makebox[2.85cm][c]{0.261} & \makebox[2.85cm][c]{0.250} & \makebox[2.85cm][c]{1.105$^{*}$} \\
 & \makebox[2.85cm][c]{(0.218)} & \makebox[2.85cm][c]{(0.198)} & \makebox[2.85cm][c]{(0.536)} \\[1pt]
\noalign{\hrule height 0.15pt}
\noalign{\vskip 1pt}
N & \makebox[2.85cm][c]{14,976} & \makebox[2.85cm][c]{10,293} & \makebox[2.85cm][c]{5,148} \\
Formation periods & \makebox[2.85cm][c]{35} & \makebox[2.85cm][c]{24} & \makebox[2.85cm][c]{12} \\
Mean adj. $R^2$ & \makebox[2.85cm][c]{0.052} & \makebox[2.85cm][c]{0.045} & \makebox[2.85cm][c]{0.034} \\
\hline
\noalign{\vskip 4pt}
\multicolumn{4}{l}{\textit{Panel B. Net Buy Share}} \\[-1pt]
\noalign{\hrule height 0.15pt}
\noalign{\vskip 2pt}
Direction signal & \makebox[2.85cm][c]{0.061} & \makebox[2.85cm][c]{0.060} & \makebox[2.85cm][c]{0.271$^{*}$} \\
 & \makebox[2.85cm][c]{(0.052)} & \makebox[2.85cm][c]{(0.048)} & \makebox[2.85cm][c]{(0.133)} \\[1pt]
\noalign{\hrule height 0.15pt}
\noalign{\vskip 1pt}
N & \makebox[2.85cm][c]{14,976} & \makebox[2.85cm][c]{10,293} & \makebox[2.85cm][c]{5,148} \\
Formation periods & \makebox[2.85cm][c]{35} & \makebox[2.85cm][c]{24} & \makebox[2.85cm][c]{12} \\
Mean adj. $R^2$ & \makebox[2.85cm][c]{0.052} & \makebox[2.85cm][c]{0.045} & \makebox[2.85cm][c]{0.036} \\
\hline
\noalign{\vskip 4pt}
\multicolumn{4}{l}{\textit{Panel C. Meaningful Tilt}} \\[-1pt]
\noalign{\hrule height 0.15pt}
\noalign{\vskip 2pt}
Direction signal & \makebox[2.85cm][c]{0.083} & \makebox[2.85cm][c]{0.095$^{*}$} & \makebox[2.85cm][c]{0.209} \\
 & \makebox[2.85cm][c]{(0.067)} & \makebox[2.85cm][c]{(0.053)} & \makebox[2.85cm][c]{(0.126)} \\[1pt]
\noalign{\hrule height 0.15pt}
\noalign{\vskip 1pt}
N & \makebox[2.85cm][c]{14,525} & \makebox[2.85cm][c]{10,190} & \makebox[2.85cm][c]{5,138} \\
Formation periods & \makebox[2.85cm][c]{35} & \makebox[2.85cm][c]{24} & \makebox[2.85cm][c]{12} \\
Mean adj. $R^2$ & \makebox[2.85cm][c]{0.052} & \makebox[2.85cm][c]{0.044} & \makebox[2.85cm][c]{0.033} \\
\hline
\noalign{\vskip 4pt}
\multicolumn{4}{l}{\textit{Panel D. Conviction-Weighted Meaningful Tilt}} \\[-1pt]
\noalign{\hrule height 0.15pt}
\noalign{\vskip 2pt}
Direction signal & \makebox[2.85cm][c]{0.026} & \makebox[2.85cm][c]{0.026} & \makebox[2.85cm][c]{0.115$^{*}$} \\
 & \makebox[2.85cm][c]{(0.023)} & \makebox[2.85cm][c]{(0.022)} & \makebox[2.85cm][c]{(0.059)} \\[1pt]
\noalign{\hrule height 0.15pt}
\noalign{\vskip 1pt}
N & \makebox[2.85cm][c]{14,525} & \makebox[2.85cm][c]{10,190} & \makebox[2.85cm][c]{5,138} \\
Formation periods & \makebox[2.85cm][c]{35} & \makebox[2.85cm][c]{24} & \makebox[2.85cm][c]{12} \\
Mean adj. $R^2$ & \makebox[2.85cm][c]{0.051} & \makebox[2.85cm][c]{0.045} & \makebox[2.85cm][c]{0.036} \\
\hline\hline
\end{tabular}
}
\end{table}


The results presented in Table \ref{tab:nonoverlap_stockpick_direction_returns} point in the same direction as those presented in Table \ref{tab:q3_stockpick_returns_rv_controls}  and Table \ref{tab:staggered_vintage_stockpick_direction_returns_main} of the main paper. The evidence is strongest in the five-day Wednesday-to-Tuesday specification. In that column, a ten-percentage-point increase in \emph{Net Buy Share} predicts $27.1$ basis points higher next-period excess return. \emph{Tilt} and \emph{Conviction-Weighted Meaningful Tilt} are also positive and statistically significant in the five-day block specification. The results of shorter block specifications are generally positive but do not attain statistical significance at conventional levels. These results show that the stock-pick direction signal is not a large-cap effect and is not driven by the correlation in dependent variables resulting from overlapping daily event windows. The evidence is strongest in equal-weighted stock selection, where smaller and larger stock-events receive comparable weight, and it continues to appear when return windows are forced to be non-overlapping within each block or vintage.


\end{document}